\documentclass[%
reprint,
 citeautoscript,
twocolumn,
 amsmath,amssymb,
 longbibliography,
aip,
floatfix,
]{revtex4-2}

\usepackage{amsmath}
\usepackage{amsthm}
\usepackage{amsfonts}
\usepackage{amssymb}
\usepackage{comment}
\usepackage{graphicx}
\usepackage{numprint}
\usepackage{mathtools}
\usepackage[colorlinks=true,linkcolor=blue,citecolor=blue,urlcolor=cyan,bookmarks,breaklinks=true]{hyperref}
\usepackage{enumitem}
\usepackage{graphicx}
\usepackage{dcolumn}
\usepackage{bm}
\usepackage{xcolor}
\usepackage[font=small,labelfont=bf,
  justification=raggedright,
   format=plain]{caption}
\usepackage{subcaption}

\begin{document}
\title{Designer Pair Statistics of Disordered Many-Particle Systems with Novel Properties}

\begin{abstract}
Knowledge of exact analytical functional forms for the pair correlation function $g_2(r)$ and its corresponding structure factor $S(k)$ of disordered many-particle systems is limited. 
For fundamental and practical reasons, it is highly desirable to add to the existing data base of analytical functional forms for such pair statistics.
Here, we design a plethora of such pair functions in direct and Fourier spaces across the first three Euclidean space dimensions that are realizable by diverse many-particle systems with varying degrees of correlated disorder across length scales, spanning a wide spectrum of hyperuniform, typical nonhyperuniform and antihyperuniform ones. 
This is accomplished by utilizing an efficient inverse algorithm that determines equilibrium states with up to pair interactions at positive temperature that precisely match targeted forms for both $g_2(r)$ and $S(k)$. 
Among other results, we realize an example with the strongest hyperuniform property among known positive-temperature equilibrium states, critical-point systems (implying unusual 1D systems with phase transitions) that are not in the Ising universality class, systems that attain self-similar pair statistics under Fourier transformation, and an experimentally feasible polymer model.
We show that our pair functions enable one to achieve systems with a wide range of translational order and self-diffusion coefficients $\cal D$, which are inversely related to one another. 
One can design other realizable pair statistics via linear combinations of our functions or by applying our inverse procedure to other desirable functional forms.
Our approach facilitates the inverse design of materials with desirable physical and chemical properties by tuning their pair statistics.
\end{abstract}

\author{Haina Wang}
\affiliation{\emph{Department of Chemistry, Princeton University}, Princeton, New Jersey, 08544, USA}
\author{Salvatore Torquato}
\email[]{Email: torquato@princeton.edu}
\affiliation{Department of Chemistry, Department of Physics, Princeton Materials Institute, and Program in Applied and Computational Mathematics, Princeton University, Princeton, New Jersey 08544, USA}
\affiliation{School of Natural Sciences, Institute for Advanced Study, 1 Einstein Drive, Princeton, NJ 08540, USA}

\maketitle
\newpage
\section{Introduction}

The determination of equilibrium and nonequilibrium properties of disordered many-particle systems in $d$-dimensional Euclidean space $\mathbb{R}^d$ based on structural information contained in correlation functions is of fundamental and practical importance in statistical and condensed-matter physics, chemistry, mathematics, biological sciences and materials engineering \cite{Dy62a, Fe98, Va06, Ha86, Ji14, He15, To02a, Hu21}.
The pair correlation function $g_2(\textbf{r})$ and the corresponding structure factor $S(\textbf{k})$ are widely used structural functions due to their computational simplicity and experimental accessibility through diffraction measurements. 
They enable one to determine key equilibrium bulk properties, such as the pressure, the isothermal compressibility \cite{To02a, Ha86} and the two-body excess entropy \cite{Ni21b}.
Moreover, pair statistics are crucial inputs for the determination of transport properties of many-particle systems, including Enskog theories for the self-diffusion coefficient \cite{Br02b, Kr09, Kr09b} and the shear and bulk viscosities \cite{Ki49}, as well as effective temperatures of structural glasses \cite{Gn10} and the time-dependent diffusion spreadability \cite{To21d}.

Despite the importance of such pair functions in direct and Fourier spaces [i.e., $g_2(\textbf{r})$ and $S(\textbf{k})$, respectively], there are only a relatively small number of disordered many-particle systems for which we know exact analytical pair functions over their entire parameter regime.
These include determinantal point processes across space dimensions \cite{Fe98, To08b, Dy62a, So00, Hough09, Cos04, Ja81, Note15}, equilibrium one-dimensional (1D) hard rods \cite{Ze27}, ``ghost'' random sequential addition (RSA) process across dimensions \cite{To06a}, as well as 2D and 3D unit-step function $g_2(\textbf{r})$ at all densities up to the ``terminal packing fraction'' $1/2^d$ \cite{Wa22b}.
\footnotetext[15]{These include $d$-dimensional ground-state free fermions [\citenum{Fe98, To08b}], eigenvalues for certain random matrices [\citenum{Dy62a, So00, Hough09}], 1D Lorentzian $g_2(r)$ [\citenum{Cos04}] and 2D Ginibre ensemble [\citenum{Ja81}].}
Given our current limited knowledge of analytical pair functions, it is highly desirable to expand the existing database of radial functional forms of $g_2(r)$ and $S(k)$ that can be achieved by statistically homogeneous and isotropic disordered  many-particle systems, where $r = |\mathbf{r}|$ and $k = |\mathbf{k}|$.
Substantially increasing such a database would enable one to probe the space of equilibrium and nonequilibrium properties analytically in comprehensive ways not heretofore possible.
Furthermore, adding to such a database would facilitate the inverse design of materials with desirable mechanical, optical and chemical properties, which could be further accelerated by machine learning techniques \cite{Ax22}.

A hypothesized functional form for the pair correlation function $g_2(r)$ is not necessarily realized by a  many-particle system at number density $\rho$, even if the corresponding structure factor $S(k)$ is nonnegative for all $k$, as is should be \cite{To02c, Cos04, To06b}.
This realizability problem has a rich and long history \cite{Ya61, Le73, To02c, Cos04, To06b, Kuna11, La15}, but it is a notoriously difficult problem to solve because an uncountable number of necessary and sufficient conditions are generally required \cite{Kuna11, La15}, which are not possible
to check in practice. 
This places great importance on the need to formulate algorithms to construct particle configurations that realize targeted hypothetical functional forms of the pair statistics \cite{Zh20, Wa22b}.

In this work, we design a plethora of exact functional forms for pair correlations in direct and Fourier spaces across the first three Euclidean space dimension that are realizable by diverse many-particle systems with varying degrees of correlated disorder across length scales with novel structural and physical properties.
As suggested by Zhang and Torquato \cite{Zh20}, we solve this realizability problem here by employing an efficient inverse algorithm \cite{To22} that determines equilibrium states with up to pair interactions $v(r)$ at positive temperature $T$ that precisely match targeted forms for both $g_2(r)$ and $S(k)$.
Our methodology minimizes an objective function that incorporates pair statistics in both direct and Fourier spaces, so that both the small-, intermediate- and large-distance correlations are very accurately captured; see Sec. \ref{meth} for details.

For purposes of illustration, we consider a broad range of functional forms that span across a wide spectrum of hyperuniform and nonhyperuniform classes.
Disordered hyperuniform systems anomalously suppress large-scale density fluctuations compared to typical disordered systems, such as liquids \cite{To03a}. 
A point pattern in $\mathbb{R}^d$ is hyperuniform if the local number variance $\sigma^2(R)=\langle N^2(R) \rangle - \langle N(R) \rangle^2$ of the number of points $N(R)$ found in a spherical window of radius $R$ grows slower than $R^d$ in the large-$R$ limit \cite{To03a, To18a}.
Equivalently, a hyperuniform state is characterized by a structure factor that vanishes in the infinite-wavelength limit, i.e., $\lim_{k\rightarrow 0} S(k) = 0$.
Often, the structure factor of hyperuniform systems takes a radial power-law form in the vicinity of the origin, i.e., $S(k) \sim k^\alpha$, where $\alpha$ is a positive exponent.
The value of $\alpha$ determines different large-$R$ scaling behaviors of the number variance \cite{To03a,Za09,To18a}, i.e., 
\begin{equation}
    \sigma^2(R)\sim \begin{cases}
          R^{d-1}, \quad \alpha > 1 \text{ (class I)} \\
          R^{d-1}\ln R, \quad \alpha=1 \text{ (class II)} \\
          R^{d-\alpha}, \quad 0<\alpha<1 \text{ (class III)} \\
     \end{cases}
    \label{alpha}
\end{equation}
Disordered hyperuniform systems are exotic states of matter that have been receiving great attention because they connect a variety of seemingly unrelated systems that arise in physics, chemistry, materials science, mathematics, and biology \cite{To18a}.
By contrast, for any \textit{nonhyperuniform} system, the local variance has the following large-$R$ scaling behaviors \cite{To21c}:
\begin{equation}
    \sigma^2(R)\sim \begin{cases}
          R^{d} \quad \alpha = 0 \text{ (typical nonhyperuniform)} \\
          R^{d-\alpha} \quad \alpha<0 \text{ (antihyperuniform).} \\
     \end{cases}
\end{equation}
For a ``typical” nonhyperuniform system, $S(0)$ is bounded. 
In antihyperuniform systems, $S(0)$ is unbounded, i.e., $\lim_{k\rightarrow 0}S(k) = +\infty$, and hence are diametrically opposite to hyperuniform systems. 
Typical nonhyperuniform states have bounded nonzero value of $S(k)$ as $k\rightarrow 0$ \cite{To21c}.
Antihyperuniform systems include those at thermal critical points (e.g., liquid-vapor and magnetic critical points) \cite{Wi65,Ka66,Fi67,Wi74,Bi92}, as well as fractals \cite{Ma82} and certain substitution tilings \cite{Og19}.

Table \ref{tab:pairfuncs} lists the target $g_2(r)$ and $S(k)$ functions studied in this work. 
The hyperuniform pair functions that we target include $d$-dimensional Gaussians, $d$-dimensional generalization of the one-component plasma (OCP) and their Fourier duals \cite{Zh20}, $d$-dimensional hyperbolic secant function $g_2(r)$, and a 2D Gaussian-damped polynomial. 
The typical nonhyperuniform functions include a 1D Hermite-Gaussian function, a hypothetical 3D hyposurficial state \cite{To03a}, as well as 1D and 2D ghost RSA packings \cite{To06a}.
These functional forms are chosen for their fundamental significance, their capability to yield systems in which particles form clusters of various sizes, and their potential applications in materials design.
It is noteworthy that the Gaussian functions, the 1D hyperbolic secant function, the 2D Gaussian-damped polynomial and the 1D Hermite-Gaussian function have the property that they are self-similar under Fourier transformations. 
Self-similar pair statistics enable one to exploit duality relations in the phase diagrams of soft-matter systems.
For example, via duality relations, information about low-density classical ground states under short-ranged interactions can be mapped to high-density ground states under long-ranged interactions \cite{To08a, To11b}.
This knowledge facilitates the determination of ground-state structures over large density ranges \cite{To11b}, which may provide good solutions to {\it covering}
and {\it quantizer} problems that arise in discrete geometry and have many applications \cite{To10d}, including wireless communication network layouts.

\newcommand{\speciallabel}[1]{
\refstepcounter{equation} \label{#1}
}

\begin{table*}[htp]
\refstepcounter{table}\label{tab:pairfuncs}

\caption*{Table \thetable. The analytical target pair statistics studied here and shown to be realizable by statistically homogeneous and isotropic disordered many-particle systems
in various spatial dimensions. The small-$k$ scaling exponents $\alpha$ in Eq. (\ref{alpha}) are shown. 
All states are at $\rho = 1$. The definitions of the special functions and symbols in the pair statistics are given in the table footnotes.}
\centering
    \begin{tabular}{|m{3cm}|m{0.4cm}|m{6cm}m{0.5cm}|m{6cm}m{0.5cm}|m{0.5cm}|}
    \hline
    Functional form & $d$ & $g_2(r)$ & & $S(k)$ & & $\alpha$ \\
    \hline
    Gaussian & 1, 2, 3 & $1 - \exp(-\pi r^2)\speciallabel{gaussiang2}$ & (\theequation) & $1 - \exp(-\frac{k^2}{4\pi}) \speciallabel{gaussianS}$ & (\theequation) & 2 \\
    \hline
    Generalized OCP\cite{pfq} & 3 & $1 - \exp(-\frac{4\pi}{3}r^3) \speciallabel{ocpg23D}$ & (\theequation) & $\begin{aligned}[t]
    &1 -\, _1F_4\left(1;\frac{1}{3},\frac{2}{3},\frac{5}{6},\frac{7}{6};-\frac{k^6}{20736 \pi ^2}\right) \\
    &+\frac{2 \pi \text{ber}_{\frac{2}{3}}\left(\frac{\left(k^2\right)^{3/4}}{3 \sqrt{\pi }}\right)}{3\sqrt{3}} \\ 
    &+ \frac{2 \pi \text{ber}_{-\frac{2}{3}}\left(\frac{\left(k^2\right)^{3/4}}{3 \sqrt{\pi }}\right)}{3 \sqrt{3}} \speciallabel{ocpS3D}
    \end{aligned}$ & (\theequation) & 2 \\
    \hline
    Fourier dual of OCP & 1 & $1 - \frac{1}{\pi r^2 + 1} \speciallabel{dualocpg21D}$ & (\theequation) & $1 - \exp(-k/\pi) \speciallabel{dualocpS}$ & (\theequation) & 1 \\
    \hline
    Fourier dual of OCP\cite{pfq} & 3 & $\begin{aligned}[t]
    &1 -\, _1F_4\left(1;\frac{1}{3},\frac{2}{3},\frac{5}{6},\frac{7}{6};-\frac{1}{324} \pi ^4 r^6\right) \\
    &+\frac{2 \pi  \text{ber}_{\frac{2}{3}}\left(\frac{2}{3} \sqrt{2} \pi  \left(r^2\right)^{3/4}\right)}{3\sqrt{3}} \\&+\frac{2\pi \text{ber}_{-\frac{2}{3}}\left(\frac{2}{3} \sqrt{2} \pi  \left(r^2\right)^{3/4}\right)}{3 \sqrt{3}}
    \speciallabel{ocpdualg22D} \end{aligned}$ & (\theequation) & $1 - \exp(- \frac{2}{3(2\pi)^2} k^3) \speciallabel{ocpdualS3D}$ & (\theequation) & 3 \\
    \hline
    Hyperbolic secant $g_2(r)$ & 1 & $1 - \operatorname{sech}(\pi r) \speciallabel{sech1Dg2}$ & (\theequation) & $1 - \operatorname{sech}(k/2) \speciallabel{sech1DS}$ & (\theequation) & 2 \\
    \hline
    Hyperbolic secant $g_2(r)$\cite{catalan} & 2 & $1 - \operatorname{sech}(2\sqrt{\pi G} r) \speciallabel{sech2Dg2}$ & (\theequation) & No closed analytical form & & 2 \\
    \hline
    Hyperbolic secant $g_2(r)$ & 3 & $1 - \operatorname{sech}(\frac{\pi^{4/3}}{2^{1/3}}r) \speciallabel{sech3Dg2}$ & (\theequation) & $\displaystyle 1-\frac{2^{2/3} \sqrt[3]{\pi } \tanh \left(\frac{k}{2^{2/3} \sqrt[3]{\pi }}\right) \operatorname{sech}\left(\frac{k}{2^{2/3} \sqrt[3]{\pi }}\right)}{k}\speciallabel{sech3DS}$ & (\theequation) & 2 \\
    \hline
    Gaussian-damped polynomial & 2 & $\displaystyle 1 - \frac{1}{3} e^{-\pi  r^2} \left(4 \pi ^2 r^4 - 8 \pi  r^2 + 3\right) \speciallabel{polygaussian2Dg2}$ & (\theequation) & $\displaystyle 1-\frac{e^{-\frac{k^2}{4 \pi }} \left(k^4-8 \pi  k^2+12 \pi ^2\right)}{12 \pi ^2}\speciallabel{polygaussian2DS}$ & (\theequation) & 2 \\
    \hline
    Hermite-Gaussian\cite{lambda} & 1 & $1 - \lambda \left(4 r^4-12 r^2+3\right) e^{-\frac{r^2}{2}} \speciallabel{hermiteg2}$ & (\theequation) & $1 - \lambda \sqrt{2 \pi }\left(-4 k^4+12 k^2-3\right) e^{-\frac{k^2}{2}} \speciallabel{hermiteS}$ & (\theequation) & 0 \\
    \hline
    Hyposurficial\cite{xx} & 3 & $\displaystyle 1 + \frac{\exp(-r^*)}{4\pi}  - \frac{\exp(-r^*)\sin(r^*)}{r^*} \speciallabel{hypog2}$ & (\theequation) & $\displaystyle \frac{6k^{*8} + 12k^{*6} + 19k^{*4} + 24k^{*2} + 16}{6(k^{*2} + 1)^2(k^{*2} - 2k^* + 2)(k^{*2} + 2k^* + 2)}\speciallabel{hypoS}$ & (\theequation) & 0\\
    \hline
    Ghost RSA\cite{cisi} & 1 & $\displaystyle\frac{2 \Theta (2 r-1)}{2-(1-r) \Theta (2-2 r)}\speciallabel{ghost1Dg2}$ & (\theequation) & $1 - 4\cos(k) \left[\text{Ci}\left(\frac{3 k}{2}\right) + \text{Ci}(2 k)\right] - 4\sin(k) \left[\text{Si}\left(\frac{3 k}{2}\right) +  \text{Si}(2 k)\right] - \frac{2 \sin (k)}{k}\speciallabel{ghost1DS}$ & (\theequation) & 0 \\
    \hline
    Ghost RSA\cite{rstar} & 2 & $\displaystyle\frac{2\Theta(r^* - 1)}{2 - \frac{2}{\pi}\left[\cos^{-1}\left(\frac{r^*}{2}\right) - \frac{r}{2}\sqrt{1-\frac{r^{*2}}{4}}\right]\Theta(2 - r^*)}\speciallabel{ghost2Dg2}$ & (\theequation) & No closed analytical form &  & 0 \\
    \hline
    Antihyperuniform\cite{bigd} & 1, 2, 3 & $\displaystyle \Theta(r/D - 1)\left(\frac{A}{(r/D)^{d-\frac{1}{2}}} + 1\right) \speciallabel{ahu1Dg2}$ & (\theequation) & See Appendix &  & $-\frac{1}{2}$ \\
    \hline
    \end{tabular}
\end{table*}

We also, for the first time, demonstrate the realizability of antihyperuniform systems across the first three space dimensions corresponding to critical points that do not belong to the Ising universality class, including a remarkable 1D critical-point state, which is to be contrasted with typical 1D systems in which phase transitions cannot exist.
Note that the pair functions in Table \ref{tab:pairfuncs} possess analytical forms in both direct and Fourier spaces in all but two cases, the 2D hyperbolic secant $g_2(r)$ and the 2D ghost RSA packing, for which $g_2(r)$ are known analytically but $S(k)$ must be numerically evaluated to high precision.
The reasons for choosing the targets in Table \ref{tab:pairfuncs} are detailed in Sec. \ref{res}.



The functional forms of our pair statistics and their corresponding potentials can be used to model soft-core polymers with varying magnitudes of short- and long-range repulsive forces [Eqs. (\ref{gaussiang2})--(\ref{ocpdualS3D}), (\ref{hypog2})--(\ref{hypoS})], ``sticky'' macromolecules with oscillatory interactions [Eqs. (\ref{polygaussian2Dg2})--(\ref{hermiteS})], as well as polymer-grafted nanoparticles with both hard- and soft-core interactions [Eqs. (\ref{ghost1Dg2})--(\ref{ghost2Dg2})].
Notably, we discover that one of our models possesses a stronger hyperuniform property (i.e., larger exponent $\alpha$) than any other positive-temperature equilibrium state known previously.
To illustrate that our procedure can be used to guide experimental polymer design, we also study a case of prescribed pair statistics whose functional forms are motivated by the polymer reference interaction site model (PRISM) \cite{Sc97, Ya04} and show that they can be realized.
It is important to note that one can realize many designed pair functions at different densities and temperatures other than those explicitly shown in Table \ref{tab:pairfuncs}, either by creating linear combinations of our pair functions or by applying our inverse procedure to other desirable functional forms.
We further show that our pair functions exhibit distinctly different translational order metric $\tau$ (defined in Sec. \ref{trans}) \cite{To15} and self-diffusion coefficients $\mathcal{D}$, and that $\tau$ and $\mathcal{D}$ are inversely related for our 3D models.

Section \ref{meth} provides a comprehensive description of our inverse methodology.
In Sec. \ref{res}, we present our results for the target and optimized pair statistics and the corresponding effective potentials.
In Sec. \ref{hypothetical}, we probe the realizability of the prescribed polymer pair statistics motivated by the PRISM model.
In Sec. \ref{trans}, we show that the translational order and the self-diffusion coefficients are inversely correlated to one another for our 3D designs.
In Sec. \ref{disc}, we discuss the implications of our work for outstanding problems in statistical mechanics and for materials design.

\section{Inverse Methodology}
\label{meth}
\begin{figure}
    \centering
    \includegraphics[width=90mm]{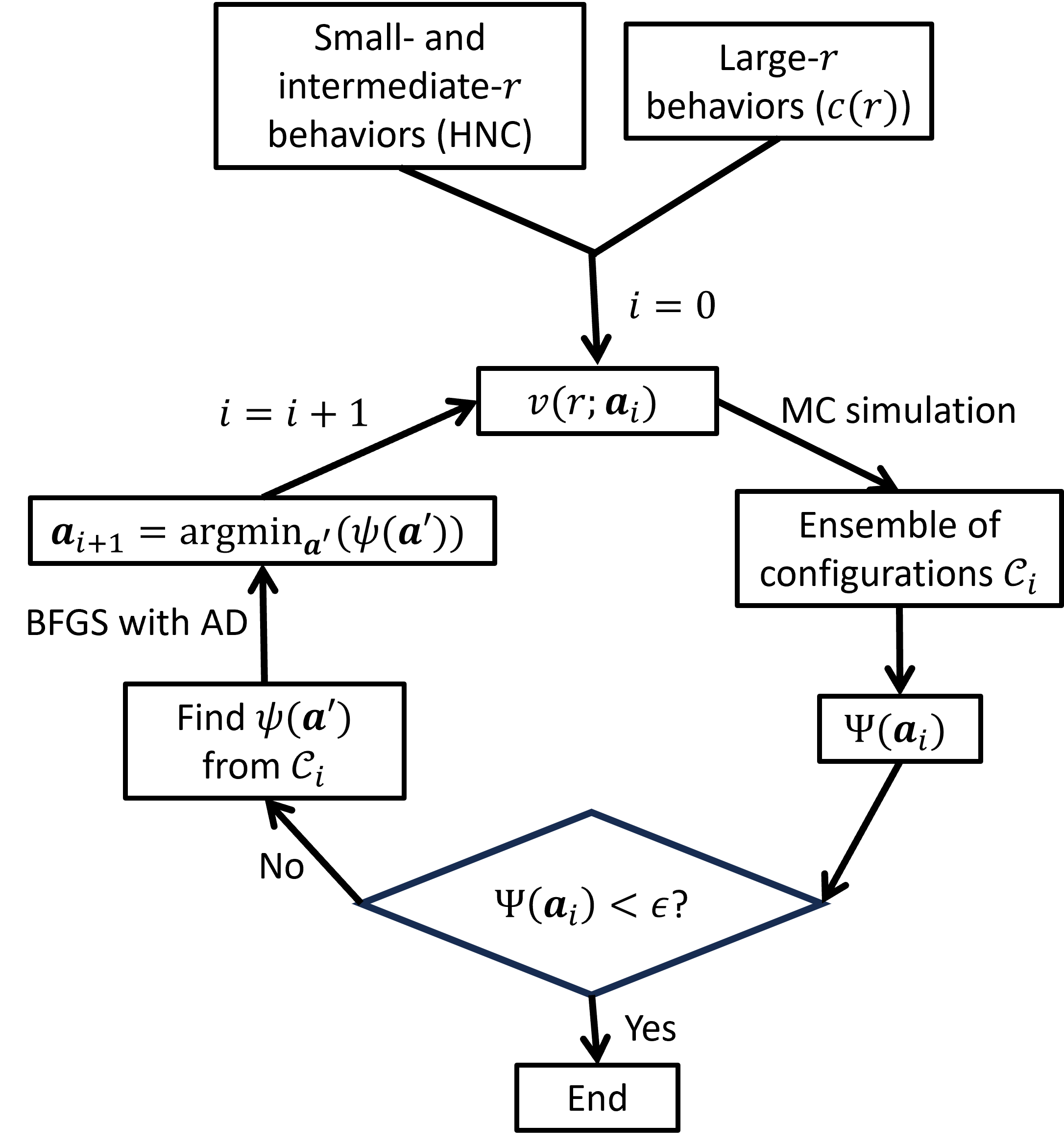}
    \caption{Flowchart of the inverse methodology developed in this work, assuming homogeneous target system.}
    \label{flowchart}
\end{figure}

To determine effective one- and two-body interactions from prescribed pair statistics, we employ an accelerated version of an inverse algorithm that we developed recently \cite{To22}, which enables one to recover the unique pair potential corresponding to given pair statistics, as dictated by Henderson's theorem \cite{He74}.
Our current methodology improves upon the original one \cite{To22} by incorporating automatic differentiation \cite{Ne10b} and canonical ensemble reweighing \cite{No08b} techniques, achieving significantly faster convergence of potentials.
A flowchart of the the algorithm is shown in Fig. \ref{flowchart}.


The methodology uses a parametrized potential $v(r;\mathbf{a})$, whose initial functional form is informed by the small- and large-distance behaviors of the target pair statistics \cite{To18a}. 
The scalar components $a_j$ of the potential parameter vector $\mathbf{a}$ are of three types: dimensionless energy scales $\varepsilon_j$, dimensionless distance scales $\sigma_j$ and dimensionless phases $\theta_j$ \cite{To22}; see Appendix B for the potential parameters in each case.
To obtain an initial form of the small- and intermediate-$r$ behavior of $v(r;\mathbf{a})$, we numerically fit the hypernetted chain (HNC) approximation for the target pair statistics; see details in Ref. [\citenum{To22}].
The large-$r$ behavior of $v(r;\mathbf{a})$ is given by the target direct correlation function $c(r)$ under mild conditions \cite{Ha86, To22}.
For hyperuniform targets, the large-$r$ asymptotic form of $v(r) \sim -c(r)$ is given by Eq. (\ref{v_long_hu}) \cite{To18a}. 
That is, the potential is long-ranged in the sense that its volume integral is unbounded. 
Thus, one requires a neutralizing background one-body potential to maintain stability \cite{To18a, Ha73, Ga79, Dy62a}, and the Ewald summation technique \cite{Ew21} is used to compute the total potential energy.
For nonhyperuniform targets, the volume integral of $c(r)$ is bounded and one only requires two-body interactions.

Once the initial form of $v(r;\mathbf{a})$ is chosen, the low-storage BFGS algorithm \cite{Liu89} is used to minimize an objective function $\Psi({\bf a})$ based on the distance between target and trial pair statistics in both direct and Fourier spaces:
\begin{equation}
\begin{split}
\Psi(\mathbf{a}) =\rho\int_{\mathbb{R}^d}
 w_{g_2}({\bf r})\left(g_{2,T}({\bf r})-g_{2}({\bf r};\mathbf{a})\right)^2
d{\bf r} \\
+ \frac{1}{\rho(2\pi)^d}\int_{\mathbb{R}^d} w_{S}({\bf k})\left(S_T({\bf k})-S(k;\mathbf{a})\right)^2 d\mathbf{k},
\label{Psi}
\end{split}
\end{equation}
where $g_{2,T}(\mathbf{r})$ and $S_T(\mathbf{k})$ are target pair statistics, $w_{g_2}({\bf r})$ and $w_{S}({\bf k})$ are weight functions, and $g_{2}({\bf r};\mathbf{a})$ and $S(\mathbf{k};\mathbf{a})$ are equilibrated pair statistics under $v(r;\mathbf{a})$ obtained from Monte-Carlo simulations.
As done in our previous works\cite{To22, Wa22b, Wa23}, we use Gaussian weight functions $w_{g_2}(r) = \exp[-(r/4)^2]$ and $w_S(k) = \exp[-(k/2)^2]$. 

The BFGS algorithm requires the gradient of the objective function $\Psi(\mathbf{a})$.
In this work, we replace the finite difference method in Ref. [\citenum{To22}] by automatic differentiation (AD), a technique that quickly generates precise gradients on input parameters of computer programs by repeatedly applying the chain rule \cite{Ne10b}. 
AD has the advantage that its computational complexity for gradients is independent of the number of parameters \cite{Ne10b}. 
Because standard implementations of AD \cite{Re16b} can only compute gradient for deterministic functions, but $\Psi(\mathbf{a})$ depends on stochastically simulated pair statistics, one requires a deterministic function $\psi(\mathbf{a})$ that approximates  $\Psi(\mathbf{a})$ in the vicinity of given $\mathbf{a}$. 
In this work, we define $\psi(\mathbf{a})$ via the canonical ensemble reweighing technique \cite{No08b}  described below.

Let $\mathcal{C}_i$ be a set of configurations equilibrated under $v(r;\mathbf{a}_i)$ drawn from a simulated canonical ensemble in iteration $i$, and let $\mathbf{a}'$ be a perturbed potential parameter in the vicinity of $\mathbf{a}_i$. 
The ensemble-averaged pair statistics under $v(r;\mathbf{a}')$ are estimated by reweighing the pair statistics of each configuration in $\mathcal{C}_i$ by a Boltzmann factor involving the difference in total configurational energies $\Phi(\mathbf{r}^N) = \sum_{i<j}^N v(|\mathbf{r}_i - \mathbf{r}_j|)$ under $v(r;\mathbf{a}_i)$ and $v(r;\mathbf{a}')$ \cite{No08b}, i.e.,
    \begin{equation}
        \varphi(x; \mathbf{a}') = \frac{\sum_{\mathbf{r}^N \in \mathcal{C}_i} \varphi(x; \mathbf{r}^N) \exp[-(\Phi(\mathbf{r}^N; \mathbf{a}') - \Phi(\mathbf{r}^N; \mathbf{a}_i))/T] }{\sum_{\mathbf{r}^N \in \mathcal{C}_i} \exp[-(\Phi(\mathbf{r}^N; \mathbf{a}') - \Phi(\mathbf{r}^N; \mathbf{a}_i))/T]},
        \label{reweigh}
    \end{equation}
where $\varphi(x; \mathbf{r}^N)$ is $g_2(r)$ or $S(k)$ computed from each configuration $\mathbf{r}^N$ in $\mathcal{C}_i$. 
The approximate objective function $\psi(\mathbf{a}')$ at the perturbed parameters $\mathbf{a}'$ can then be computed by inserting (\ref{reweigh}) into (\ref{Psi}).
For a fixed set of simulated configurations $\mathcal{C}_i$, $\psi(\mathbf{a}')$ is a deterministic function of $\mathbf{a}'$. 
Thus, one can apply BFGS optimization with AD-generated gradients to find the optimal parameters $\mathbf{a}_{i+1}$ that minimize $\psi(\mathbf{a}')$.
In the next iteration, simulations are performed under $v(r;\mathbf{a}_{i+1})$ to obtain a new set of configurations $\mathcal{C}_{i+1}$. 
The iterations repeat until $\Psi(\mathbf{a}_i)$ computed from $\mathcal{C}_i$ is smaller than some given convergence threshold $\epsilon$, set to be $10^{-3}$ in this study.
If convergence is not achieved, a different set of basis functions is chosen and the optimization process is repeated. 
Note that this inverse methodology accelerates our original algorithm \cite{To22} as it requires just one simulation per iteration for $n$ potential parameters, as opposed to $2n$ simulations using finite differences.
For systems away from the critical point, convergence can be achieved within 5 hours on an Intel 2.8 GHz CPU with 15 cores and 4GB memory per core.

For systems near the critical point, common inverse methods based on predictor-corrector approaches, such as IBI \cite{So96}, do not perform well due to critical slowing down and the large number of iterations required for the trial potential to converge \cite{To22}. 
IBI updates a binned trial potential $v(r)$  based on the difference between the target and simulated $g_2(r)$ in each iteration.
By contrast, our inverse method uses an analytical potential form dictated by statistical-mechanical theory.
This form is robust to variations in the large-$r$ behavior of the simulated $g_2(r)$, which can cause undesirable long-ranged oscillation in $v(r)$ in the IBI approach. 
Thus, the potential in our method converges in much fewer iterations.
For the 1D antihyperuniform state, which is the most challenging target among the ones studied here, convergence was achieved within 24 hours.

\section{Results for Designed Pair Correlation Functions}
\label{res}
In this section, we describe the targeted analytical forms
for pair statistics in direct and Fourier spaces in various space dimensions summarized in Table \ref{tab:pairfuncs} and then show their realizability by equilibrium states with up to pair interactions via the inverse algorithm described in Sec. \ref{meth}.
For each target, we show that the optimized pair statistics precisely match the prescribed functions. 
The initial functional forms of the effective potentials are chosen based on the small- and large-$r$ behaviors of the target pair statistics, as dictated by statistical mechanical theory \cite{To22}.
The asymptotic decay rate of the direct correlation function $c(r)$, which is linked to the total correlation function $h(r) = g_2(r) - 1$ via the Ornstein-Zernike equation \cite{Or14}, determines whether one needs both one-body and two-body interactions (for hyperuniform states) or only two-body interactions (for nonhyperuniform states); see Sec. \ref{meth} for details.
Note that certain pair interactions for hyperuniform states are not square integrable, but the total energy is nevertheless bounded due to the background one-body interaction that maintains overall charge neutrality \cite{To18a, To22}.
Unless otherwise stated, we take $\rho = 1$, and the lengths are expressed in units of $\rho^{-1/d}$. 
We also set the reduced temperature $k_BT$ to be unity for all the equilibrium states, where $k_B$ is the Boltzmann constant.

\subsection{$d$-dimensional Hyperuniform Gaussian Pair Statistics}
\label{sec:gaussian}
We consider $d$-dimensional generalizations of Gaussian pair statistics given by Eqs. (\ref{gaussiang2}) and (\ref{gaussianS}) in Table \ref{tab:pairfuncs}.
These targets are hyperuniform with $\alpha = 2$ in Eq. (\ref{alpha}).
Note that the pair functions are self-similar in under Fourier transformations up to scaling, and are independent of the dimensionality.
Gaussian pair statistics incorporate soft-core short-ranged repulsion found in macromolecules, and thus are commonly used to model polymer systems \cite{Fl58, Ko83, Ya04, Be10}.
The Gaussian pair functions represent the low-density, high-temperature limit of the pair statistics for the popular Gaussian-core model \cite{St76, Lo00b, Za08, Kr09b}, which is known to exhibit anomalous phase behaviors such as reentrent melting \cite{Pr05, Za08}.

\begin{figure*}
    \centering
    \subfloat[]{\includegraphics[width = 60mm]{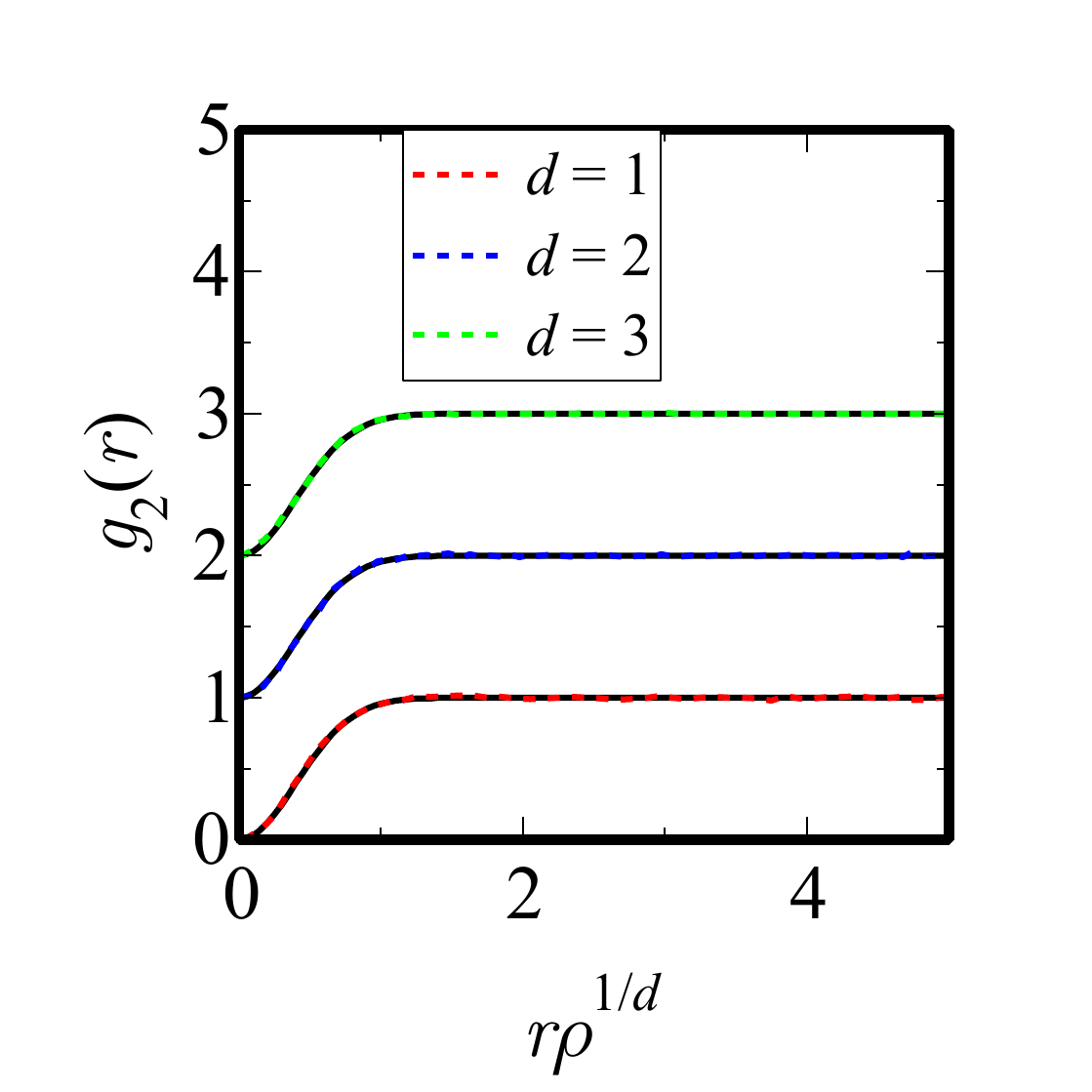}}
    \subfloat[]{\includegraphics[width = 60mm]{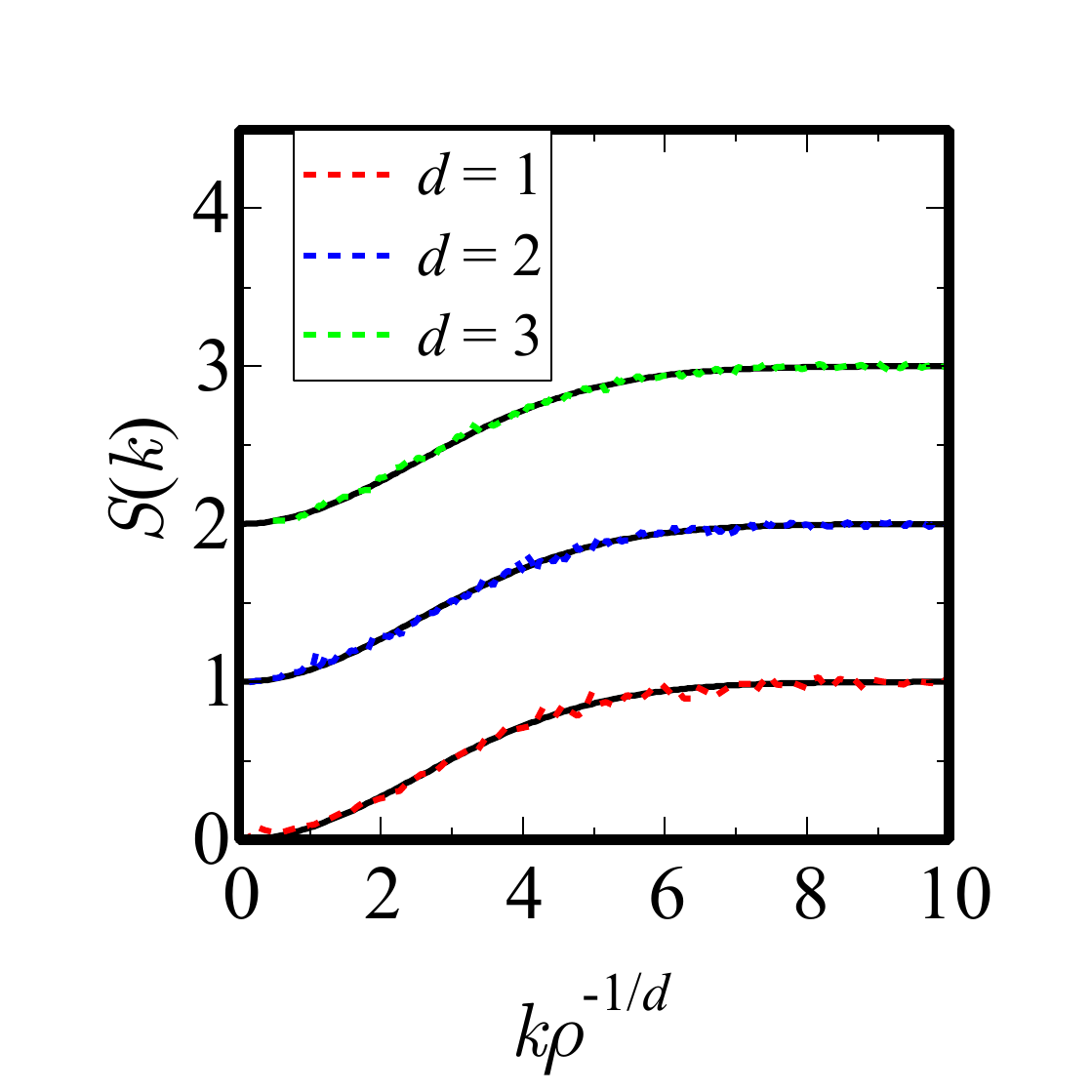}}

    \subfloat[]{\includegraphics[width = 60mm, trim={4cm -1cm 4cm 0cm},clip]{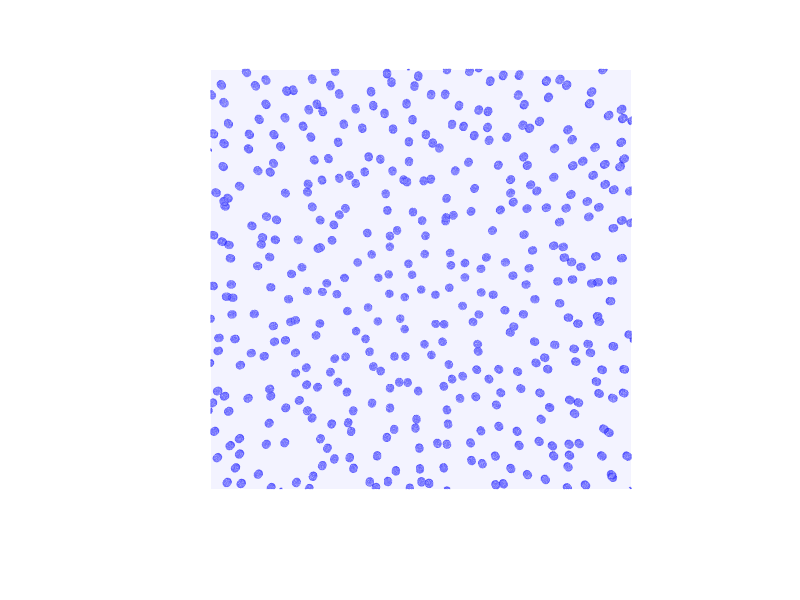}}
    \subfloat[]{\includegraphics[width = 72mm, trim={4cm -3cm 4cm 0cm},clip]{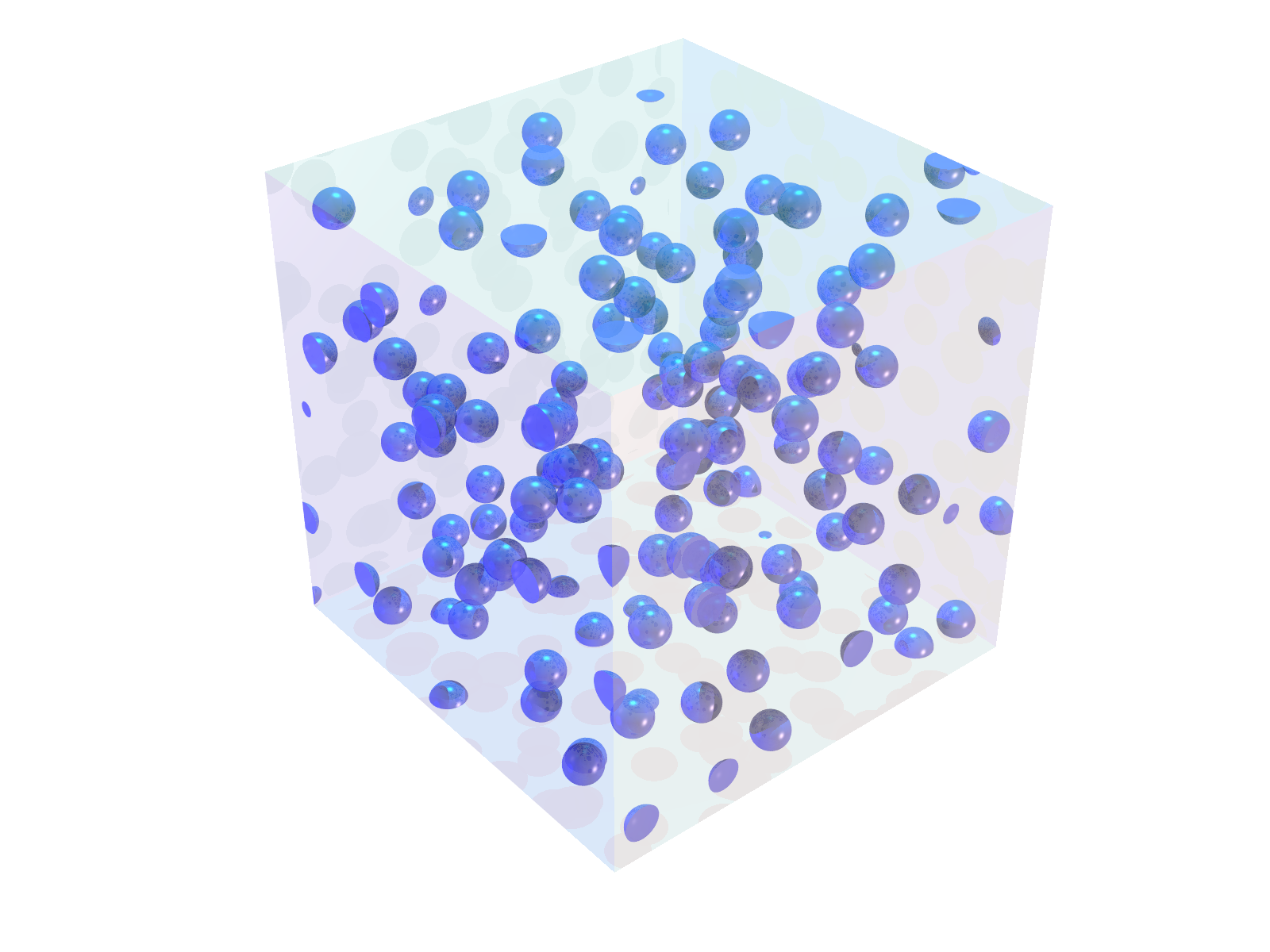}}
    \caption{(a)--(b) Target (solid curves) and optimized (dashed curves) $g_2(r)$ and $S(k)$ for Gaussian pair statistics. For clarity, the pair statistics for $d=2$ and $d=3$ are shifted up by 1 and 2, respectively. (c) A 400-particle configuration of the 2D equilibrium state with Gaussian pair statistics.
    (d) A 125-particle configuration of the 3D equilibrium state with Gaussian pair statistics.
    The point particles are shown as spheres for visualization purposes.}
    \label{fig:gaussian}
\end{figure*}

Figure \ref{fig:gaussian}(a) and (b) show the target and optimized $g_2(r)$ and $S(k)$, respectively, for the hyperuniform Gaussian pair statistics in one, two and three dimensions.
It is clear that the target and optimized pair statistics agree very well with one another.
The precise form of the corresponding effective potential is given in the Appendix and plotted in Fig. \ref{fig:potentials}(a).
Torquato \cite{To18a} showed that for hyperuniform targets, the long-ranged asymptotic form of $v(r)$ is given by 
\begin{equation}
    v(r)\sim 
    \begin{cases}
    r^{-(d-\alpha)}, \quad d \ne \alpha\\
    -\ln(r), \quad d = \alpha,
    \end{cases}
    \label{v_long_hu}
\end{equation}
which can be regarded as a generalized Coulombic interaction of ``like-charged'' particles.
For $\alpha = 2$, the long-ranged parts of the potentials are given by the Coulomb interactions $v(r)\sim r^{-(d-2)}$ in the respective dimensions.
Note that for the 2D Gaussian pair statistics, it is known that the effective pair potential is exactly given by the 2D Coulombic interaction $v(r) = -2\ln(r)$ \cite{Ja81}. 
Our inverse procedure has accurately recovered this potential.

Figure \ref{fig:gaussian}(c) and (d) show representative configurations of the 2D and 3D equilibrium states with Gaussian pair statistics.
Note that while particles can get arbitrarily close, both configurations do not contain large clusters or large spherical holes that are void of particles.
Indeed, the particles tend to form short chains that are well separated.
These observations are consistent with the soft repulsive interactions in the effective potentials.

\subsection{$d$-dimensional Hyperuniform Generalized OCP and Their Fourier Duals}
\label{sec:ocp}
We consider 3D generalizations \cite{Zh20} of the hyperuniform OCP given by Eqs. (\ref{ocpg23D})--(\ref{ocpS3D}) as well as the 1D and 3D Fourier duals of OCP given by Eqs. (\ref{dualocpg21D})--(\ref{ocpdualS3D}) in Table \ref{tab:pairfuncs}.
A OCP is an equilibrium system of identical point particles of charge $e$ interacting via the Coulomb potential and immersed in a rigid uniform background of opposite charge to ensure overall charge neutrality \cite{Ja81}.
OCPs are classical analogs of the quantum jellium model and are used as as a first approximation to dense plasmas \cite{Ba80}.
The pair statistics for the 2D OCP, which is equivalent to the Ginibre ensemble, are exactly given by the Gaussian functions (\ref{gaussiang2}) and (\ref{gaussianS}) \cite{Ja81}.

Zhang and Torquato \cite{Zh20} proposed a generalization of the OCP pair correlation function to arbitrary dimensions, given by
\begin{equation}
    g_2(r) = 1 - \exp[-v_1(r)],
    \label{ocpg2}
\end{equation}
where $v_1(r)$ is the volume of a $d$-dimensional sphere of radius $r$.
The corresponding structure factors at $\rho = 1$ are hyperuniform with $\alpha = 2$.
For $d=1$, the generalized OCP pair statistics are identical to a Lorenztian target
\begin{equation}
    g_2(r) = 1 - \exp(-2r),
\end{equation}
which has been shown to be realizable by a determinantal point process \cite{Cos04}.
Thus, we study the 3D case here, whose realizability is unknown.

For the $d$-dimensional Fourier duals of OCP, the structure factors assume the same form as the OCP pair correlation function (\ref{ocpg2}) up to scaling, i.e.,
\begin{equation}
    S(k) = 1 - \exp[-v_1(k/2\pi)],
\end{equation}
for which the exponent in Eq. (\ref{alpha}) is given by $\alpha = d$.
This implies that $S(k)$ is nonanalytic at the origin for any odd dimension, and thus the corresponding $g_2(r)$ has power-law asymptotic $r^{-2d}$ \cite{To22}.

\begin{figure*}
    \centering
    \subfloat[]{\includegraphics[width = 55mm]{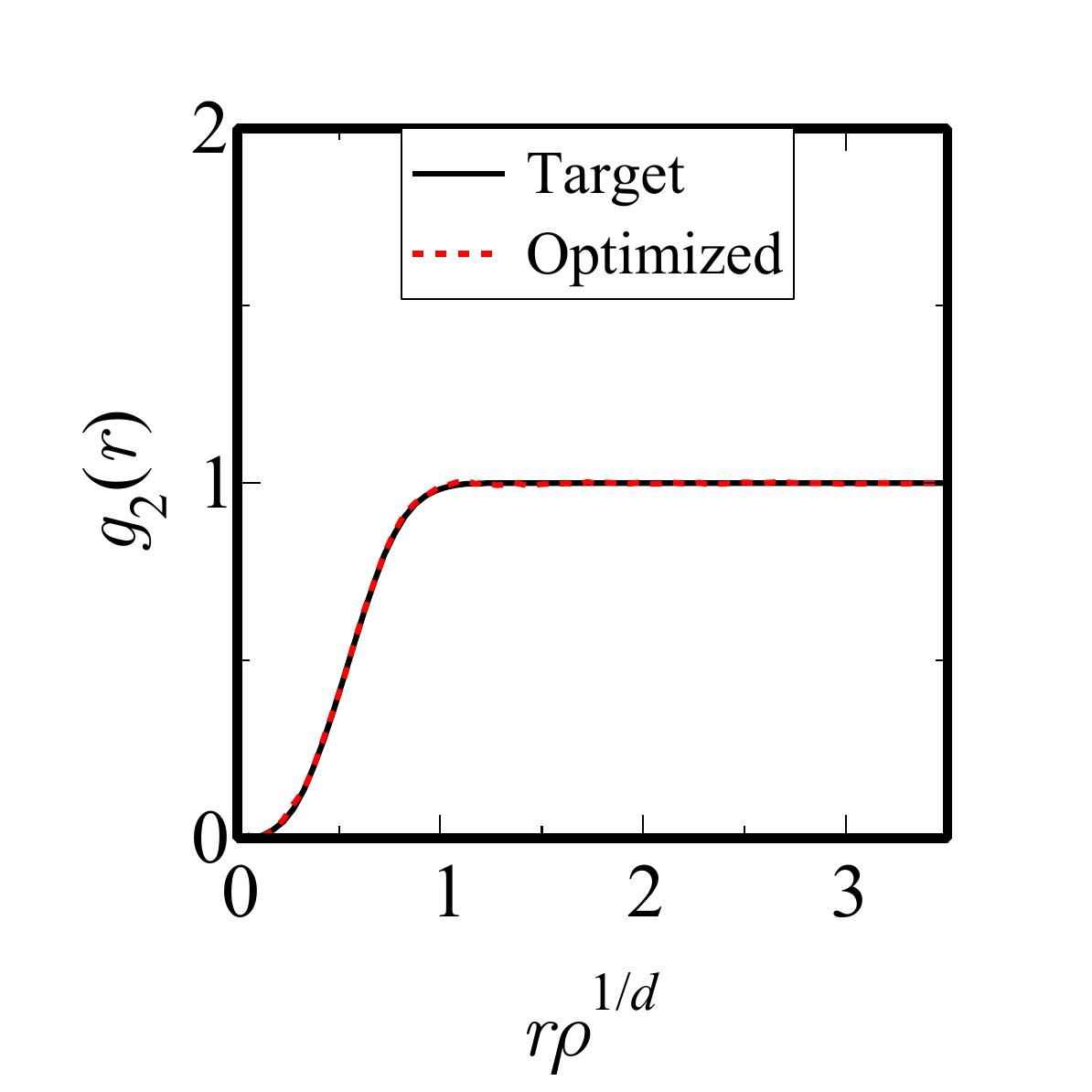}}
    \subfloat[]{\includegraphics[width = 55mm]{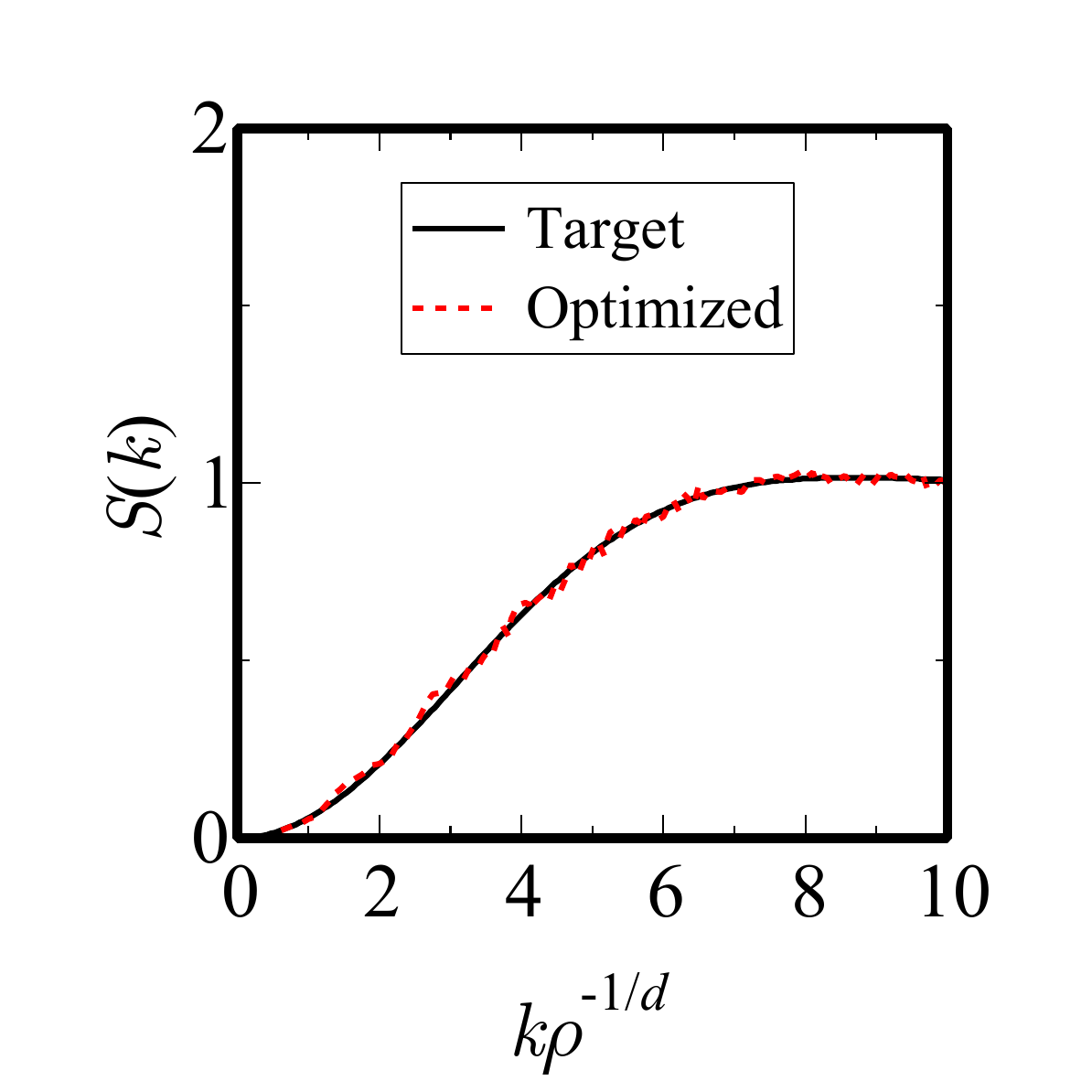}}
    \subfloat[]{\includegraphics[width = 65mm, trim={8cm -0cm 4cm 0cm},clip]{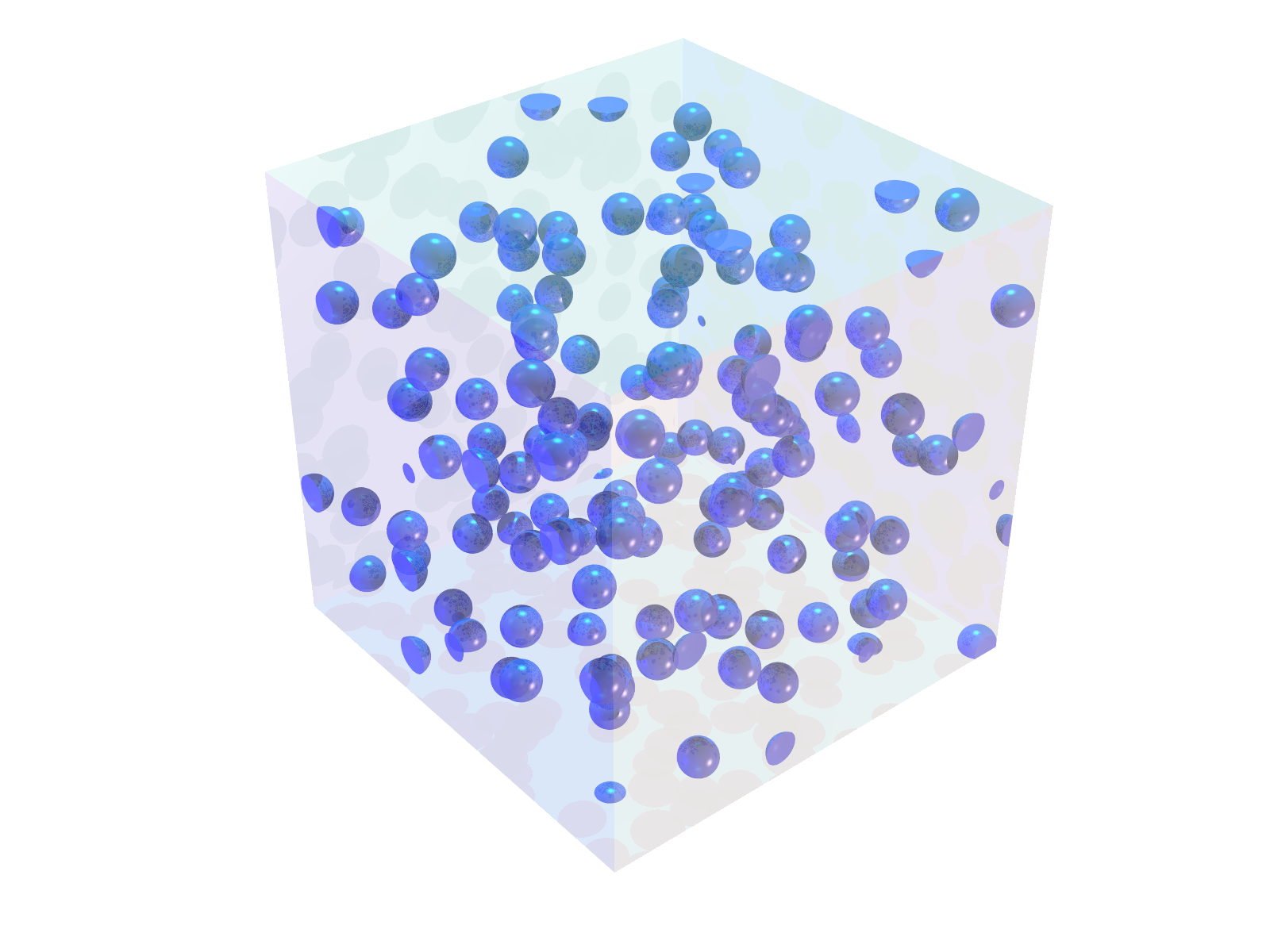}}
    
    \subfloat[]{\includegraphics[width = 55mm]{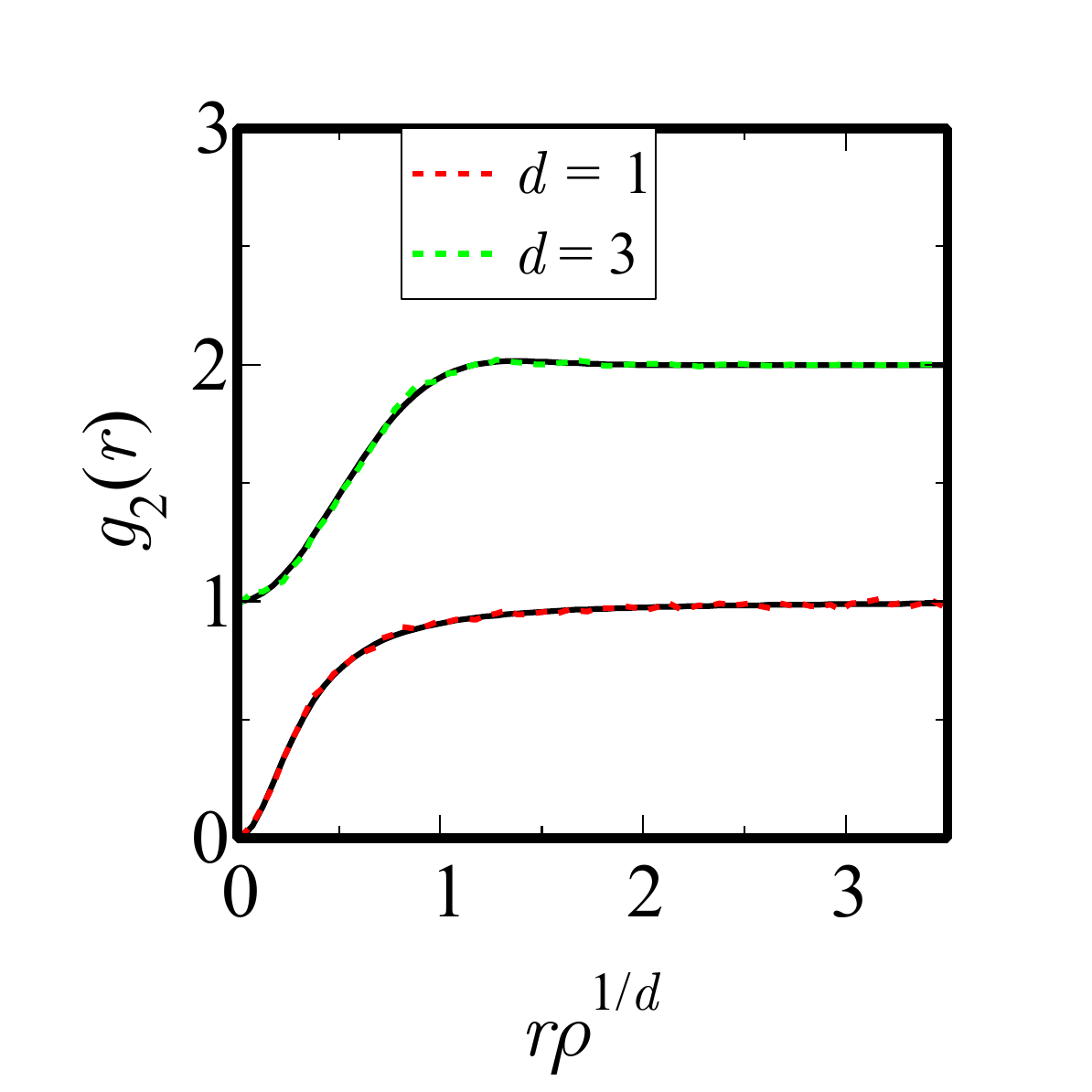}}
    \subfloat[]{\includegraphics[width = 55mm]{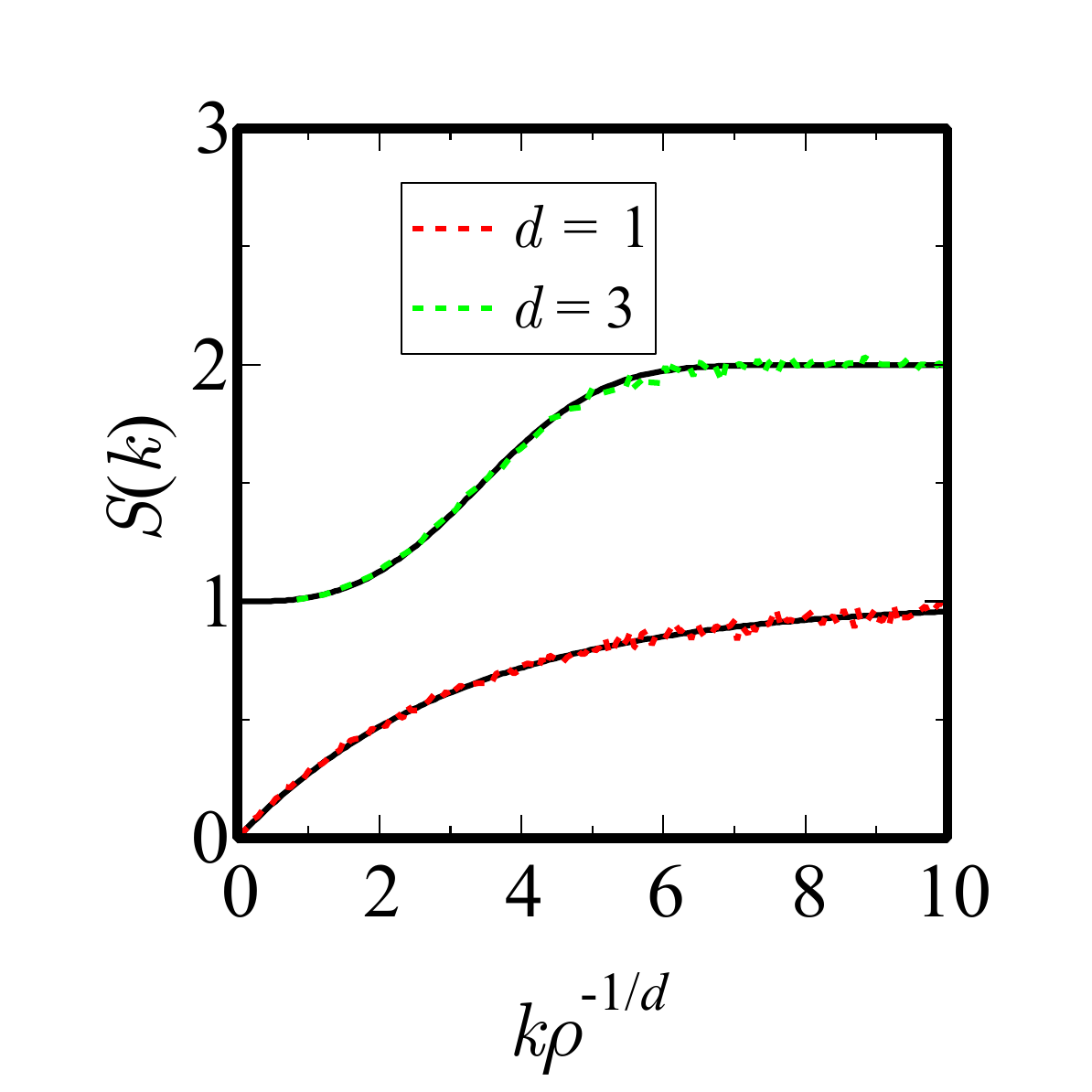}}
    \subfloat[]{\includegraphics[width = 65mm, trim={8cm -0cm 4cm 0cm},clip]{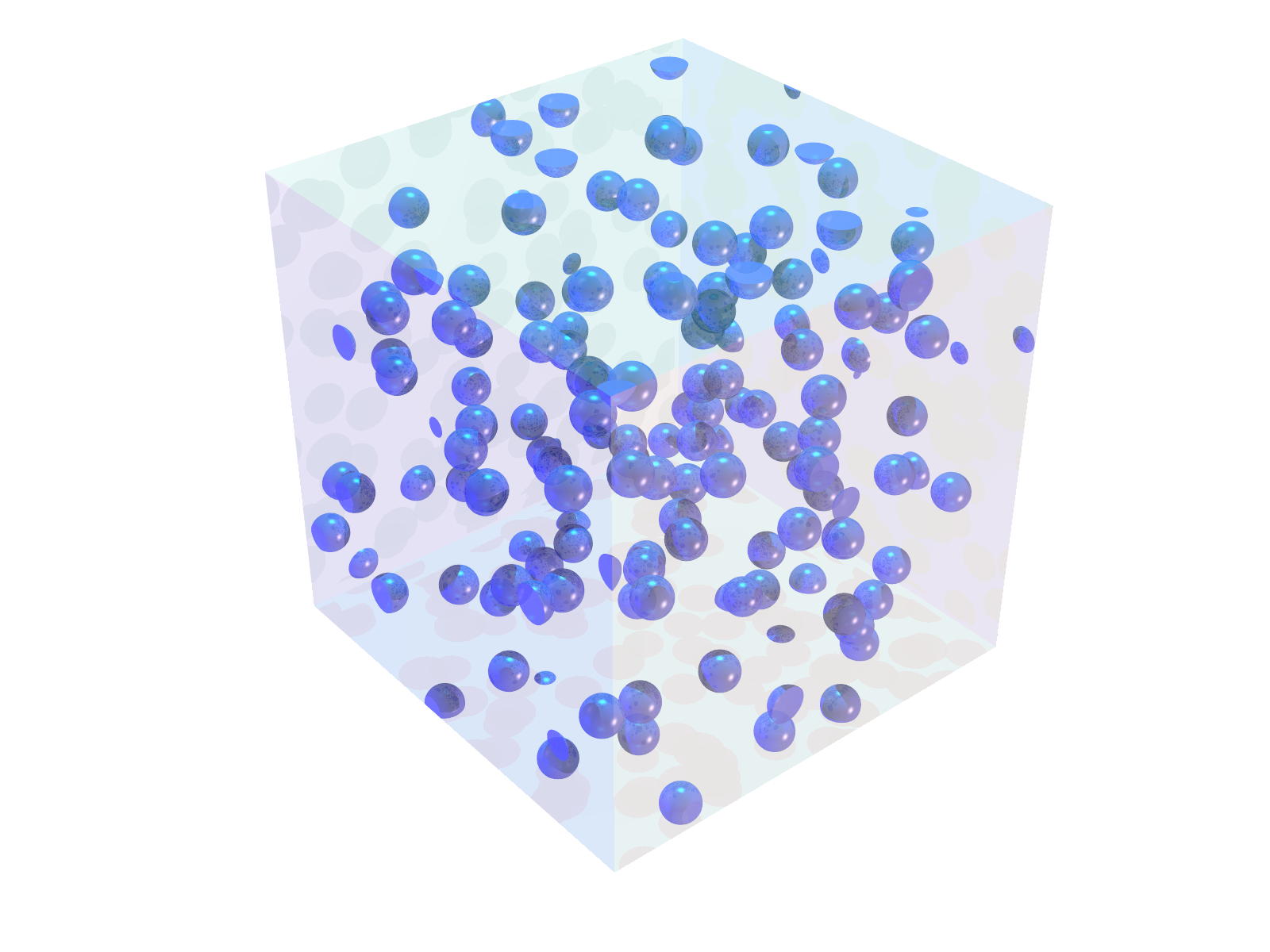}}
    \caption{(a)--(b) Target and optimized $g_2(r)$ and $S(k)$ for 3D generalized OCP. 
    (c) A 125-particle configuration of the 3D equilibrium state with the generalized OCP pair statistics.
    (d)--(e) Target (solid curves) and optimized (dashed curves) for 1D and 3D Fourier dual of OCP. For clarity, the pair statistics for the 3D case are shifted up by 1.
    (f) A 125-particle configuration of the 3D equilibrium state with the Fourier dual OCP pair statistics.
    In all 3D configurations, the point particles are shown as spheres for visualization purposes.}
    \label{fig:ocp}
\end{figure*}

We achieved excellent agreement between target and optimized pair statistics for both OCP and its Fourier dual.
Figure \ref{fig:ocp}(a) and (b) show the target and optimized pair statistics for the 3D OCP. 
The corresponding configuration [Fig. \ref{fig:ocp}(c)] is similar to that for the 3D Gaussian pair statistics, as both systems have soft-core repulsive interactions.
However, the OCP has the larger force magnitudes as $r\rightarrow 0$, as shown in Fig. S1 in the Supplemental Information (SI), which plots the forces $-d v(r)/dr$ corresponding to the effective potentials for our 3D hyperuniform models.
The difference in the small-$r$ repulsion is due to the fact that the OCP $g_2(r)$ increases as $r^d$ in the vicinity of the origin, rather than $r^2$ for the Gaussian. 
Our capacity to design systems with varying magnitudes of short-range repulsive forces enables coarse-grained modeling of polymers with different levels of chain rigidity and penetrability \cite{Mu02, Sl22}.

Figure \ref{fig:ocp}(d)--(f) present the results for the Fourier dual of OCP in the first three spatial dimensions. 
Note that the exponent $\alpha$ in (\ref{alpha}) for the simulated 3D Fourier dual of OCP, i.e., $\alpha = 3$, is larger than that of any other positive-temperature equilibrium states known previously.
The largest known exponent prior to this work was $\alpha = 2$ for systems under Coulombic interactions \cite{To18a}.
Our result here shows that  states with $\alpha > 2$ can be attained by pair potentials that are longer-ranged than the $d$-dimensional Coulombic interactions.
While it is experimentally challenging to create nanoparticles with such long-ranged molecular interactions, they are hypothesized to be attainable by topological defects in the gauge fields in the form of space-time instantons \cite{Kr04, Kr05}.
Hyperuniform systems with large $\alpha$ are desirable for optical purposes, because they tend to be effectively transparent for a wide range of wave numbers compared to nonhyperuniform ones \cite{To21a}.
We note further that despite having strong large-$r$ repulsion, particles in the 3D Fourier dual of OCP are less repulsive at small $r$ than those in the 3D OCP, as shown in Fig. S1.

\subsection{$d$-dimensional Hyperuniform Hyperbolic Secant Function $g_2(r)$}
We consider the hyperuniform states characterized by the hyperbolic secant function $g_2(r)$ given by Eqs. (\ref{sech1Dg2})--(\ref{sech3DS}) in Table \ref{tab:pairfuncs}.
As in the case of the Gaussian pair statistics, these are soft-core models and the structure factors scale as $k^2$ at small $k$. 
Moreover, in the 1D case, $g_2(r)$ and $S(k)$ are self-similar up to scaling. 
We have chosen this target for its capability to model polymers with softer repulsive interactions than the Gaussian core \cite{Fa17} and to generate configurations with a higher degree of clustering than the Gaussian pair statistics.
A hyperbolic secant probability distribution is \textit{leptokurtic}, i.e., it has smaller probability density around the mean but has a heavier tail than the corresponding Gaussian distribution with the same mean and variance \cite{Fa17}.
For a hyperuniform state at unit number density, the hyperbolic secant $g_2(r)$ rises more rapidly at small $r$ but decay to unity more slowly at large $r$ compared to the Gaussian $g_2(r)$.
Thus, we propose that hyperbolic secant pair functions and other leptokurtic functions \cite{Fa17} can be applied to describe polymer systems with ``ultrasoft'' particles, such as star polymers \cite{Li98b, St00}.

\begin{figure*}
    \centering    
    \subfloat[]{\includegraphics[width = 55mm]{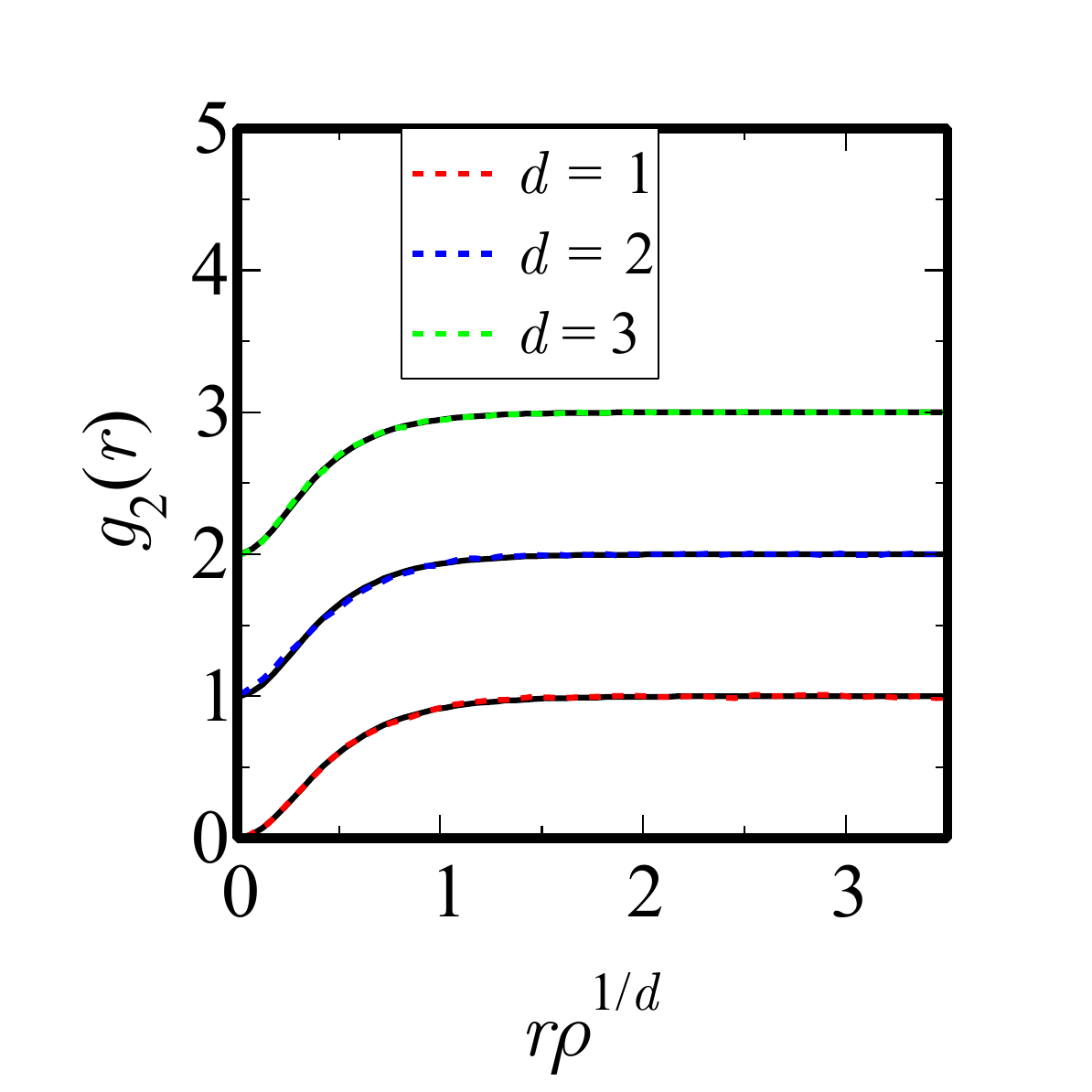}}
    \subfloat[]{\includegraphics[width = 55mm]{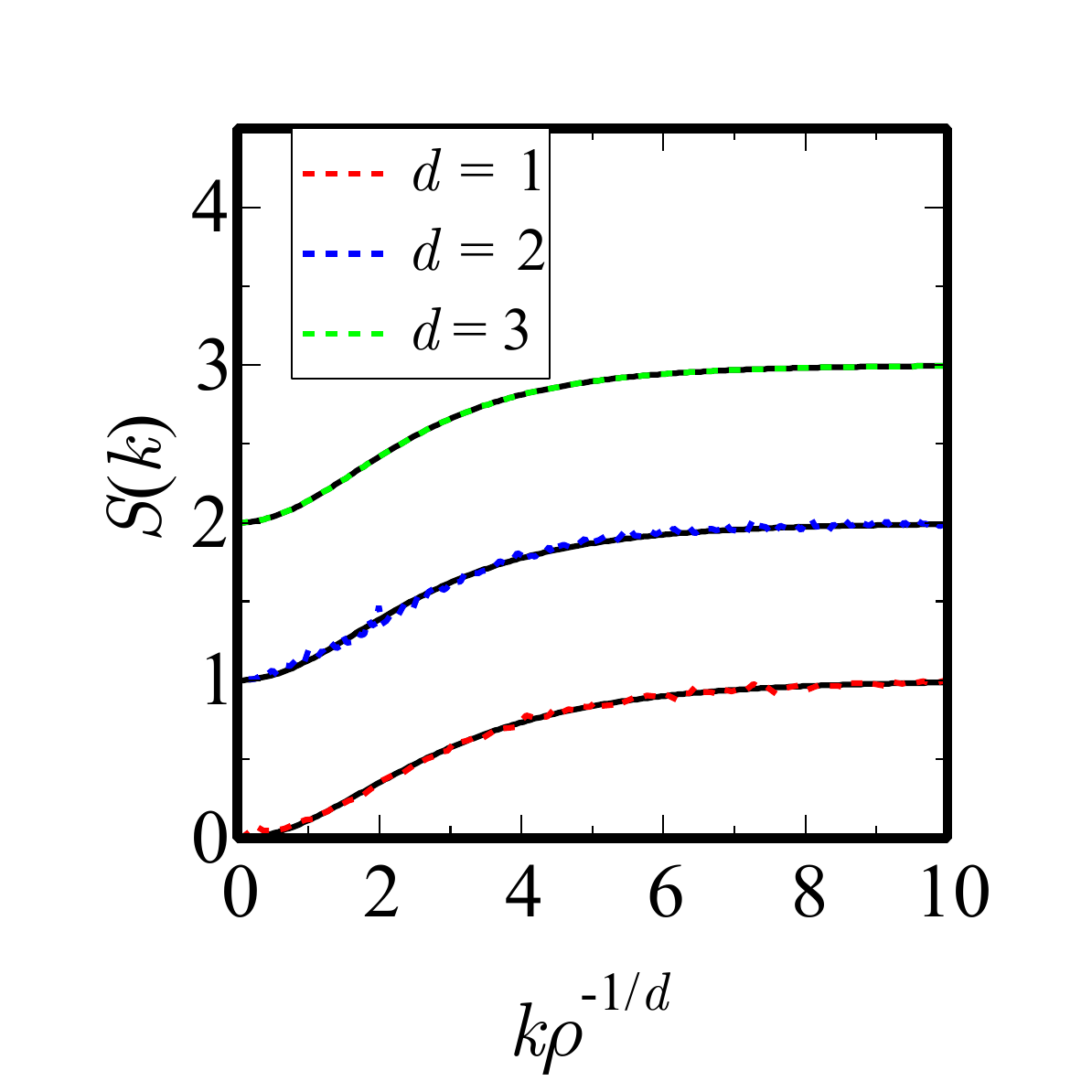}}
    \subfloat[]{\includegraphics[width = 65mm, trim={8cm 0cm 4cm 0cm},clip]{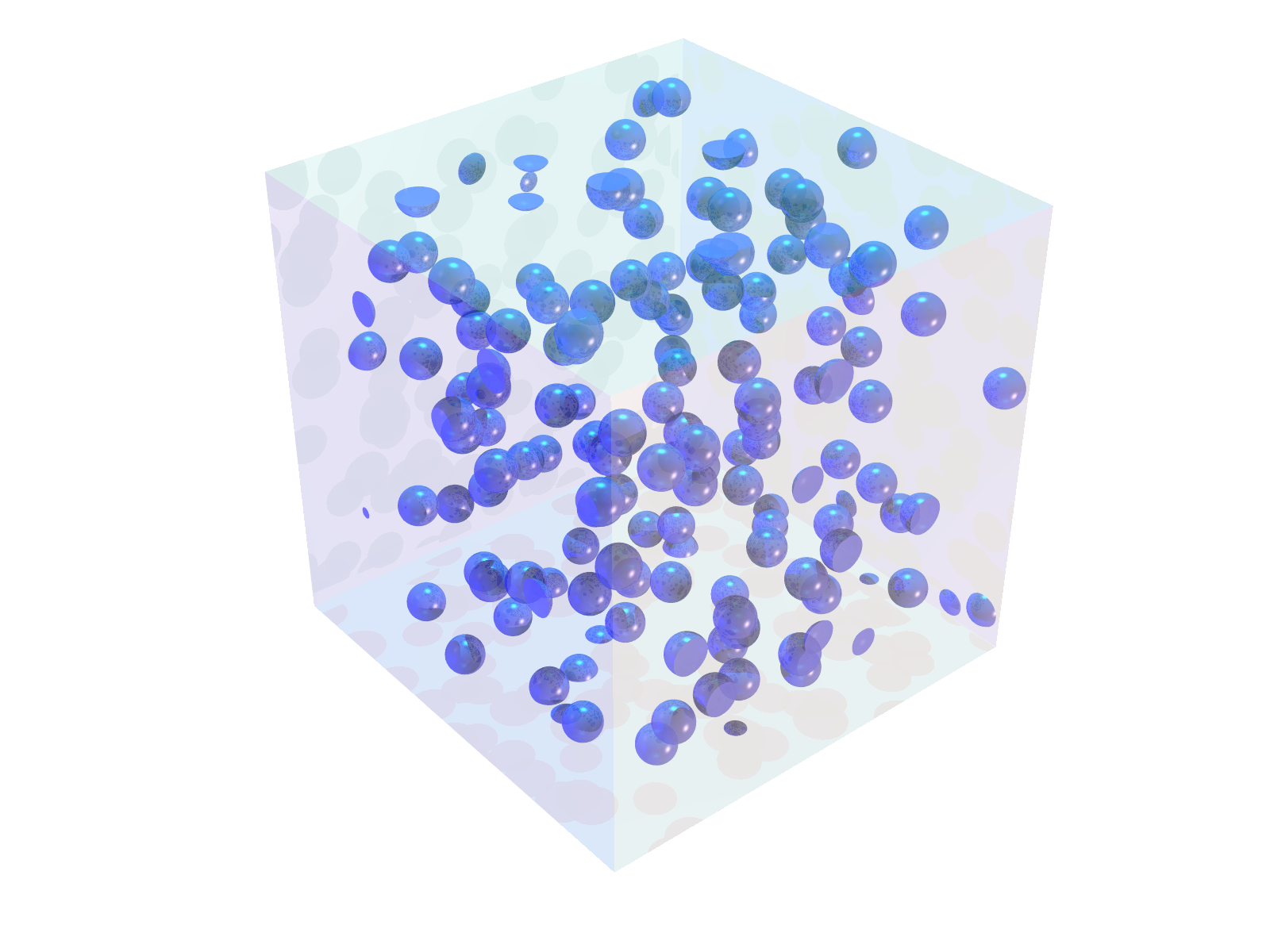}}
    \caption{(a)--(b) Target (solid curves) and optimized (dashed curves) for 1D--3D hyperbolic secant $g_2(r)$. For clarity, the pair statistics for $d=2$ and $d=3$ are shifted up by 1 and 2, respectively. 
    (c) A 125-particle configuration of the 3D equilibrium state with hyperbolic secant $g_2(r)$.
    The point particles are shown as spheres for visualization purposes.}
    \label{fig:sech}
\end{figure*}

Figure \ref{fig:sech} (a) and (b) show the target and optimized pair statistics for the hyperbolic secant $g_2(r)$.
Because these pair functions imply softer core interactions than those for the Gaussian pair statistics, the corresponding 3D configuration [Fig. \ref{fig:sech}(c)] contains more clusters with four or more closely spaced particles with pair distances $r\leq 0.4\rho^{-1/3}$ compared to the Gaussian case.
Clusters in latter system are primarily dimer and trimer; see Fig. \ref{fig:gaussian}(d).
Cluster statistics can be quantified by mapping the point configurations to a two-phase medium generated by decorating each point particle by a sphere of a certain radius and then using theoretical tools from percolation theory \cite{To88c, To02a, Xu05}.
Figure S1 shows that the effective potential for the hyperbolic secant $g_2(r)$ is the least repulsive at small $r$ among the 3D hyperuniform models.
As remarked in Sec. \ref{sec:ocp}, softer interactions are useful in describing polymers molecules with high penetrability.

\subsection{2D Hyperuniform Gaussian-Damped Polynomial and 1D Nonhperuniform Hermite-Gaussian Pair Statistics}
\begin{figure*}
    \centering
    \subfloat[]{\includegraphics[width = 55mm]{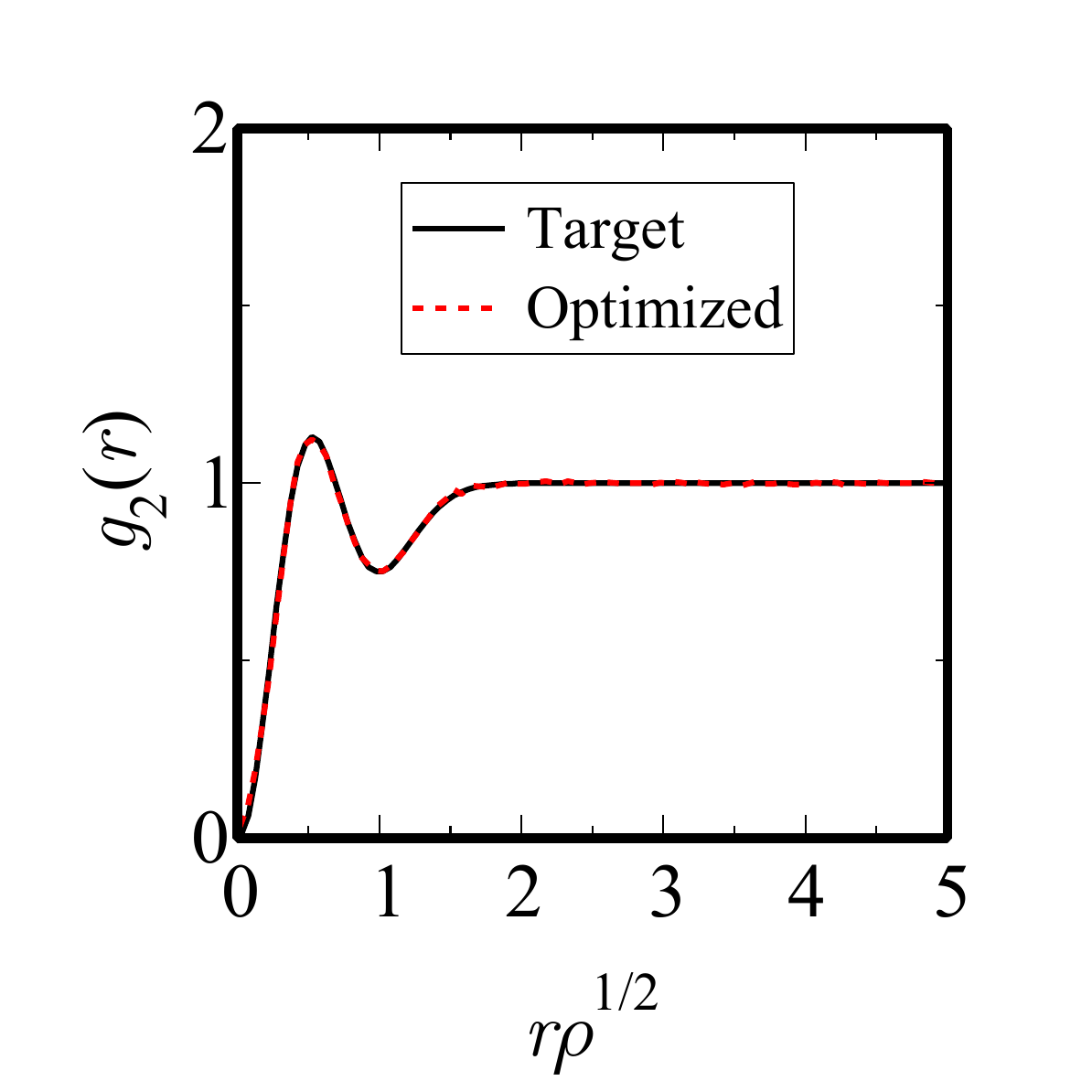}}
    \subfloat[]{\includegraphics[width = 55mm]{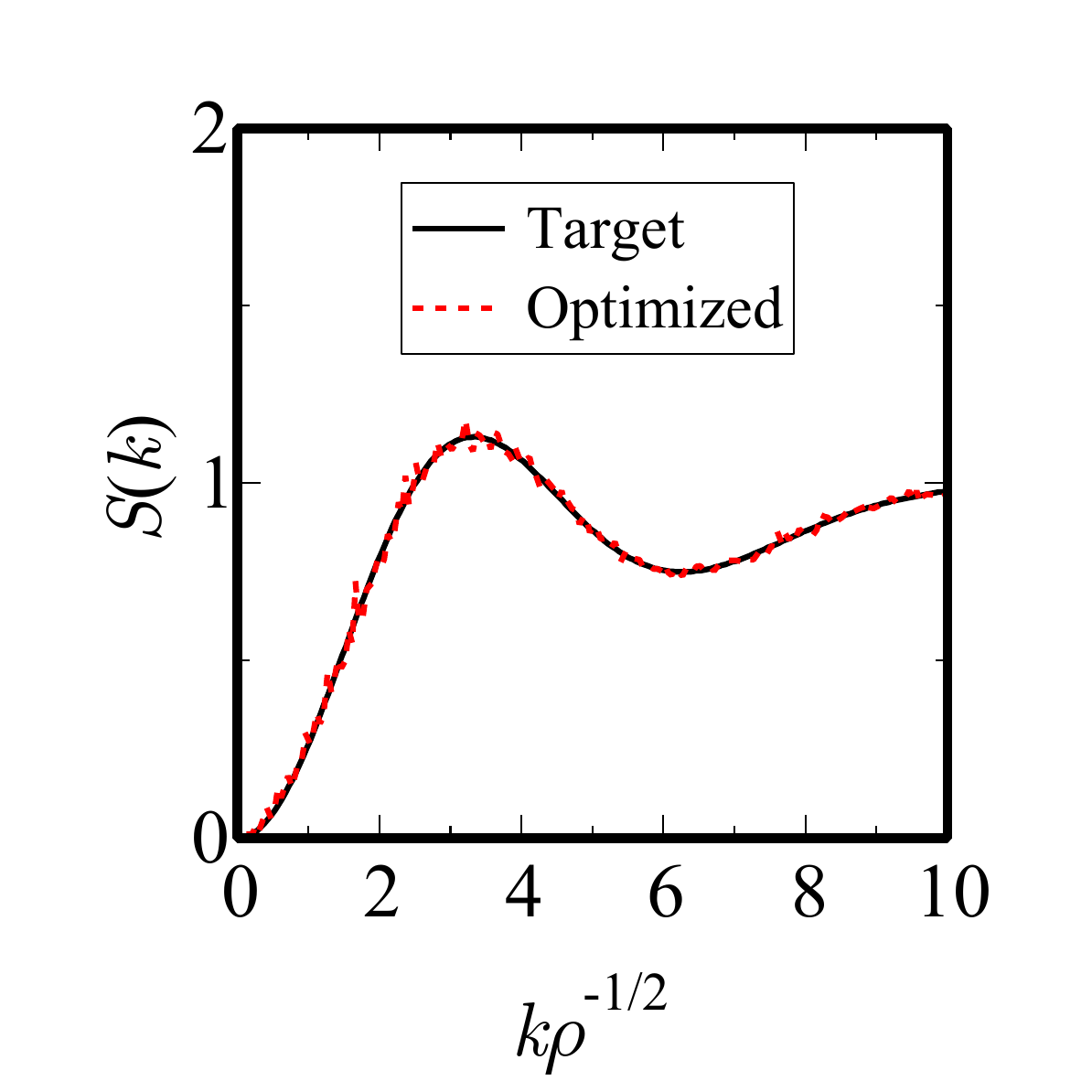}}
    \subfloat[]{\includegraphics[width = 55mm, trim={5cm 0cm 4cm 0cm},clip]{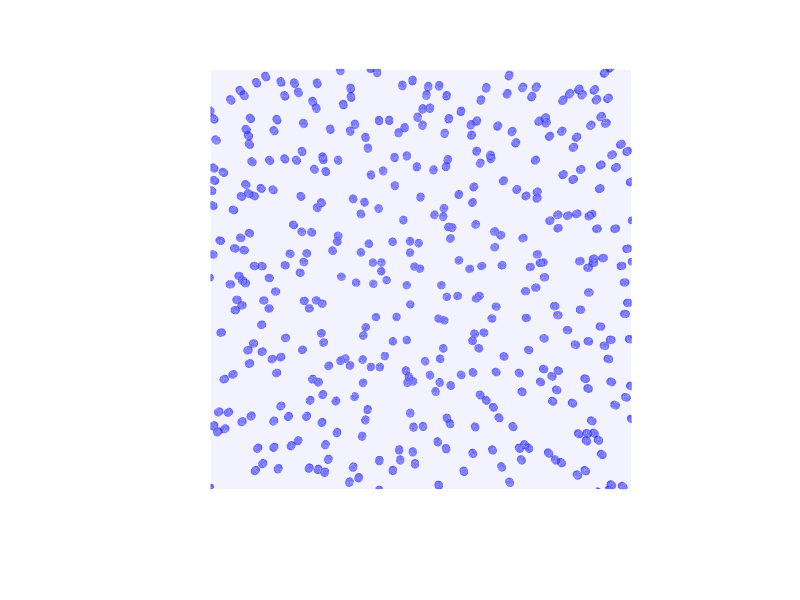}}
    \caption{(a)--(b) Target and optimized $g_2(r)$ and $S(k)$ for the 2D Gaussian-damped polynomial pair statistics.
    (c) A 400-particle configuration of the 2D equilibrium state with the Gaussian-damped polynomial pair statistics.}
    \label{fig:poly2D}
\end{figure*}

\begin{figure*}
    \centering
    \subfloat[]{\includegraphics[width = 55mm]{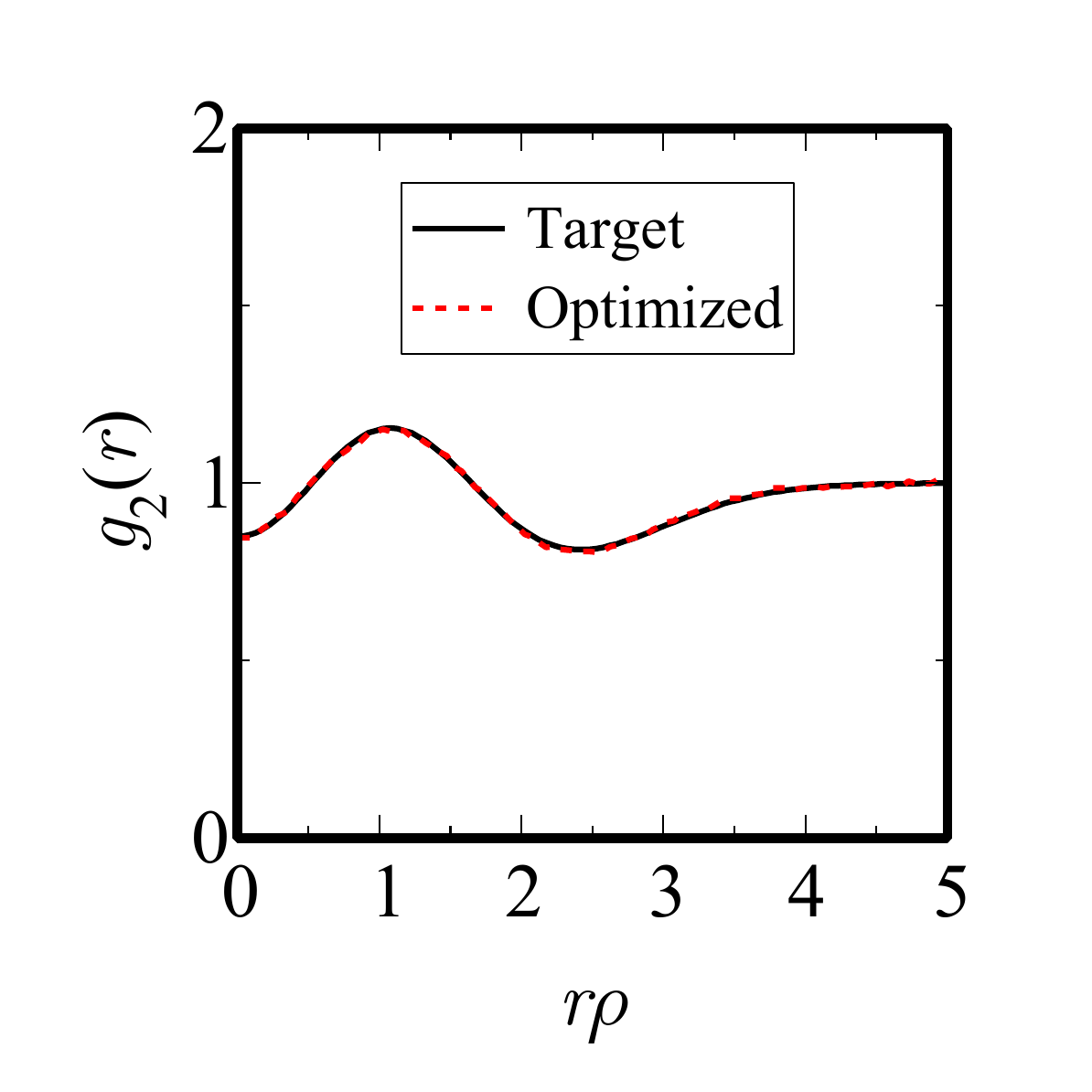}}
    \subfloat[]{\includegraphics[width = 55mm]{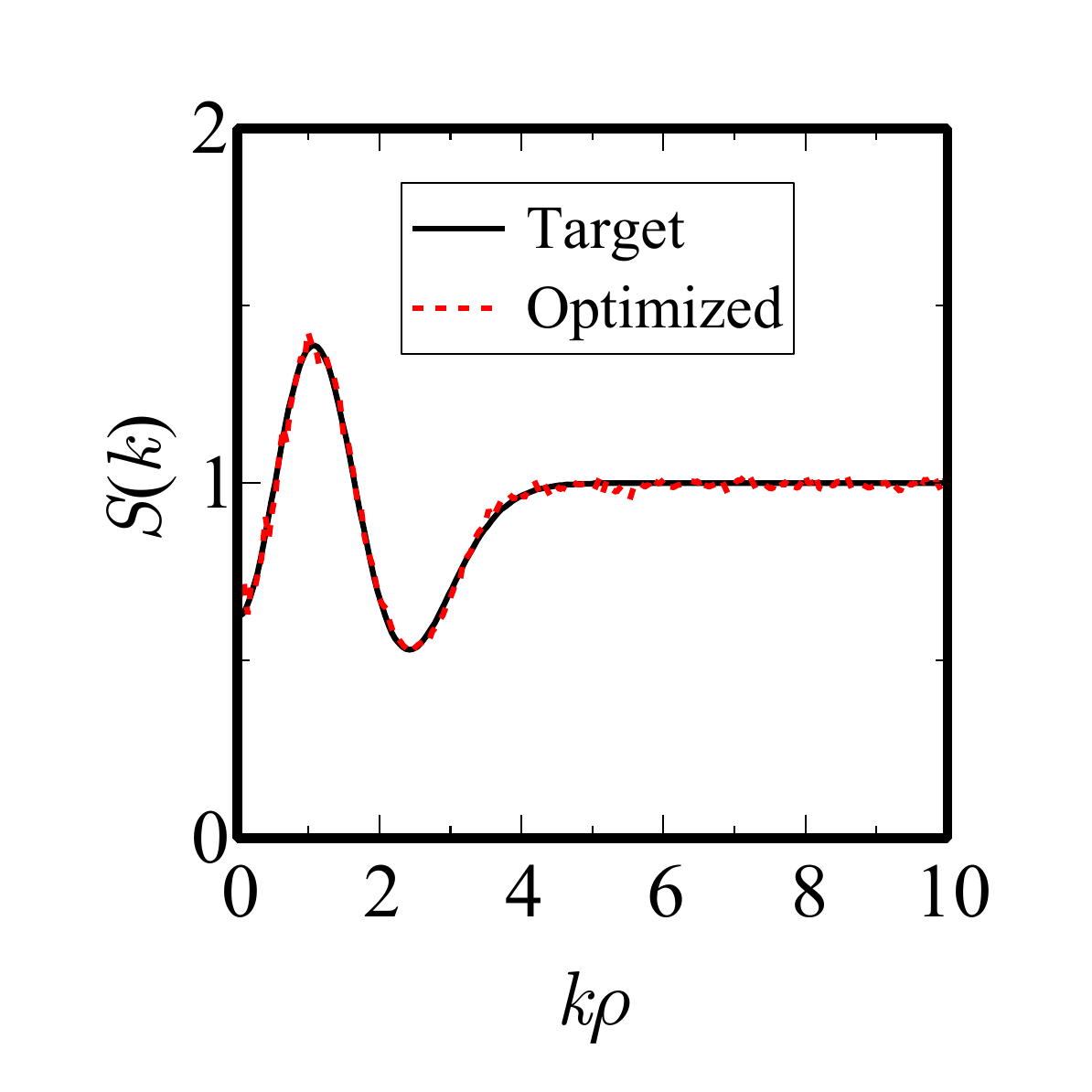}}
    \caption{Target and optimized (a) $g_2(r)$ and (b) $S(k)$ for the 1D hyposurficial Hermite-Gaussian $g_2(r)$ with $\lambda = 1/20$.}
    \label{fig:hermite1D}
\end{figure*}

We consider the 2D hyperuniform Gaussian-damped polynomial pair statistics given by Eqs. (\ref{polygaussian2Dg2}) and (\ref{polygaussian2DS}) in Table \ref{tab:pairfuncs}, as well as the 1D nonhyperuniform Hermite-Gaussian pair statistics given by (\ref{hermiteg2}) and (\ref{hermiteS}).
These pair functions mimic those found in dense polymer systems, which often exhibit oscillations due to the coordination shells \cite{Do98, Ye01}.
In both cases, the pair functions are self-similar under Fourier transformations up to scaling.
Notably, the forms of the 2D pair functions are chosen to closely resemble the pair statistics of polymer chains modeled by freely rotating fused hard spheres \cite{Ye01}.
The oscillatory $g_2(r)$ function [Fig. \ref{fig:poly2D}(a)] dictates that the corresponding many-particle system must have significant particle clustering. 
Also note that the form of the 1D total correlation function $h(r)$ is a Hermite-Gaussian function of order 4.
Hermite-Gaussian functions have been used to describe semiflexible polymers such as DNA and actin \cite{Wi96}.
We can precisely realize the pair functions in both cases, as shown in Fig. \ref{fig:poly2D} and \ref{fig:hermite1D}.
The effective potentials for both cases are oscillatory, and may be realized by ``sticky'' macromolecules, such as DNA-coated nanoparticles, as one can achieve precise control of attraction and repulsion at specific radial distances for these particles \cite{Ml13, Ku17}; see Appendix for the precise functional forms.
Compared to the configuration for the 2D Gaussian pair function [Fig. \ref{fig:gaussian}(c)], the configuration for the 2D Gaussian-damped polynomial [Fig. \ref{fig:poly2D}(c)] contains considerably fewer chains.
Instead, it contains many clusters with more than four closely spaced particles due to the peak in $g_2(r)$ at $r\rho^{1/2} \sim 0.6$. 
Despite this heterogeneity on the small to intermediate scale, Fig. \ref{fig:poly2D}(c) is homogeneous on the large scale as a result of the prescribed hyperuniformity, i.e., there are no large clusters or spherical holes void of particles on the order of the system size.

\subsection{3D Hyposurficial State}
\begin{figure*}
    \centering
    \subfloat[]{\includegraphics[width = 55mm]{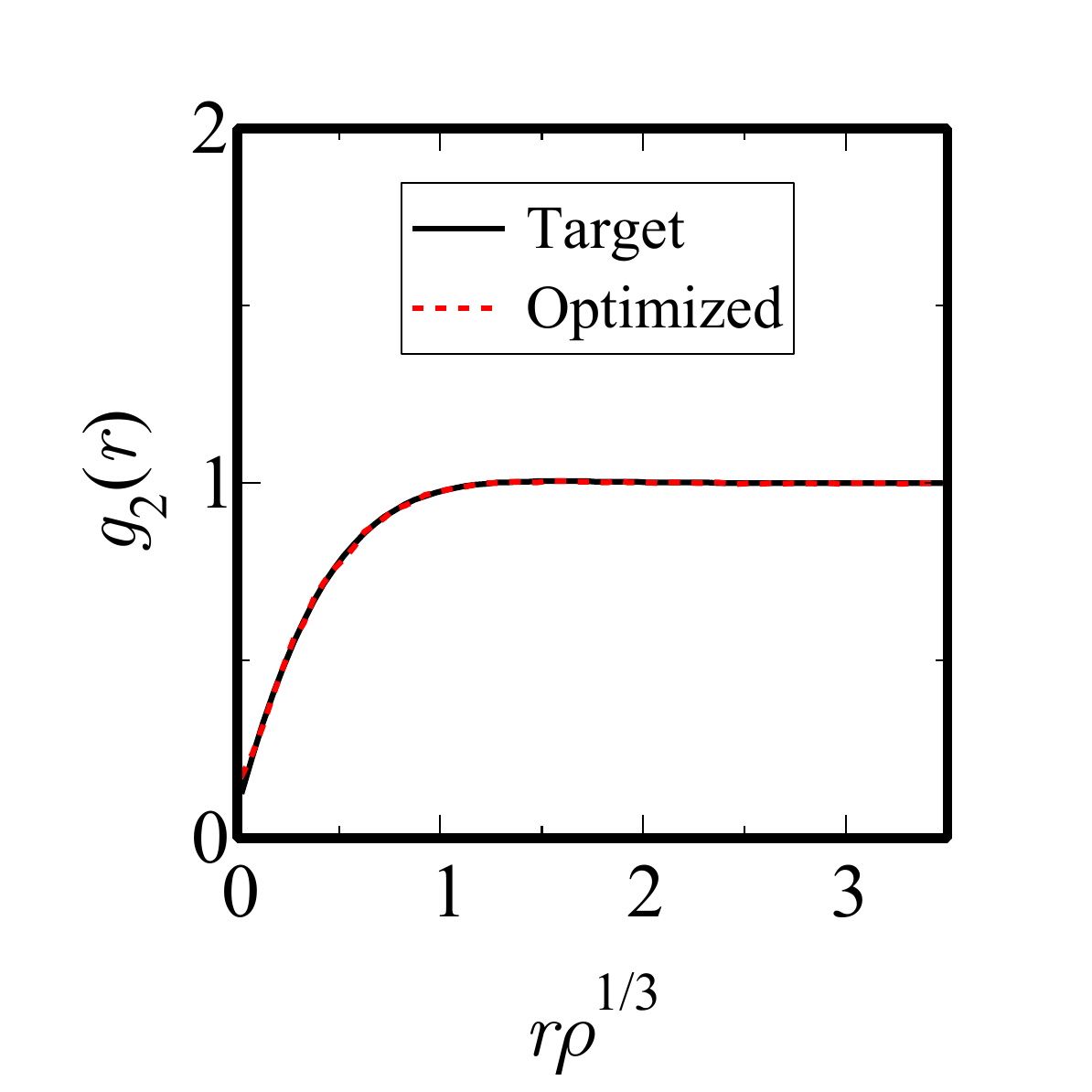}}
    \subfloat[]{\includegraphics[width = 55mm]{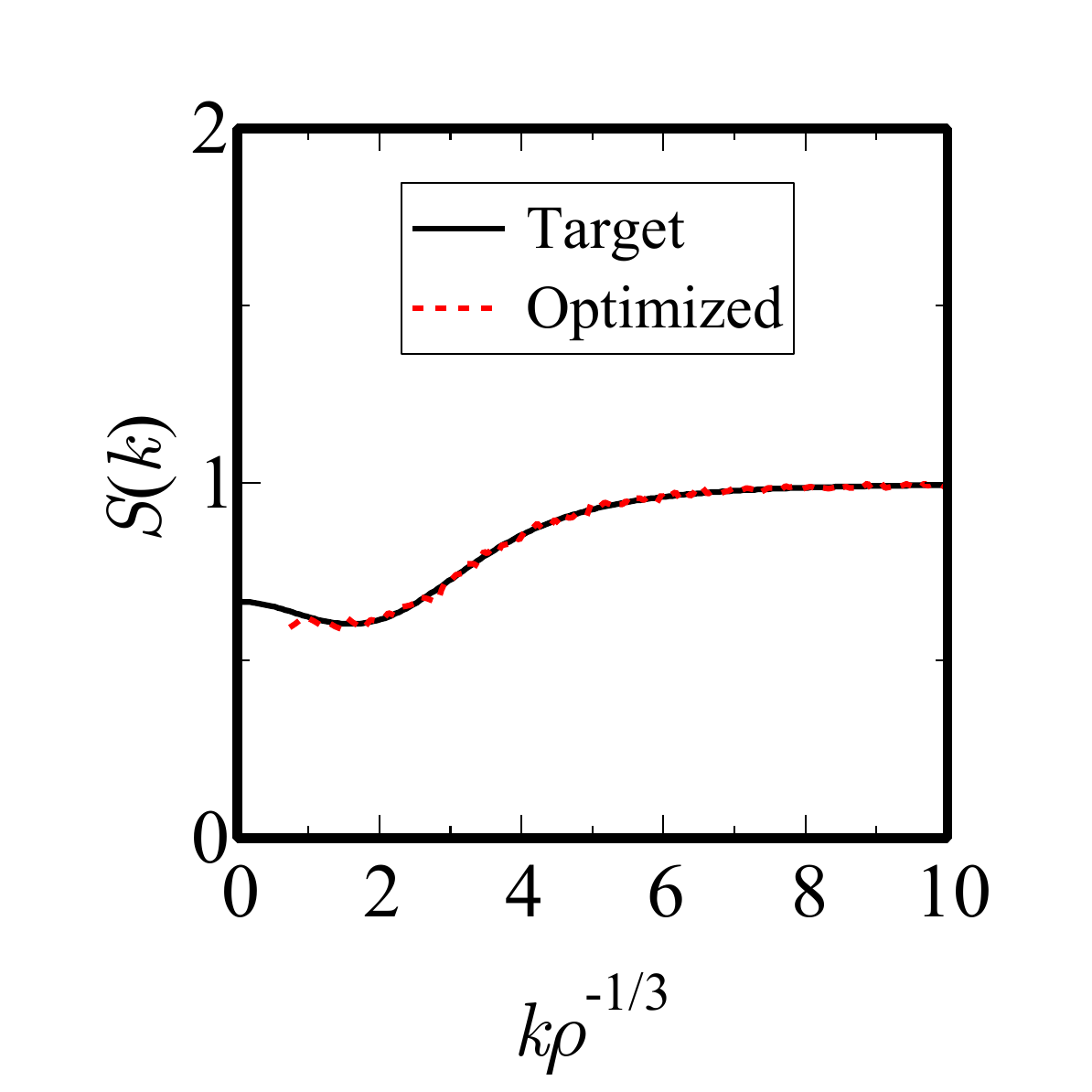}}
    \subfloat[]{\includegraphics[width = 65mm, trim={8cm 0cm 4cm 0cm},clip]{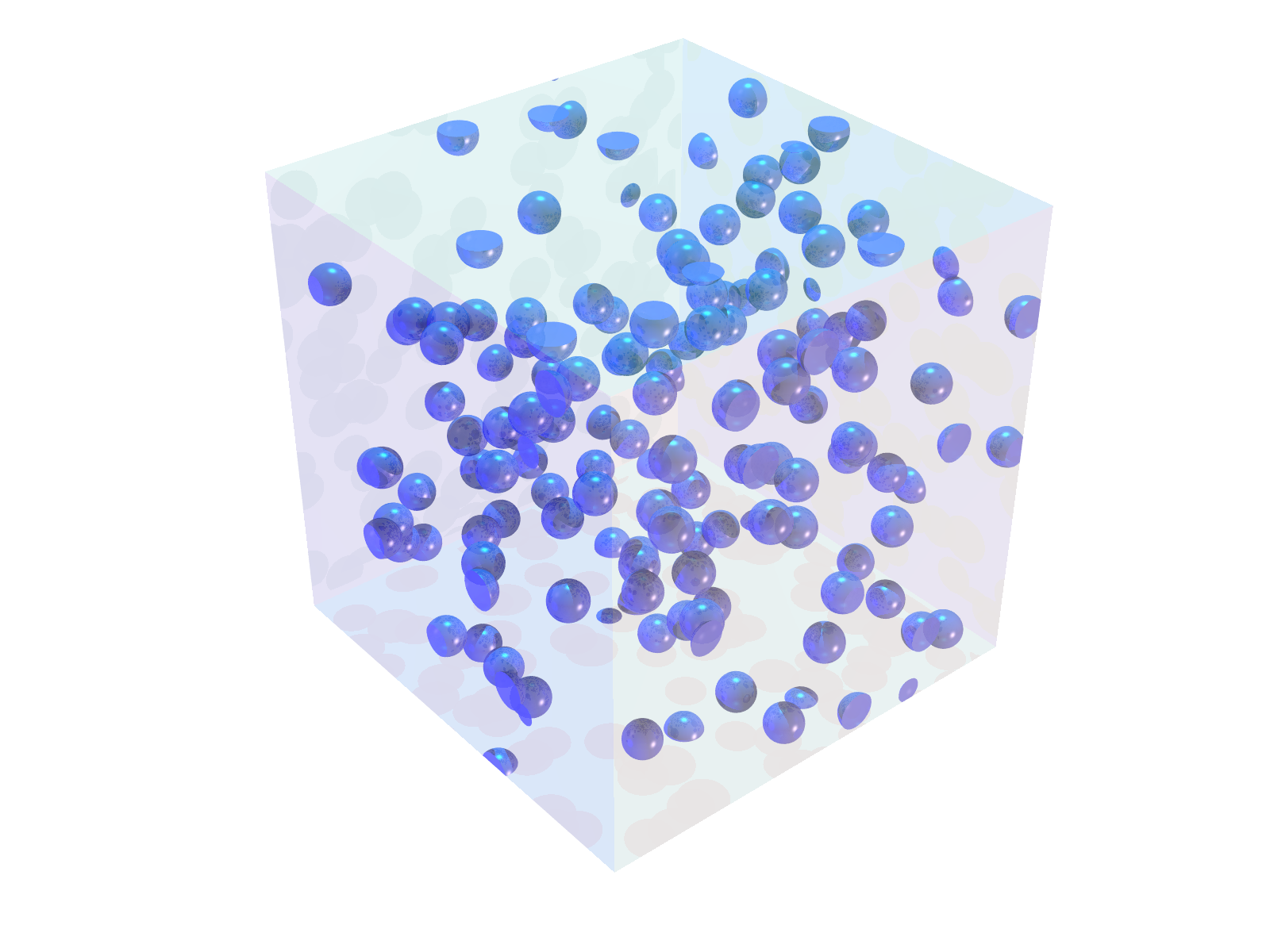}}
    \caption{(a)--(b) Target and optimized $g_2(r)$ and $S(k)$ for the 3D hyposurficial state. (c) A 125-particle configuration of the 3D equilibrium hyposurficial state. The point particles are shown as spheres for visualization purposes.}
    \label{fig:hypo3D}
\end{figure*}

We consider, for the first time, the possible realizability of the nonhyperuniform hyposurficial state specified by Eqs. (\ref{hypog2}) and (\ref{hypoS}) in Table \ref{tab:pairfuncs}, which applies in three dimensions.
Hyposurficial systems are a special class of nonhyperuniform states that can be observed in nonequilibrium phase transitions in amorphous ices \cite{Mar17}, supercritical equilibrium liquids \cite{To22} and certain hard-core systems \cite{To03a}.
They behave like ideal gases in their large-scale local number variance $N(R)$, because they lack a ``surface-area'' term proportional to $R^{d-1}$ in the growth of $N(R)$, i.e., $\sigma^2(R)\sim AR^d + o(R^{d-1})$ as $R\rightarrow\infty$, where $A$ is a positive constant.
Hyposurficial pair statistics obey the following sum rule \cite{To03a}:
\begin{equation}
    \int_0^\infty r^d h(r) dr = 0,
    \label{hypodef}
\end{equation}
which implies that they must generally contain both negative and positive correlations.
Ideal gases are trivially hyposurficial.
Figure \ref{fig:hypo3D} shows that we can precisely realize the designed hyposurficial target. 
The configuration [Fig. \ref{fig:hypo3D}(c)] shows spatial heterogeneities, i.e., there exists significant particle clustering as well as large spherical holes void of particles.

\subsection{1D and 2D Ghost RSA}
We consider the nonhyperuniform pair statistics corresponding to the 1D and 2D ghost RSA packings, given by Eqs. (\ref{ghost1Dg2})--(\ref{ghost2Dg2}) in Table \ref{tab:pairfuncs}.
 The nonequilibrium ghost RSA process \cite{To06a, To06b} is a special case of a generalization of the standard RSA process \cite{Re63, wi66, To02a} and is the only known model for which all $n$-particle correlation functions are exactly solvable \cite{To06a}. 
 In the ghost RSA process, spherical ``test'' particles of diameter $D$ are added continually to $\mathbb{R}^d$ during time $t\geq 0$ according to a translationally invariant Poisson process of density $\eta$ per unit time, i.e., $\eta$ is the number of points per unit volume and time, here taken to be unity without loss of generalization. 
 A test sphere centered at position $\mathbf{r}$ at time $t$ is retained if and only if it does not overlap with any test sphere added in the time interval $[0,t)$ \cite{To06a}. 
 The saturation packing fraction for the ghost RSA process is $\phi(t=+\infty)=1/2^d$ \cite{To06a}. 
 
 We have previously shown that the 3D ghost RSA pair statistics at the terminal packing fraction is realizable by an equilibrium state with up to pair interactions \cite{Wa22b}. 
 Here, we investigate whether the same is true for $d=1$ and 2.
 It is not obvious that the ghost RSA processes are realizable in these lower dimensions because, generally speaking, it is  known that the lower the space dimension, the more difficult it is to satisfy realizability conditions \cite{To06b}. 
 This dimensionality effect arises on account of the decorrelation principle \cite{To06b}, which states that spatial correlations that exist for a particular model in lower dimensions diminish as the space dimension becomes larger.

\begin{figure*}
    \centering
    \subfloat[]{\includegraphics[width = 55mm]{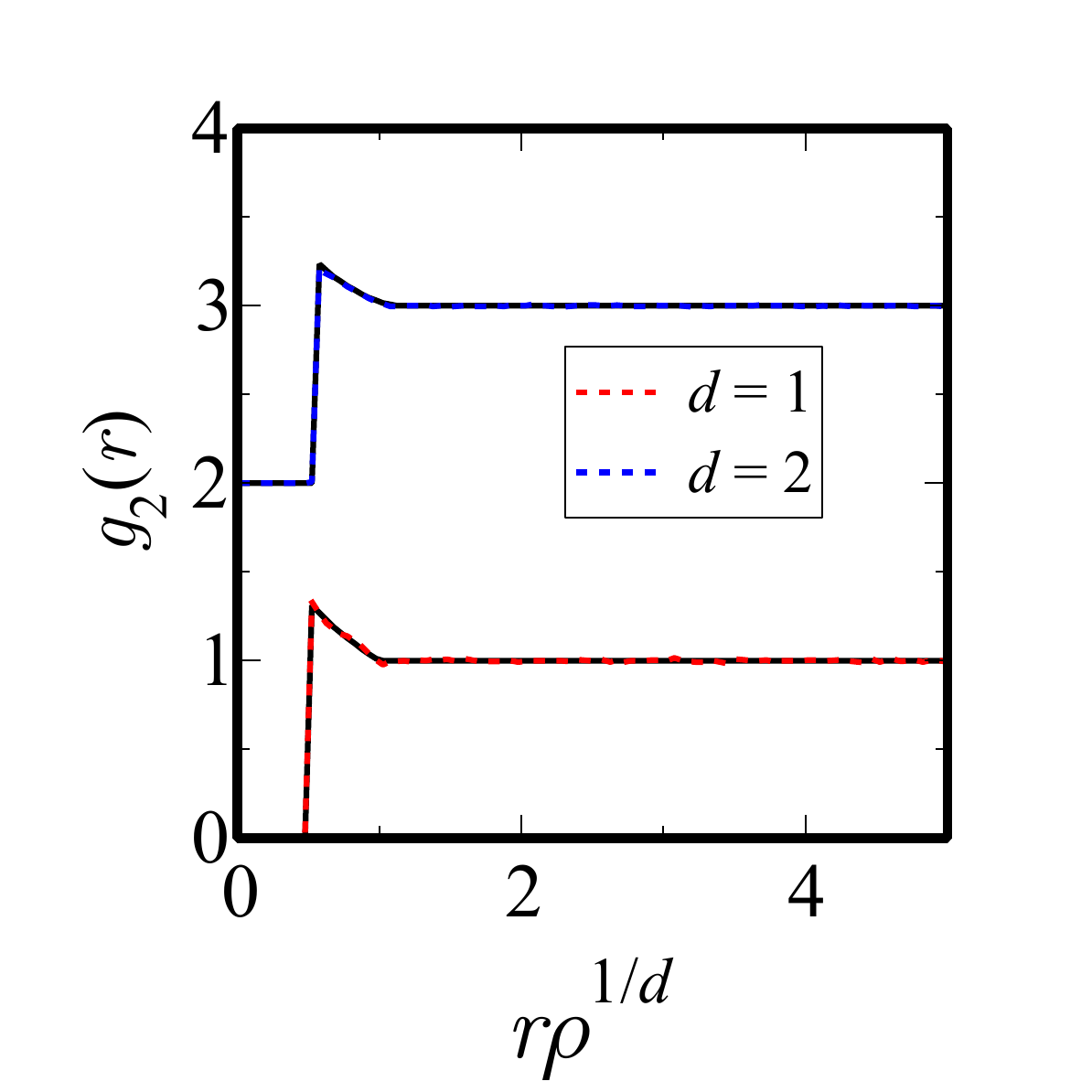}}
    \subfloat[]{\includegraphics[width = 55mm]{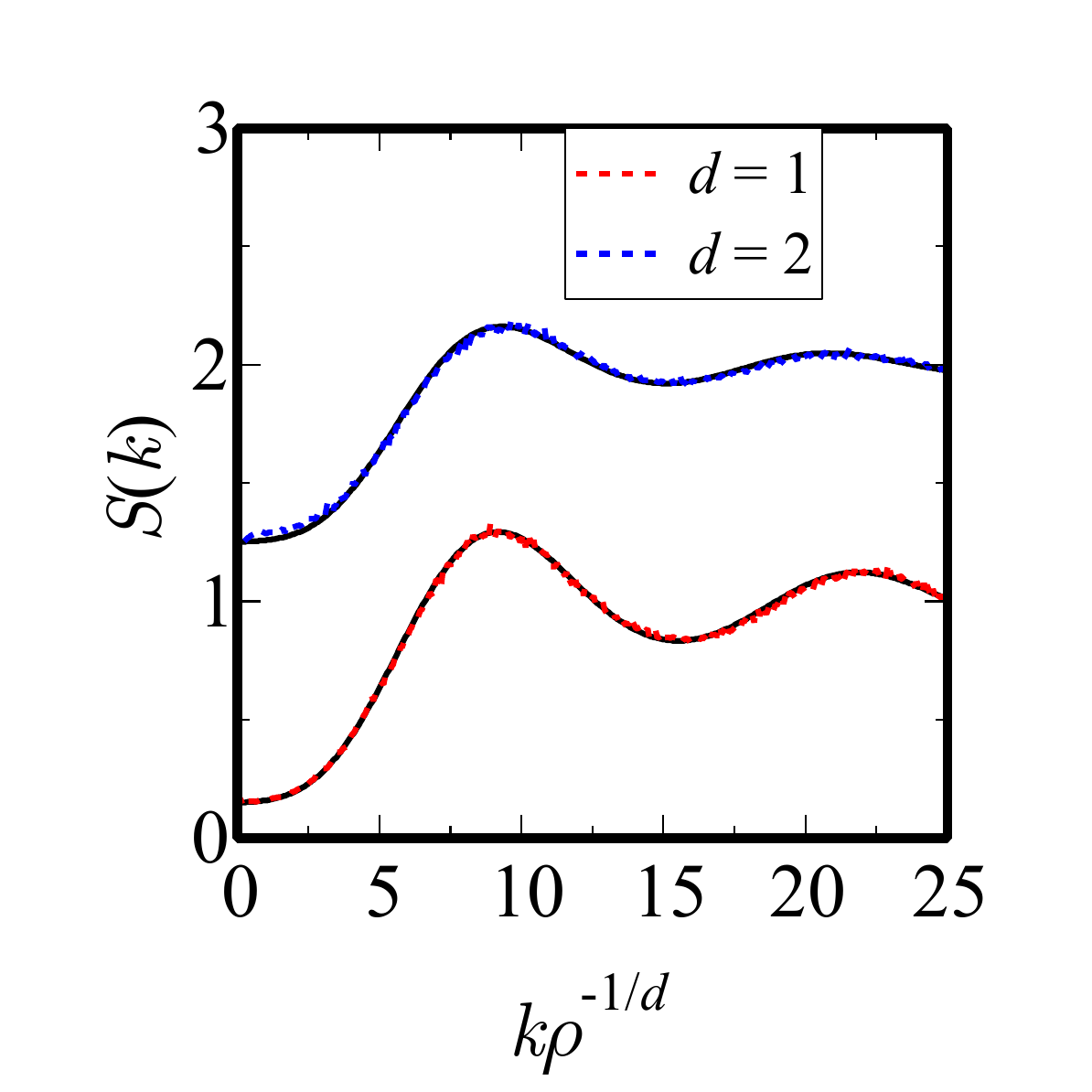}}
    \caption{Target (solid curves) and optimized (dashed curves) (a) $g_2(r)$ and (b) $S(k)$ for the 1D and 2D ghost RSA packings. For clarity, the $g_2(r)$ curve for $d = 2$ is shifted up 2 and the $S(k)$ curve for $d=2$ is shifted up by 1.}
    \label{fig:ghost}
\end{figure*}

Figure \ref{fig:ghost} shows that we can achieve the nonequilibrium 1D and 2D ghost RSA packings with the effective potentials in Fig. \ref{fig:potentials}(i), which contain both a hard-core and soft repulsion beyond the diameter.
The effective soft repulsion ensures that particles are well-separated, and corresponds to the exclusion of real particles by the ``ghost'' test particles in the nonequilibrium process. 
This combination of hard core and soft repulsion in $v(r)$ can be found in colloidal suspensions of highly charged nanoparticles \cite{Li04}.
 Note that the soft repulsion is more significant for the 1D case than the 2D case, reflecting the aforementioned decorrelation principle \cite{To06b}. 
For $d=3$, the ghost RSA $g_2(r)$ \cite{To06a} is nearly  equal to unity for $D \leq r \leq 2D$, and the soft repulsion in its effective potential \cite{Wa22b} is weaker than that in the 2D case. 
 In the limit that $d\rightarrow \infty$, the ghost RSA $g_2(r)$ approaches the unit-step $g_2(r)$ \cite{To06b}, and the effective potential would consequently converge to that of the hard-sphere potential.
 The effective potentials also exhibit weak attractive wells [Fig. \ref{fig:ghost}(a)], which offset the oscillations in $g_2(r)$ that would occur in pure equilibrium hard-sphere system, ensuring that $g_2(r)$ is flat beyond two diameters.

\subsection{$d$-dimensional Antihyperuniform Pair Statistics}
We study the antihyperuniform pair statistics (\ref{ahu1Dg2}) in one- two and three dimensions.
Note that the target $g_2(r)$ functions imply a hard core with diameter $D$. 
The functional forms of $S(k)$ as well as the packing fraction $\phi = \rho v_1(D/2)$ and the parameter $A$ are given in the Appendix A.
The values of $\phi$ and $A$ chosen ensure the necessary condition that $S(k)$ must be nonnegative at all $k$.
As mentioned in the Introduction, $S(k)$ for an antihyperuniform state has the small-$k$ scaling behavior $S(k) \sim k^{\alpha}$, where $\alpha$ is negative.
This divergence behavior of $S(k)$ at small $k$ are characteristic of fluids at thermal critical points, which can have fractal-like structures \cite{Wi65, Fi67}.
For our antihyperuniform targets in any dimension, we take $\alpha = -1/2$, implying that $h(r)$ has a long-ranged power-law decay behavior at large $r$ as $h(r) \sim r^{-(d-1/2)}$ \cite{To18a}.
Realizable systems with such pair statistics would correspond to a different universality class than the Ising class, which gives the large-$r$ behavior $h(r) \sim r^{-(d-2 + \eta)}$, where $\eta = 1/4$ for $d = 2$ and $\eta = 0.036$ \cite{Bi92}. 

\begin{figure*}[htp]
    \centering
    \subfloat[]{\includegraphics[width = 55mm]{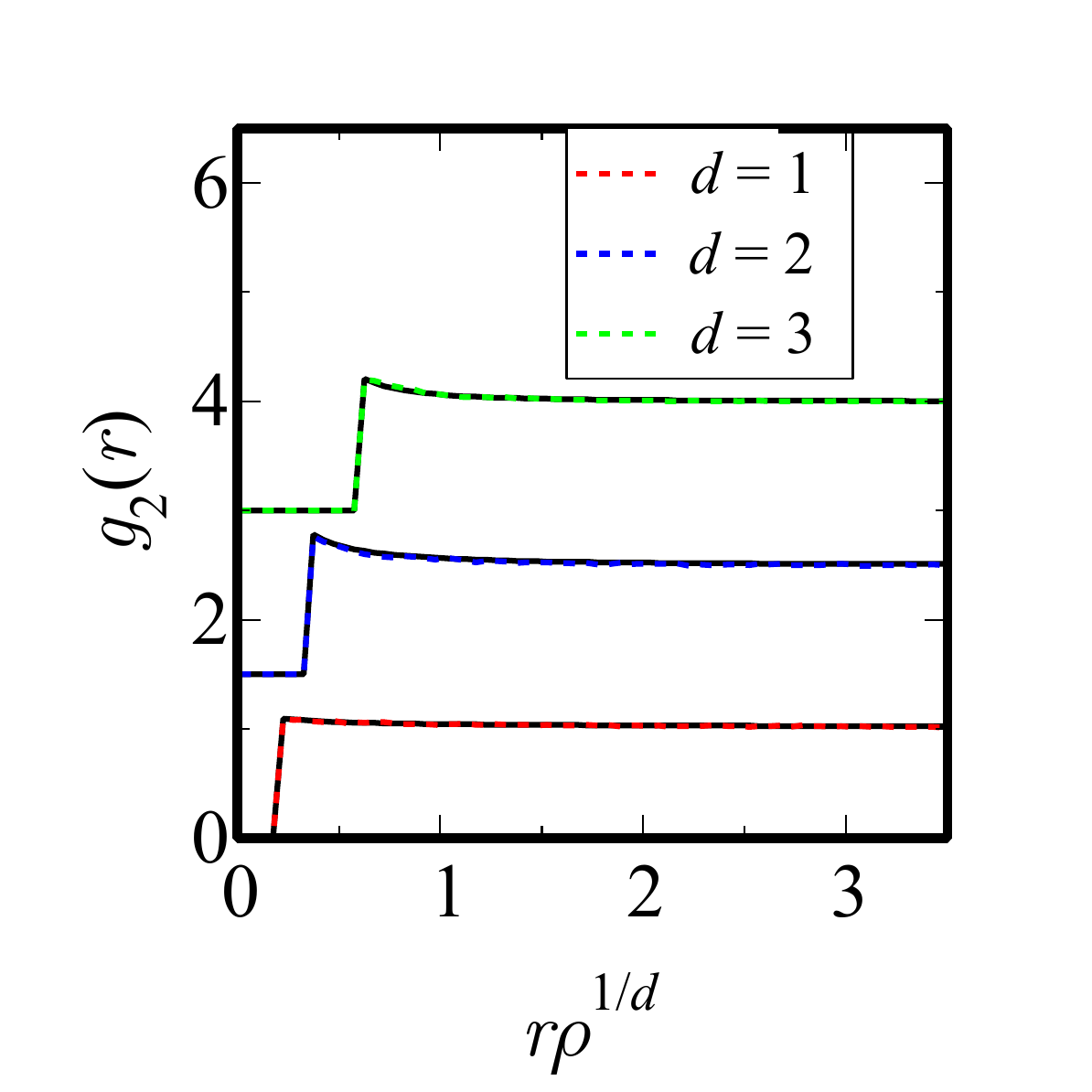}}
    \subfloat[]{\includegraphics[width = 55mm]{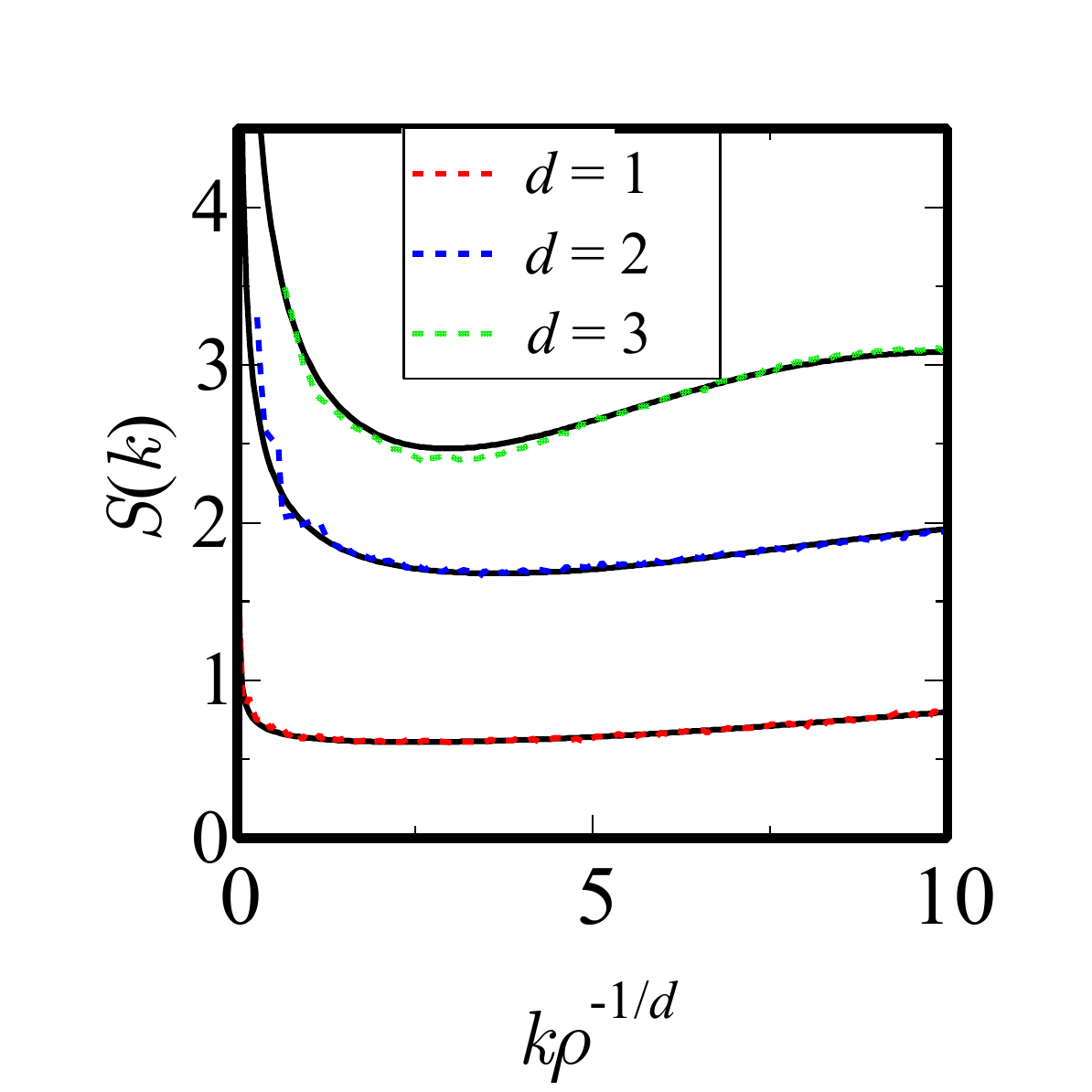}}

    \subfloat[]{\includegraphics[width = 60mm, trim={4cm 1cm 4cm 0cm},clip]{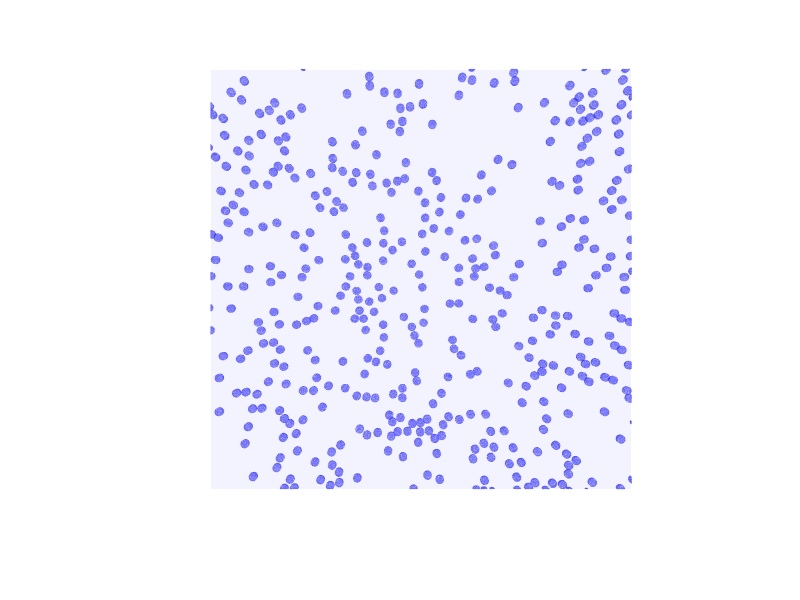}}
    \subfloat[]{\includegraphics[width = 60mm, trim={4cm -1cm 4cm 0cm},clip]{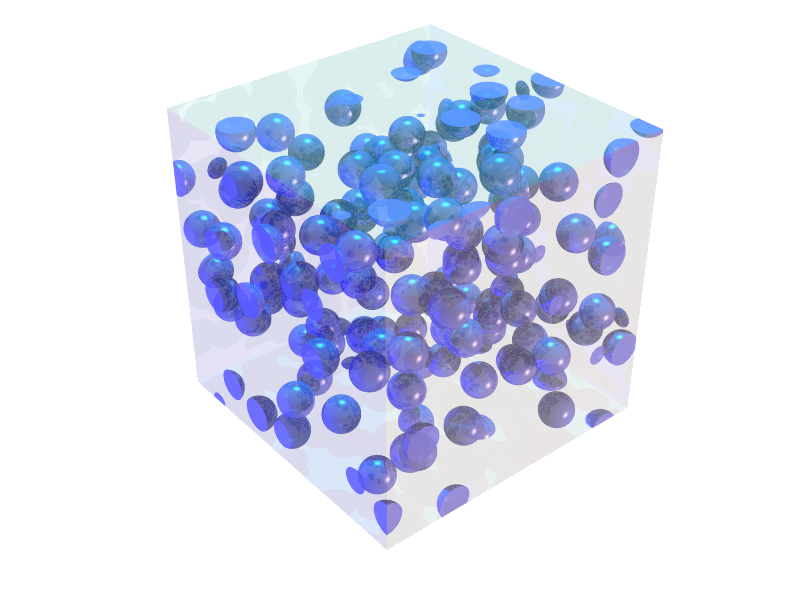}}
    \caption{(a)--(b) Target (solid curves) and optimized (dashed curves) $g_2(r)$ and $S(k)$ for $d$-dimensional antihyperuniform pair statistics (\ref{ahu1Dg2}).
    For clarity, $g_2(r)$ for $d = 2$ and $d = 3$ are shifted up by 1.5 and 3, respectively, and $S(k)$ for $d = 2$ and $d = 3$ are shifted up by 1 and 2, respectively.
    (c) A 400-particle configuration of the 3D equilibrium antihyperuniform state.
    (d) A 125-particle configuration of the 3D equilibrium antihyperuniform state.
    The disk and sphere diameters in the configurations are the hard-core diameter $D$.}
    \label{fig:ahu}
\end{figure*}

Computing equilibrium properties at critical points using Monte-Carlo or molecular dynamics simulations is known to be challenging due to the critical slowing down phenomena, i.e. the relaxation time to the equilibrium state diverges at the critical point \cite{Wi65, Wi84}.
Our inverse method reduces the time required to reach the targeted critical-point pair statistics compared to previous methods, as it uses parametrized analytical pair potentials rather than binned potentials \cite{To22}; see Sec. \ref{meth} for details.
We have realized the antihyperuniform critical-point pair statistics in one, two and three dimensions, as shown in Fig. \ref{fig:ahu}.
The 2D and 3D critical-point states (Fig. \ref{fig:ahu}(c) and (d)] clearly show large clusters and large empty spherical holes on the order of the system size, confirming that antihyperuniformity corresponds to divergent long-range number density fluctuations.

The realizability of the 1D critical state is unusual because it is commonly thought that phase transitions cannot exist in 1D systems, but this only true when the interactions are sufficiently short-ranged \cite{Dy69, Ci87}.
Indeed, in the Appendix, we show that the effective pair potential has a long-ranged attractive part, i.e., $v(r)\sim -r^{-(d-1/2)}$ as $r\rightarrow\infty$.
This long-range behavior in the interaction explains why our 1D, 2D and 3D critical-point states do not belong to the Ising university class that are characterized by short-ranged interactions \cite{Ka66, Fi67, Wi74, Bi92}.

\section{Realizability of Pair Statistics from a 3D Polymer Model}
\label{hypothetical}
\begin{figure*}[htp]
    \centering
    \subfloat[]{\includegraphics[width = 55mm]{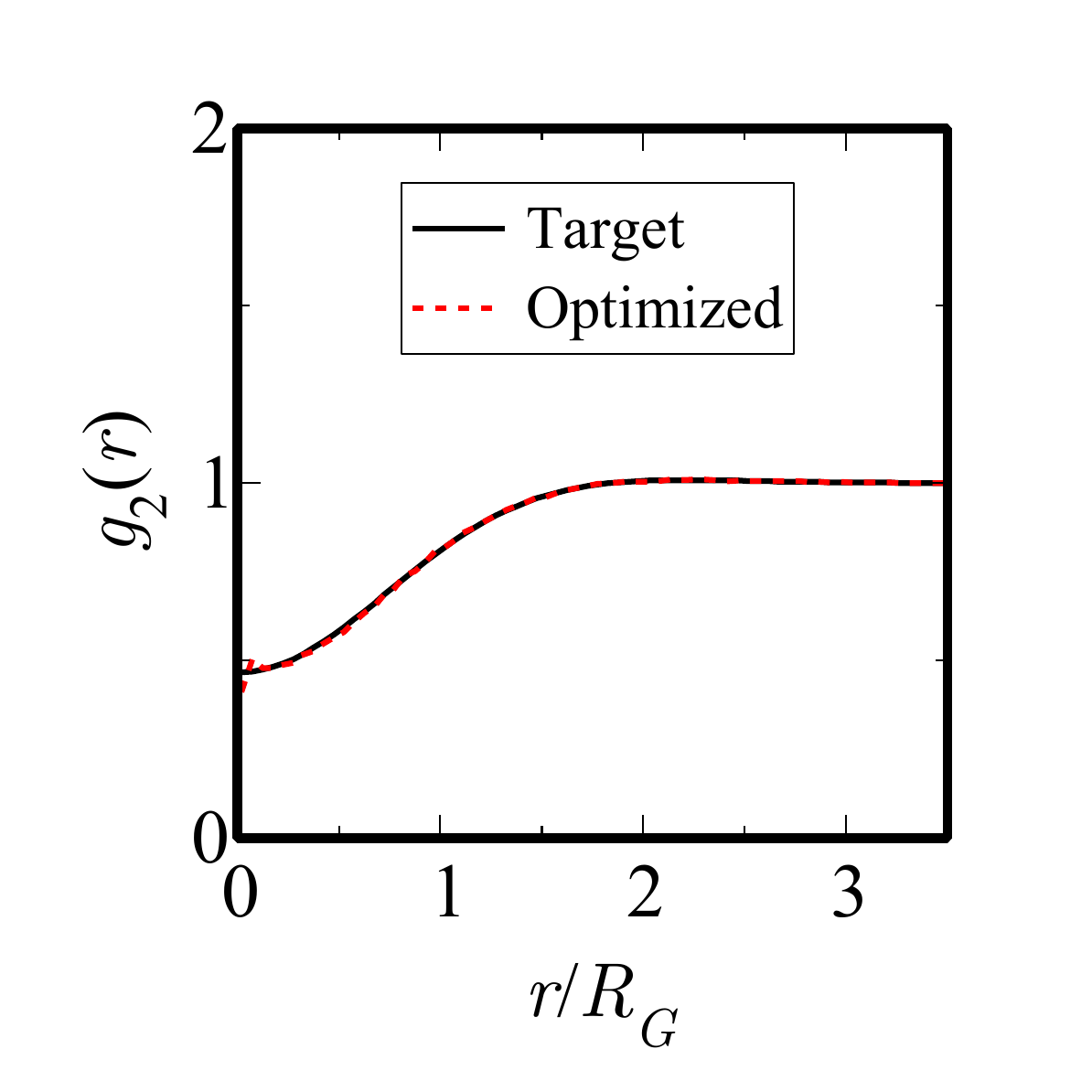}}
    \subfloat[]{\includegraphics[width = 55mm]{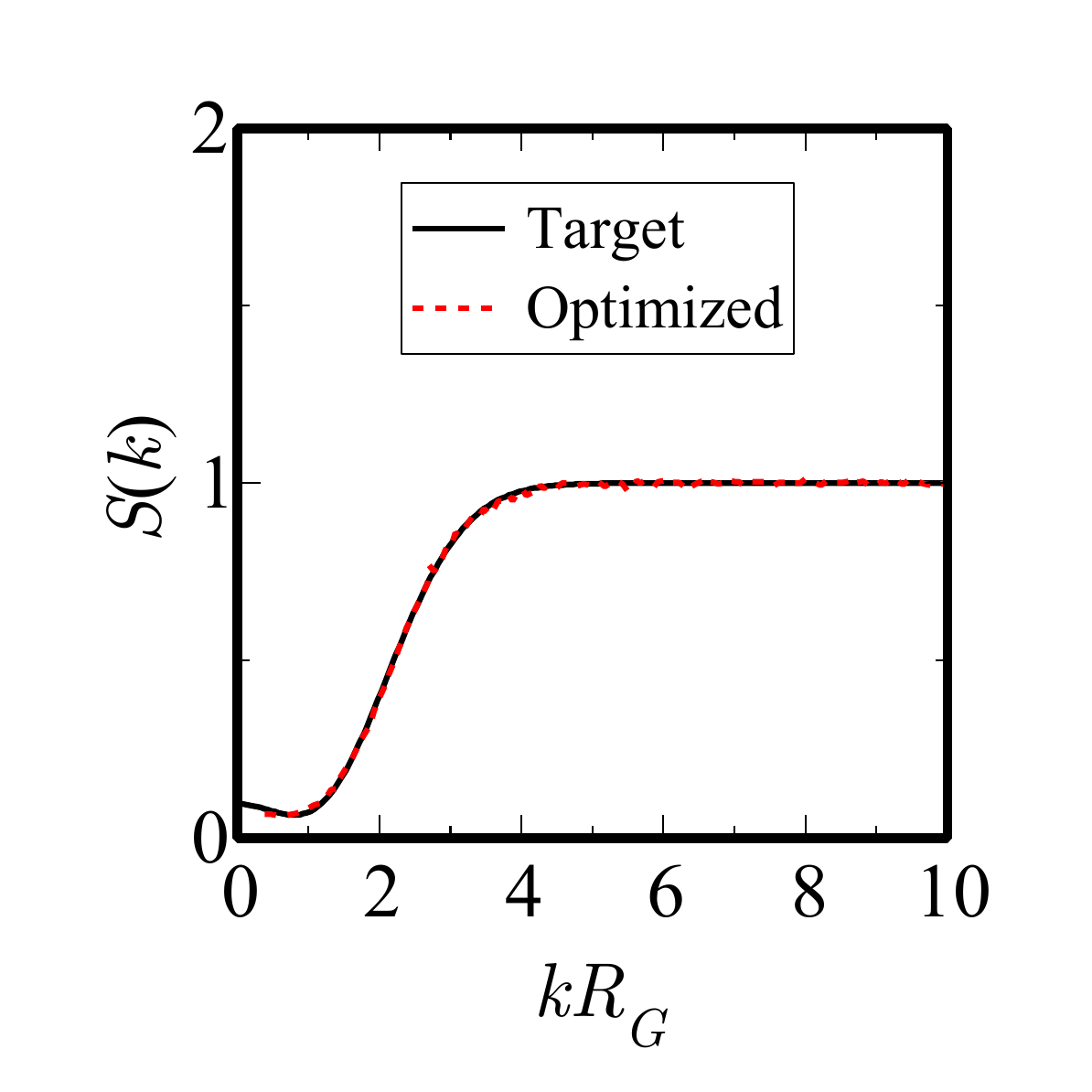}}
    \caption{Target and optimized (a) $g_2(r)$ and (b) $S(k)$ for a hypothetical 3D polymer motivated by the PRISM model with the total correlation function (\ref{polymerhr}) at $\rho = 0.5$.}
    \label{fig:guenza3D}
\end{figure*}

To illustrate that our procedure can be used to guide experimental polymer design, we now examine a case of prescribed pair statistics whose functional forms are motivated by the PRISM model \cite{Sc97, Ya04}.
According to this model, the total correlation function for polymer chains  is given by \cite{Ya04} 
\begin{widetext}
\begin{equation}
    h(r; \rho) = \frac{3}{4} \sqrt{\frac{3}{\pi }} e^{-\frac{1}{4} \left(3 r^2\right)} \tilde{\xi}'_{\rho} \left(1-\frac{1}{2 \tilde{\xi}_{\rho}^2}\right)
    -\frac{\tilde{\xi}'_{\rho} \left(1-\frac{1}{2 \tilde{\xi}_{\rho}^2}\right)^2 e^{\frac{1}{3 \tilde{\xi}_{\rho}^2}} \left[e^{\frac{r}{\tilde{\xi}_{\rho}}} \text{erfc}\left(\frac{1}{\sqrt{3} \tilde{\xi}_{\rho}}+\frac{\sqrt{3} r}{2}\right)-e^{-\frac{r}{\tilde{\xi}_{\rho}}} \text{erfc}\left(\frac{1}{\sqrt{3} \tilde{\xi}_{\rho}}-\frac{\sqrt{3} r}{2}\right)\right]}{2 r},
    \label{polymerhr}
\end{equation}
\end{widetext}
where lengths are measured in units of the radius of gyration $R_G$, $\tilde{\xi}_\rho = 1/[\sqrt{2} \left(\pi  \sqrt{2} \rho +1\right)]$ and $\tilde{\xi}'_\rho = 1/(2\pi \rho)$.
The corresponding $S(k)$ is also known analytically  \cite{Ya04}.
Yatsenko et al. \cite{Ya04} showed that for $0.86 < \rho < 1.62$, Eq. (\ref{polymerhr}) accurately describes the experimental radial distributions of polyethylene and polypropylene melts at small to intermediate $r$. 
However, experimental realizability of (\ref{polymerhr}) outside of this density range has not been tested.
We have applied our inverse algorithm to Eq. (\ref{polymerhr}) at $\rho = 0.5$ and have achieved excellent agreement between the target and optimized pair functions by a soft-core effective potential, as shown in Fig. \ref{fig:guenza3D}. 
These hypothetical pair statistics might be experimentally achievable by polymer melts at low density, e.g. via polyethylene molecules with a smaller number of monomers per chain than those considered in Ref. \cite{Ya04}.

\section{Translational Order Metric and Self-Diffusion Coefficient}
\label{trans}
To demonstrate that our designed analytical pair functions enable one to achieve a wide range of structural order and physical properties, we compute the translational order metric $\tau$ \cite{To15} and the self-diffusion coefficient $\mathcal{D}$ \cite{Kr09b} for the 3D states studied here.
The $\tau$ order metric is defined by \cite{To15}
\begin{equation}
    \tau = \rho \int h^2(r) d\mathbf{r} = \frac{\rho}{(2\pi)^d}\int_{\mathbb{R}^d} \tilde{h}^2(k) d \mathbf{k}, 
    \label{tau}
\end{equation}
where $\tilde{h}(k)$ is the Fourier transform of $h(r)$.
This order metric measures deviations of pair statistics from that of the uncorrelated (Poisson) distribution $g_2(r) = 1$. 
Since both positive and negative correlations contribute to the integral, due to the fact that $h(r)$ is squared, $\tau$ measures the degree of translational order across all length scales. 
It vanishes for the Poisson distribution and diverges for an infinite crystal\footnote[20]{The metric $\tau$ can still be used to analyze crystals by examining its growth with the system size for finite configurations [\citenum{To22}].} and is a positive bounded number for correlated disordered systems without long-range order (i.e., Bragg peaks) \cite{Ph23}.

To compute the self-diffusion coefficients, we utilize the semi-empirical scaling laws developed by Rosenfeld \cite{Ro99c}. 
The reduced self-diffusion coefficient $\mathcal{D}\rho^{1/3}T^{1/2}$ for 3D fluids in the dilute limit is given by \cite{Ro99c}
\begin{equation}
    \mathcal{D}\rho^{1/3}T^{1/2} = 0.37(-s_2)^{-2/3},
    \label{diffus}
\end{equation}
where 
\begin{equation}
    s_2 = -(\rho/2)\int_{\mathbb{R}^d} \{g_2(r) \ln [g_2(r)] -g_2(r) + 1\} d\mathbf{r}
\end{equation}
is the two-body excess entropy. 
Equation (\ref{diffus}) provides a good approximation for soft-core models with $-s_2 \leq 0.5$ \cite{Kr09b}, which is satisfied by all the soft-core pair statistics in this study.
For the ghost RSA and the antihyperuniform packings, which have hard-core interactions, we use the corresponding scaling law \cite{Ro99c}
\begin{equation}
   \mathcal{D}\rho^{1/3}T^{1/2} = 0.58 \exp(0.78 s_2).
\end{equation}

\begin{figure}
    \centering
    \includegraphics[width =90mm]{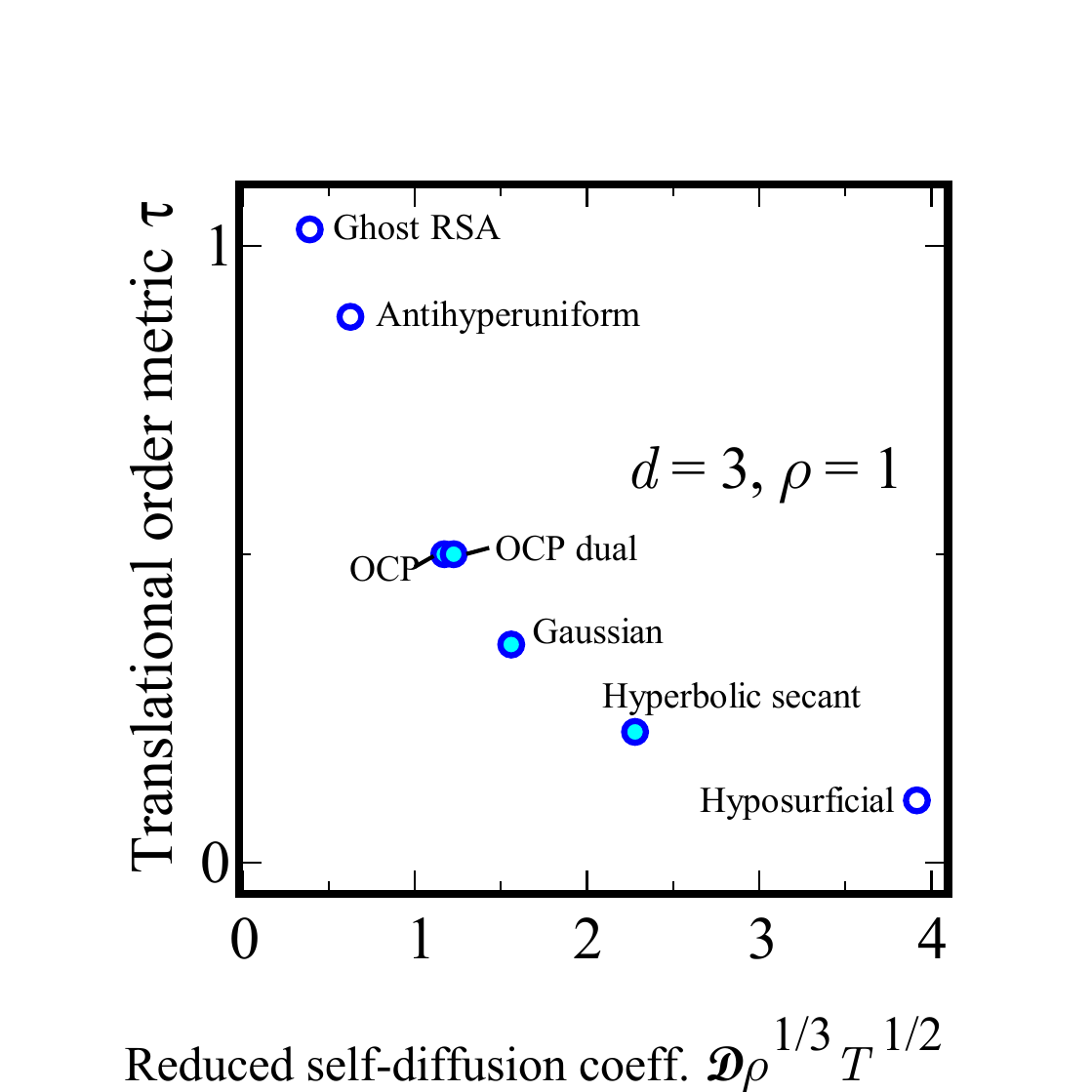}
    \caption{Translational order metric $\tau$ and the reduced self-diffusion coefficient $\mathcal{D}\rho^{1/3}T^{1/2}$ for the 3D pair statistics studied in this work and in Ref. \cite{Wa22b}. The filled circles refer to hyperuniform states and the unfilled ones refer to nonhyperuniform states.}
    \label{fig:tauAndD}
\end{figure}

Figure \ref{fig:tauAndD} plots $\tau$ against the reduced self-diffusion coefficient, where hyperuniform states are indicated with filled circles. 
It is evident that $\tau$ and $\mathcal{D}$ are inversely correlated to one other. 
Pair functions with smaller translational order metrics are closer to the Poisson pair statistics.
Thus, in equilibrium states with smaller $\tau$, particles encounter less hindrance from neighboring particles, resulting in a higher self-diffusivity.
Importantly, $\tau$ has a strong positive correlation to $-s_2$ \cite{Lo17}, and in the limit that $g_2(r)$ is an infinitesimal perturbation from unity (i.e., from an ideal gas), one has exactly $\tau \sim -4 s_2$ \cite{Lo17}.
Consequently, $\tau$ is inversely related to $\mathcal{D}$ via Eq. (\ref{diffus}).

The ghost RSA and the antihyperuniform packings have hard-core interactions and achieve higher $\tau$ and lower $\mathcal{D}$, because the hard particles are more locally confined due to the frequent collisions with their neighbors \cite{Dz96}.
The soft-core hyperuniform states have intermediate $\tau$ and $\mathcal{D}$. 
Figure S1 shows the forces corresponding to the effective potentials for these states, from which it is clear that models with larger magnitudes of small-$r$ repulsion give larger $\tau$ and smaller $\mathcal{D}$.
The OCP, which has the strongest small-$r$ repulsion among the 3D hyperuniform states, is least diffusive.
This is followed by the Fourier dual of OCP, the Gaussian pair statistics and the hyperbolic secant $g_2(r)$. 
Interestingly, while OCP and its Fourier dual both have $\tau = 1/2$, the former has a slightly lower self-diffusivity because its $g_2(r)$ scales as $r^3$ at small $r$, representing a stronger small-$r$ repulsion than the $r^2$ scaling for the Fourier dual of OCP.
Our observation that particles with soft repulsive interactions are more diffusive than hard-core particles is consistent with the findings for macromolecules via molecular dynamics simulations \cite{Sl22} and for lattice gases via Monte-Carlo simulations \cite{Cz96}.

Finally, the hyposurficial state, which lacks both hard-core interactions and hyperuniformity, has the smallest $\tau$ and largest $\mathcal{D}$, as its pair statistics are closest to those of an ideal gas.
Its low translational order is consistent with the observation in Ref. \cite{Ma17} that hyposurficiality appears to be associated with spatial heterogeneities.

\section{Discussion}
\label{disc}
We have designed a diverse set of pair functions with varying degree of disorder across length scales that are realizable by disordered many-particle systems, many of which have novel structural and physical properties.
We achieved this via an efficient algorithm that determines equilibrium states with up to pair interactions at positive temperatures.
The realizability of the prescribed pair statistics via equilibrium states with up to pair interactions is rather strong, because it does not exclude the possibility that the same pair statistics can be realized by equilibrium states involving higher-body interactions or by nonequilibrium states.

We have chosen our pair functions based on their capability to generate a wide spectrum of hyperuniform and nonhyperuniform systems with various short-range correlations, including hyperuniform soft-core systems with different small-$r$ repulsive forces, a nonhyperuniform 3D hyposurficial state, 1D and 2D ghost RSA packings and antihyperuniform packings.
These forms of pair functions dictate that the corresponding systems have different cluster sizes from one another.
Notably, the 3D Fourier dual of OCP possesses a larger hyperuniformity exponent $\alpha$ than any other positive-temperature equilibrium state known previously, which can be useful in optical applications \cite{To21a}.
We have designed a diverse family of pair functions that are self-similar under Fourier transformations, which means that their low-density and high-density ground-state structures under the effective potentials can be mapped to one another via duality relations \cite{To08a, To11b}. 
Furthermore, we have realized a case of pair statistics whose functional forms are motivated by the PRISM model \cite{Ya04}, indicating that it may be experimentally achieved.
Our designed pair functions exhibit distinctly different translational order metric $\tau$ and self-diffusion coefficients $\mathcal{D}$. 
We have shown that for our 3D models, $\tau$ and $\mathcal{D}$ are inversely correlated, the same trend of which should be true for the corresponding one- and two-dimensional designs as well.
This relationship enables one to easily achieve desired diffusivity through designed pair functions.

The pair potentials for many states can be experimentally achieved by polymer or nanoparticle systems with tunable chemical components.
For example, Yatsenko et al. \cite{Ya04} showed that in the limit of long polymer chains, the total distribution function of polyethylene melts with radius of gyration $R_G$ can be approximated by a Gaussian-damped polynomial,
\begin{equation}
    h(\Tilde{r}, \xi) = -\frac{39}{16}\sqrt{\frac{3}{\pi}}\xi(1 + \sqrt{2}\xi)(1 - \frac{9 \tilde{r}^2}{26})\exp(-\frac{3\Tilde{r}^2}{4}),
    \label{polyethyl}
\end{equation}
where $\tilde{r} = r/R_G$ and $\xi$ is a parameter inversely proportional to the number density of polymer chains. 
At small to intermediate $\tilde{r}$, Eq. (\ref{polyethyl}) resembles our Gaussian pair statistics.
Interestingly, it has been suggested that hyperuniform pair statistics with soft-core interactions can be achieved by star or bottlebrush polymers \cite{Ch18e}.
Note that the effective potentials of many of our hyperuniform models\footnote[10]{These include Gaussian pair statistics, OCP and its Fourier dual and the hyperbolic secant $g_2(r)$.} contain a Gaussian-damped logarithmic term for their short-ranged parts; see the Appendix for the precise functional forms.
Such logarithm-Gaussian pair potential is commonly used to describe of intermolecular forces of star polymers \cite{Ar02}. 
We also find that the oscillatory soft potentials in the cases of polynomial-damped pair statistics can be achieved by ``sticky'' macromolecules such as DNA-coated nanoparticles \cite{Ml13}, and that the combination of hard core and soft repulsion in the interactions for the ghost RSA packings can be found in colloidal suspensions of highly charged nanoparticles \cite{Li04}. 

The excellent agreement between the designed and simulated pair statistics in all cases studied in this work demonstrates the power of our inverse procedure to tackle realizability problems.
It can be applied to study many prescribed pair functions at different densities and temperatures other than those explicitly shown in this work, provided that $g_2(r)$ and $S(k)$ at least meet necessary nonnegativity conditions.
For example, we have shown that antihyperuniform critical-point systems can be designed that do not belong to the Ising universality class.
Remarkably, 1D critical-point states, which are impossible in 1D models with Ising-like short-ranged interactions, can be achieved as a result of the long-ranged pair potentials determined via our inverse procedure.
Knowledge of the critical point is essential in the construction of phase diagrams fluids \cite{Fi67}.
Previous studies on critical points typically utilize the direct approach of statistical mechanics, i.e., simulations with known potentials. 
Our inverse approach to generate antihyperuniform systems of different universality classes can significantly accelerate the exploration of the phase diagram and the fabrication of systems with fractal-like structures.
One can also systematically study the realizability of $g_2(r)$ with different asymptotic decay rates at large $r$, including power-law, exponential and super-exponential rates.
It is known that the decay behavior of $g_2(r)$ for a fluid determines the Fisher-Widom line that separates liquid-like and gas-like behaviors in the supercritical region \cite{Wi65, Fi69}.
Thus, $g_2(r)$ with prescribed decay rates may enable one to create many-particle systems with desired phase behaviors.

Another important future project is to target disordered hyperuniform structure factors with $\alpha \geq 3$ for their potential optical applications \cite{To21a}.
Furthermore, we remark that the pair functions in Table \ref{tab:pairfuncs} can be used as basis functions to create new designs at unit density, i.e.,
\begin{equation}
    g^*_2(r) = \Sigma_{i=1}^n \lambda_i g_{2,i}(r), \quad 0\leq \lambda_i \leq 1, \Sigma_i^n \lambda_i = 1.
    \label{newg2}
\end{equation}
Equation (\ref{newg2}) is expected to be realizable because it is realizable in all of the extreme cases $\lambda_i = 1$. 
Finally, note that if a given $g_2(r)$ is realizable at $\rho = 1$, then it is also realizable at all lower densities, because the realizable density range for a given $g_2(r)$ is always a continuous interval \cite{St05}.

Because pair statistics based on our designed functions exhibit distinctly different translational order and self-diffusivity, they can have potential applications in the design of many-particle systems with tailored mass transport properties.
Importantly, while Fig. \ref{fig:tauAndD} shows discrete values of $\tau$ and $\mathcal{D}$ for the 3D models in Table \ref{tab:pairfuncs}, values between the plotted points can be achieved by linear combinations of our pair functions in the form of Eq. (\ref{newg2}), and their corresponding potentials can be readily determined via our inverse procedure.
Thus, one can flexibly design pair functions that yield a continuous spectrum of $\tau$ and $\mathcal{D}$ values.
In the contexts of diffusion-controlled reactions, the designed pair statistics would guide the fabrication of nanoparticles, micelles or vesicles loaded with reactants or catalysts to achieve desired reaction rates \cite{Wa62, Ol88}.
Moreover, since our procedure can realize a wide range of short- and long-range order, the equilibrium states can be modified into multi-phase media with tailored long-time diffusion spreadability \cite{Wa22a} for controlled drug delivery purposes \cite{La81}.

Our study also has important implications for the Zhang-Torquato conjecture, which states that any realizable set of pair statistics, whether from a nonequilibrium or equilibrium system, can be achieved by equilibrium systems involving up to two-body interactions \cite{Zh20}.
We have previously tested this conjecture by accurately determining effective potentials for various nonequilibrium pair statistics, including the unit-step function $g_2(r)$ from zero density to the dimension-dependent maximum realizable densities across the first three space dimensions \cite{Wa22b}, 3D ghost RSA \cite{Wa22b}, ``cloaked'' uniformly randomized lattices \cite{To22}, perfect glasses and critical absorbing states \cite{Wa23}.
The realizability of all the prescribed pair statistics in this study lends further support to this conjecture.
In the future, it is important to further test this conjecture with other nonequilbrium pair statistics, including those of quantum states \cite{To08a}. 

\section*{Supplementary Material}
The supplemental material provides a plot of the intermolecular force $d v(r)/dr$ corresponding to the effective pair potentials for our 3D hyperuniform models.

\section*{Acknowledgements}
This work is supported in part by the National Science Foundation CBET-2133179 and the U.S. Army Research Office 171 under Cooperative Agreement No. W911NF-22-2-0103.

\section*{Author Contributions}
S.T. conceived of the project and research questions; H. W. and S.T. performed the research and H.W. carried out all numerical calculations; and  H. W. and S.T.  wrote the paper.

\section*{Conflicting Interest}
The authors declare no competing interests.

\section*{Data Availability}

Analytical pair functions that are known to be realizable, including the ones in this and previous studies, are available in the PairFunctions repository, reference number [\citenum{Wa23b}].

\begin{widetext}
\section*{Appendix}
\renewcommand\thesubsection{\Alph{subsection}}
\setcounter{equation}{0}
\renewcommand\theequation{\thesubsection\arabic{equation}}
\subsection{Structure factors for antihyperuniform states}
\label{sec:ahuS}
\begin{itemize}

\item 1D case:
\begin{equation}
  S(k) = 1 + \phi A \sqrt{\frac{2 \pi}{kD}} \left(-2 \text{FresnelC}\left(\sqrt{\frac{2kD}{\pi }}\right) + 1\right) - 2\phi\frac{\sin (k)}{kD},
\end{equation}
where we use $\phi = 0.2, A = 0.1$ in this work.
\item 2D case
\begin{equation}
    S(k) = 1 - 16 A \phi  \, _1F_2\left(\frac{1}{4};1,\frac{5}{4};-\frac{k^2}{4}\right)+\frac{4 \sqrt{2} A \phi  \Gamma \left(\frac{1}{4}\right)}{\sqrt{k} \Gamma \left(\frac{3}{4}\right)}-\frac{8 \phi  J_1(k)}{k},
\end{equation}
where we use $\phi = 0.1, A = 0.3$.
\item 3D case:
 \begin{equation}
     S(k) = 1 + \frac{24 \sqrt{2 \pi } A \phi -48 \sqrt{2 \pi } A \phi  \text{FresnelC}\left(\sqrt{\frac{2kD}{\pi }}\right)}{\sqrt{kD}}+\frac{48 A \phi  \sin (kD)}{kD}-\frac{24 \phi  \sin (kD)}{(kD)^3}+\frac{24 \phi  \cos (kD)}{(kD)^2},
 \end{equation}
where we use $\phi = 0.1, A = 0.25$.
 \end{itemize}
 
\subsection{Effective potentials}
\setcounter{equation}{0}
\label{eff_pot}
\begin{figure*}
    \centering
    \subfloat[]{\label{gaussiana}\includegraphics[width = 45mm]{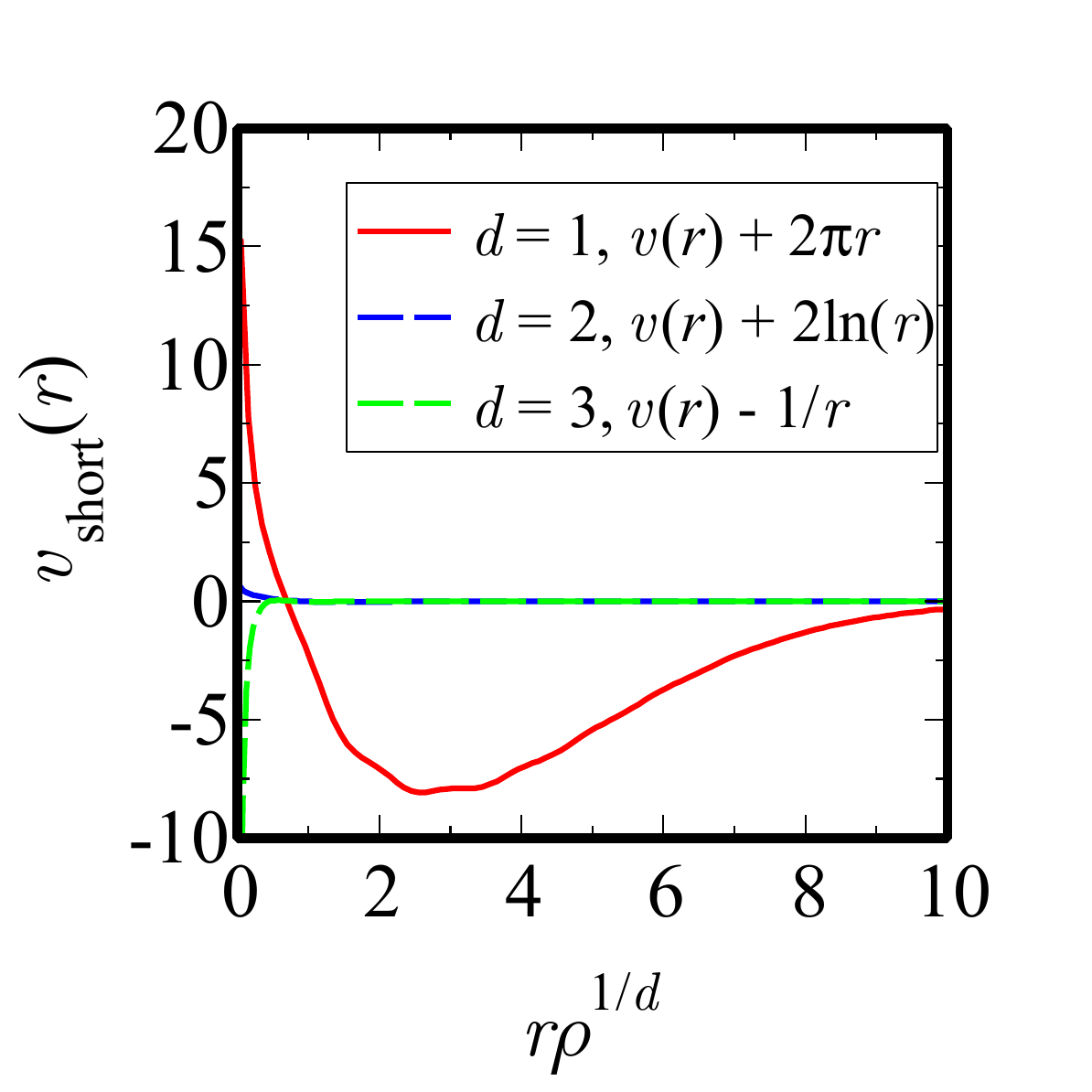}}
    \subfloat[]{\label{}\includegraphics[width = 45mm]{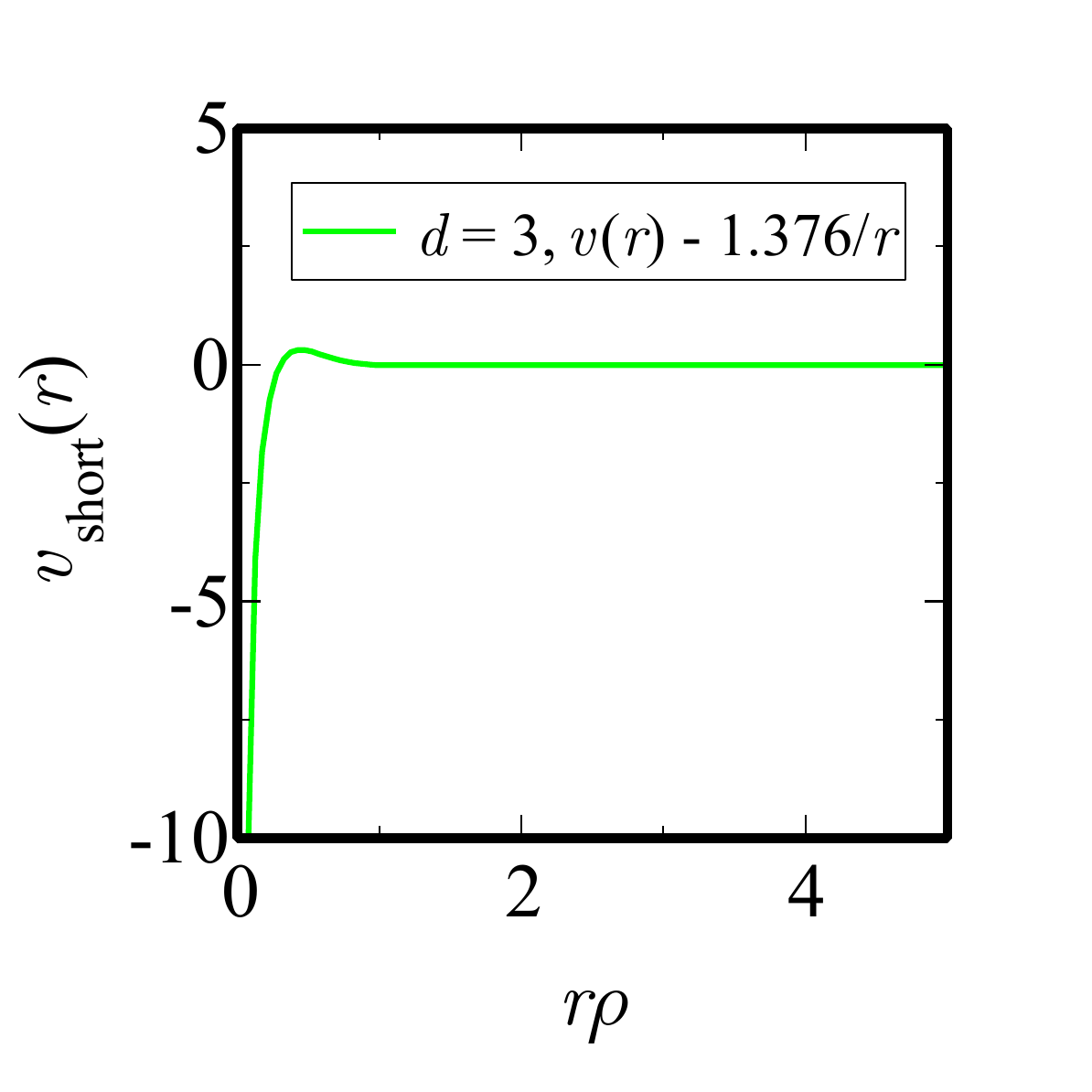}}
    \subfloat[]{\label{}\includegraphics[width = 45mm]{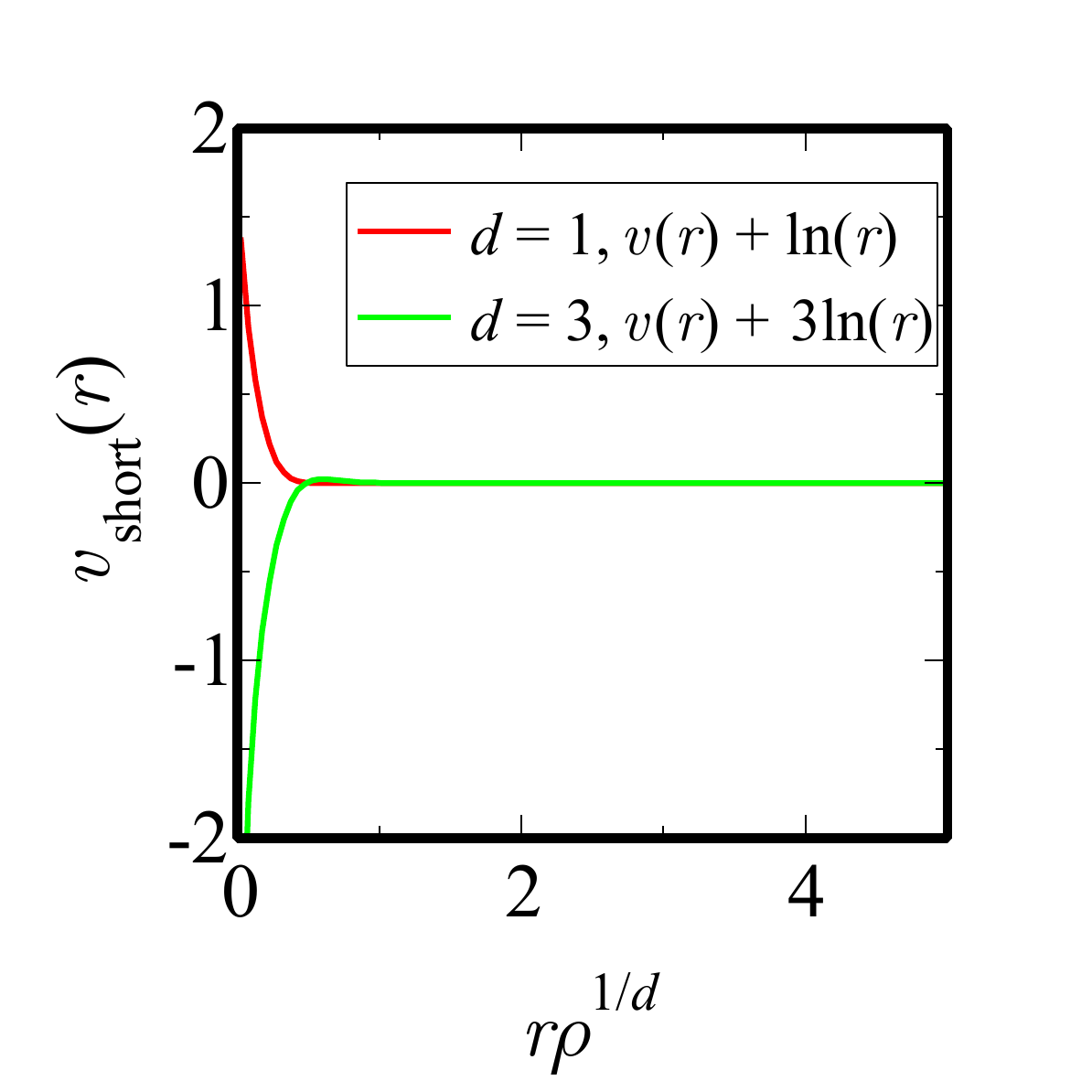}}
    
    \subfloat[]{\label{}\includegraphics[width = 45mm]{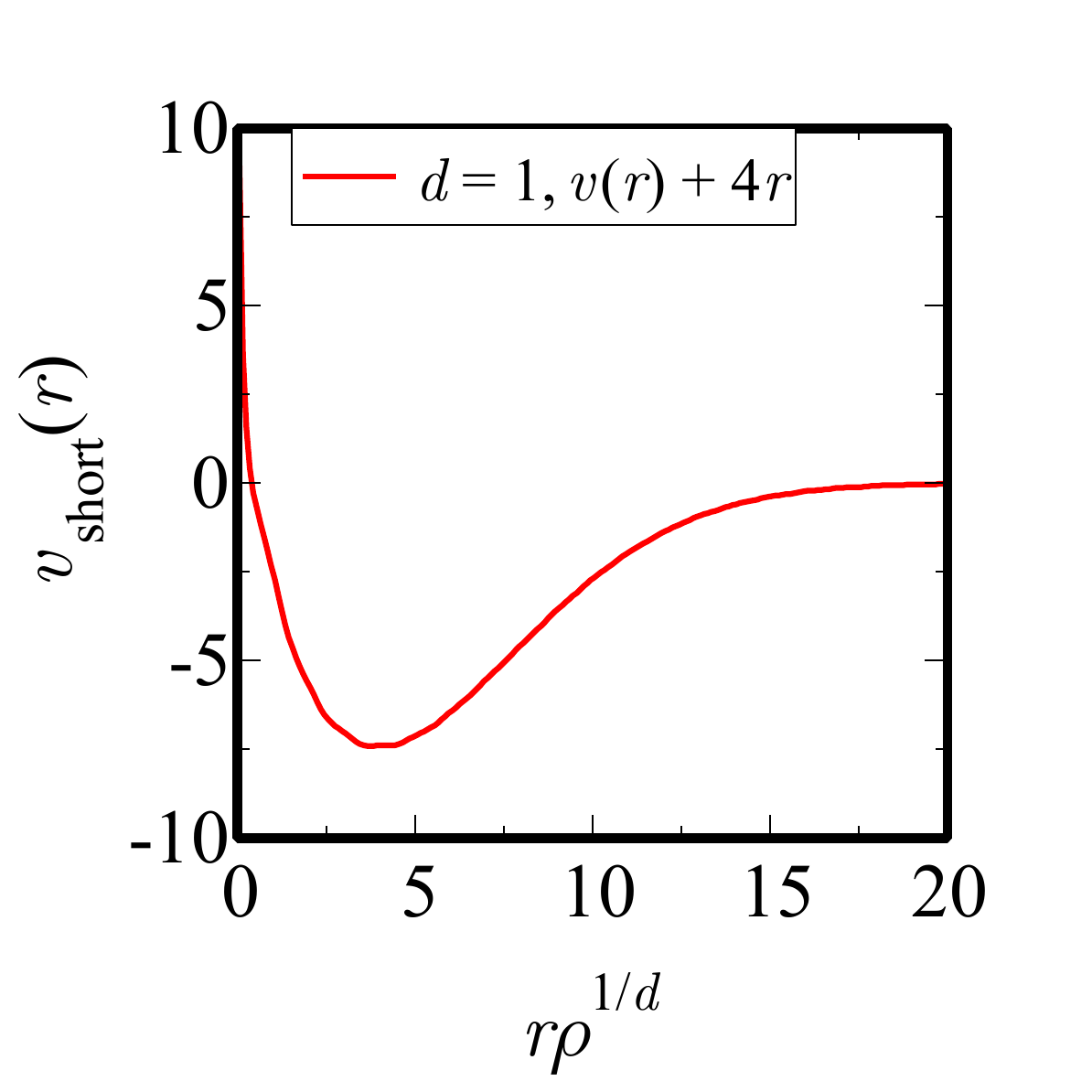}}
    \subfloat[]{\includegraphics[width = 45mm]{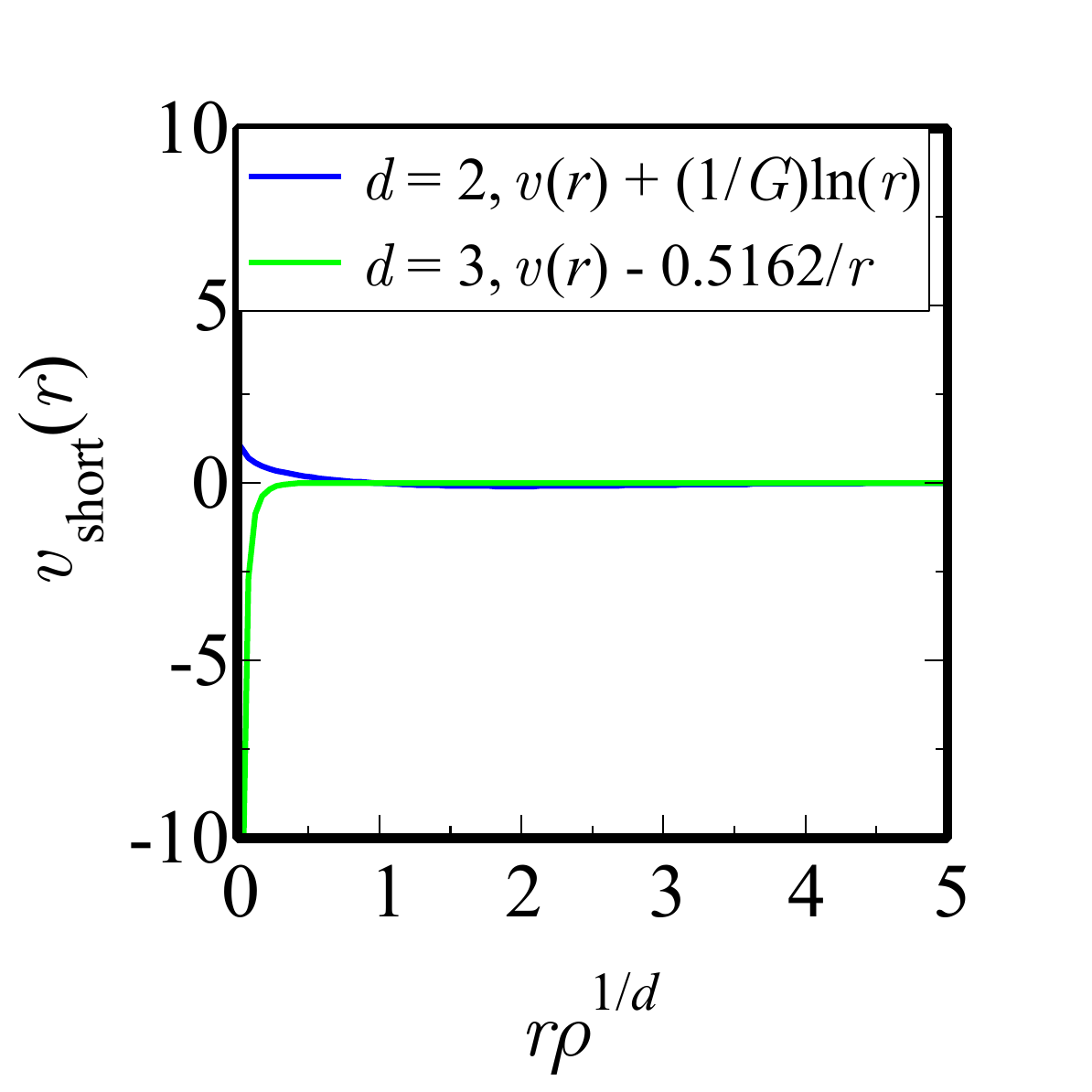}}
    \subfloat[]{\label{}\includegraphics[width = 45mm]{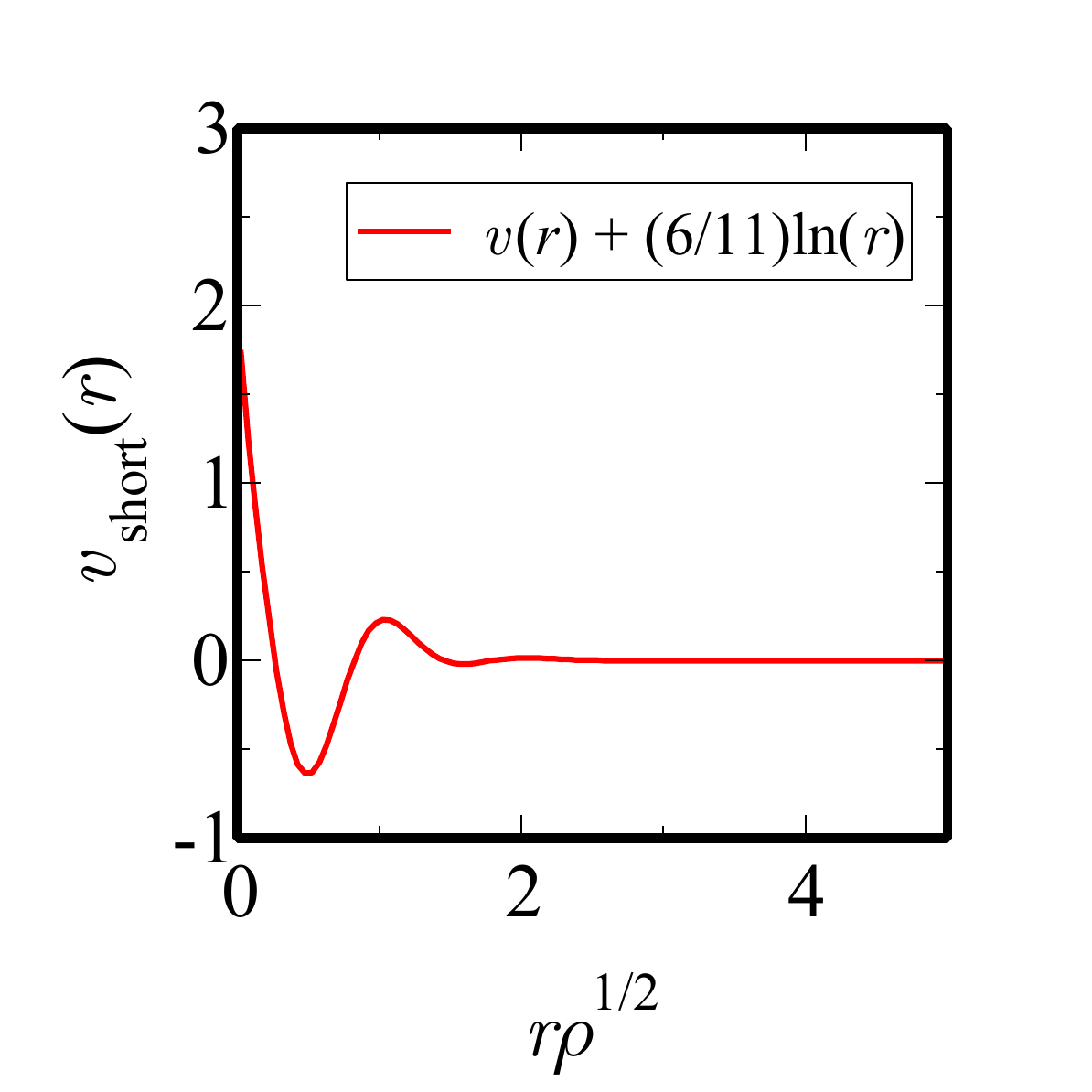}}
    
    \subfloat[]{\label{}\includegraphics[width = 45mm]{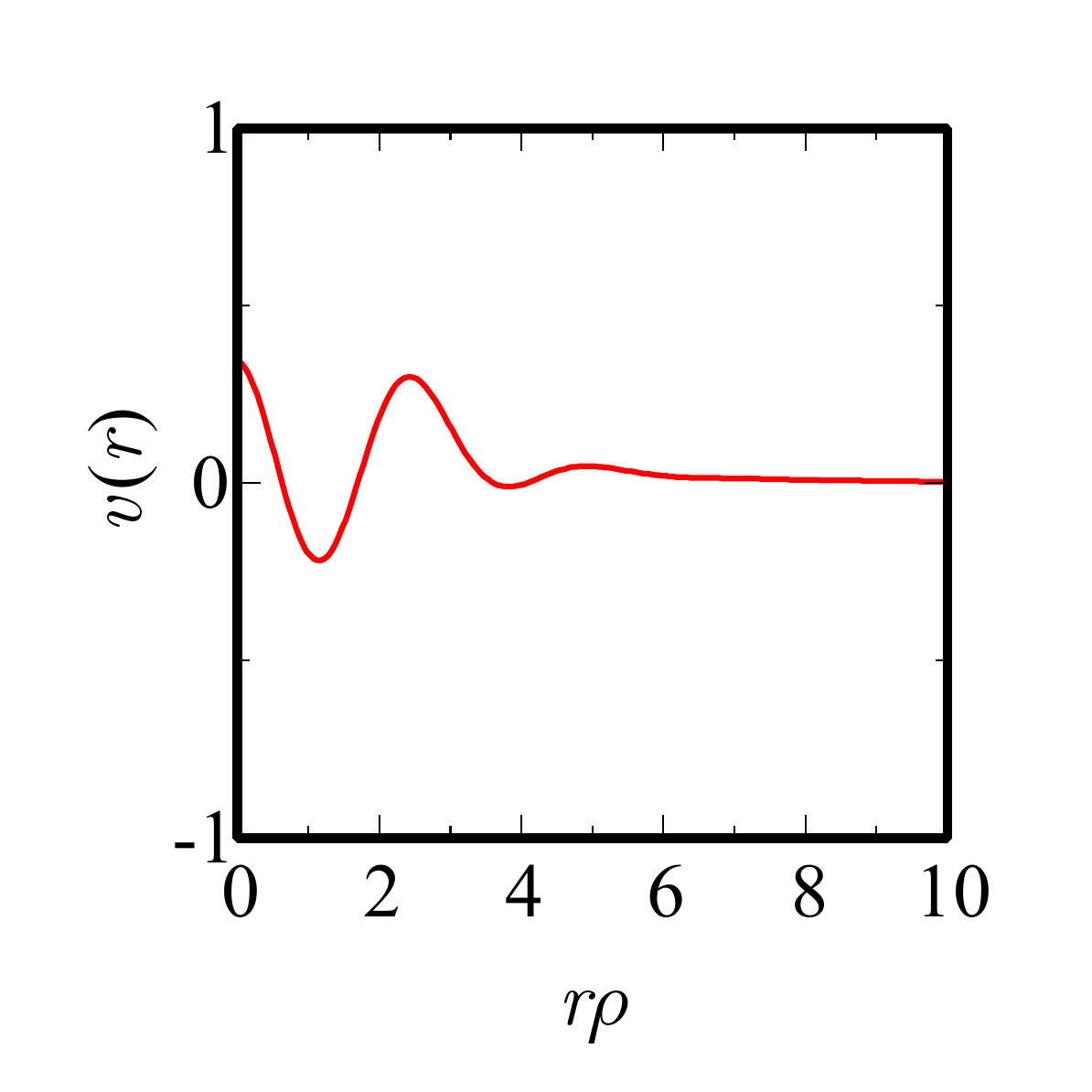}}
    \subfloat[]{\label{}\includegraphics[width = 45mm]{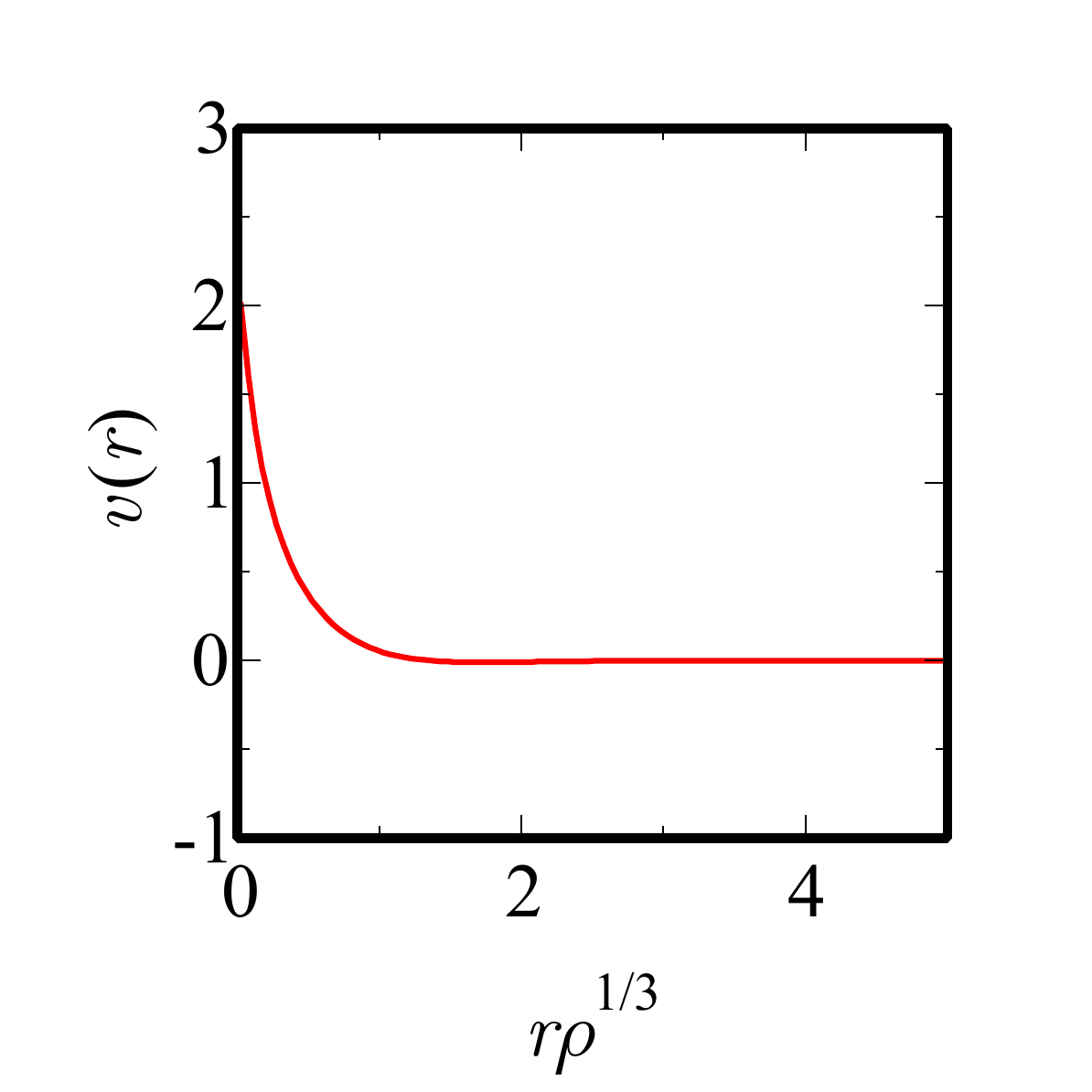}}
    \subfloat[]{\label{}\includegraphics[width = 45mm]{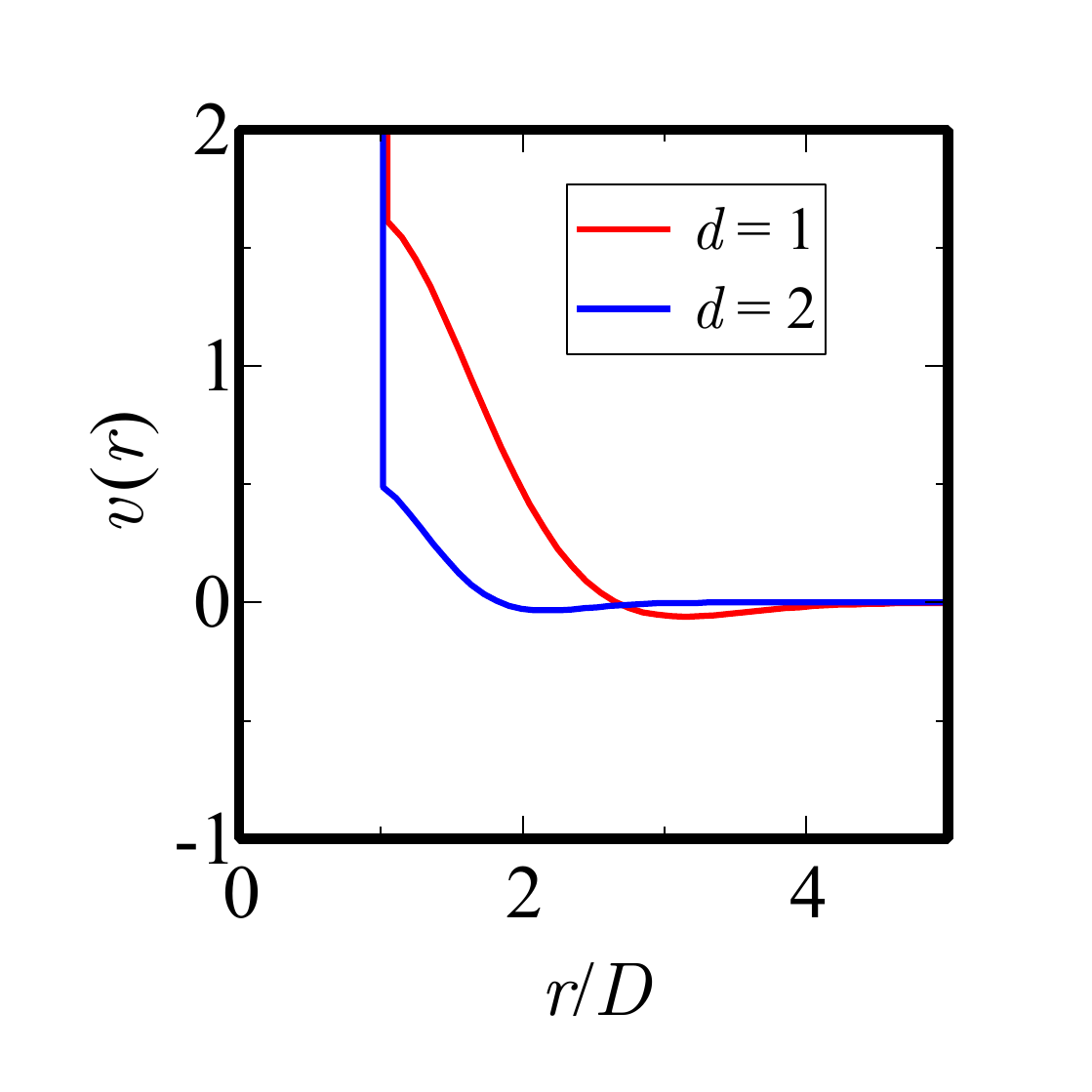}}
    
    \subfloat[]{\label{}\includegraphics[width = 45mm]{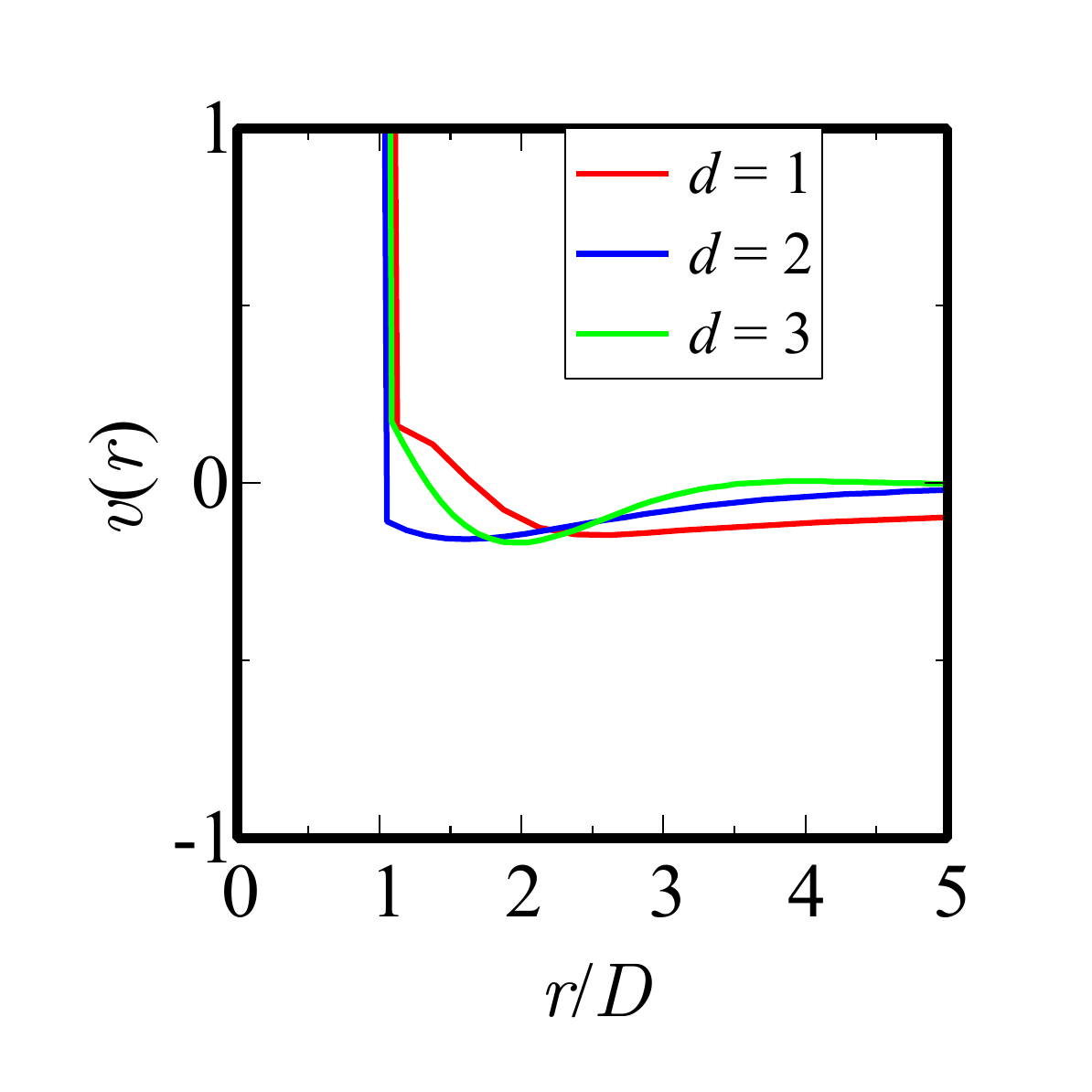}}
    \subfloat[]{\label{}\includegraphics[width = 45mm]{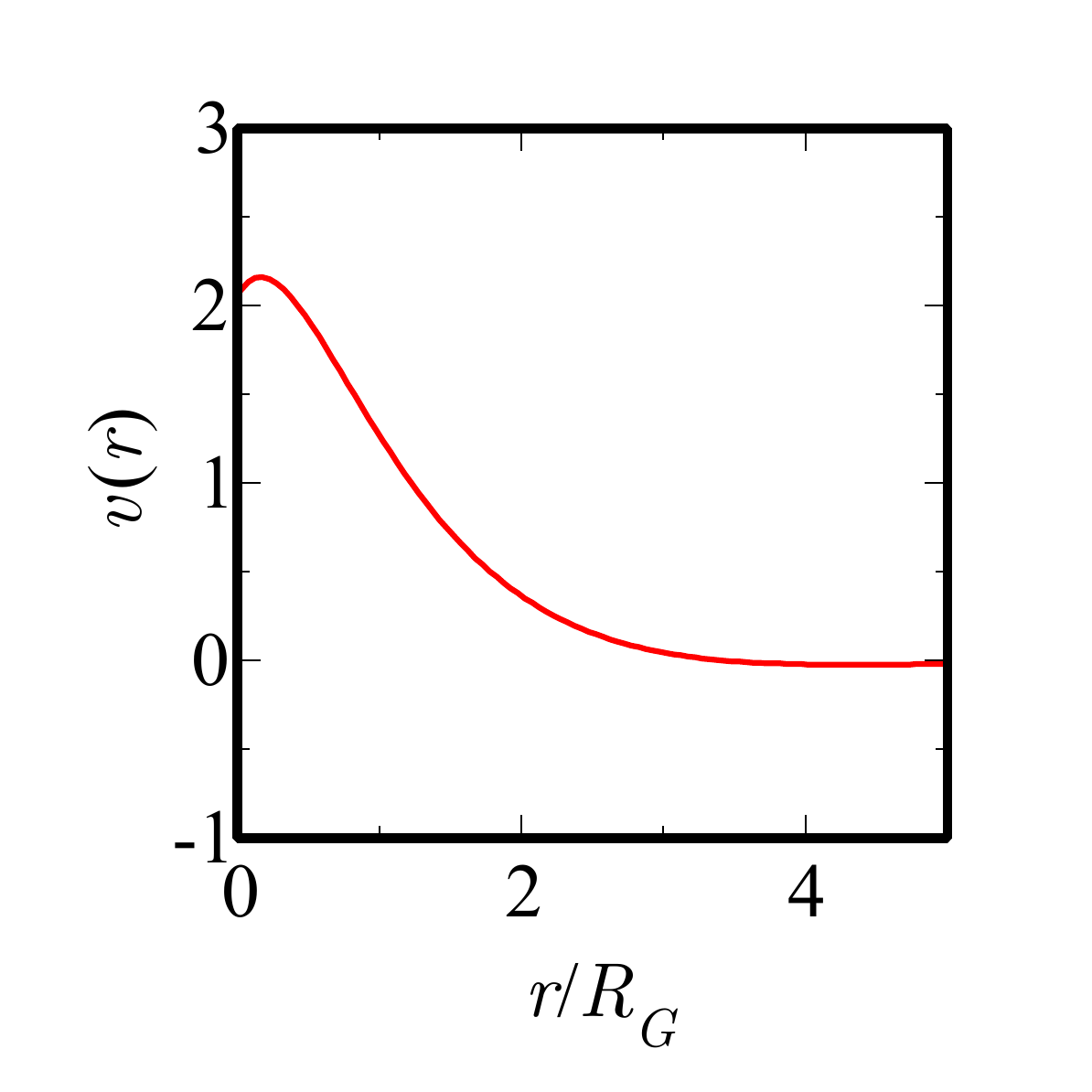}}
    \caption{(a) Short-ranged part of the effective pair potentials (\ref{1Dgaussian})--(\ref{3Dgaussian}) for the 1D, 2D and 3D Gaussian pair statistics. 
    (b)  Short-ranged part of the effective pair potential (\ref{3docpv}) for the 3D OCP pair statistics.
    (c) Short-ranged part of the effective potentials (\ref{dualocp1Dv})--(\ref{dualocp3Dv}) for 1D and 3D Fourier dual of OCP pair statistics.
    (d) Short-ranged part of the effective potential (\ref{1Dsechv}) for the 1D hyperbolic secant $g_2(r)$. 
    (e) Short-ranged part of the effective potentials, Eqs. (\ref{2dsechv}) and (\ref{3dsechv}), for 2D and 3D hyperbolic secant $g_2(r)$, respectively.
    (f) Short-ranged part of the effective potential (\ref{poly2D}) for the 2D Gaussian-damped polynomial pair statistics.
    (g) Effective potentials (\ref{1Dhermite}) for the 1D hyposurficial Hermite-Gaussian $g_2(r)$ with $\lambda = 1/20$.
    (h) Effective potential (\ref{3Dhypo}) for the 3D hyposurficial state.
    (i) Effective potentials (\ref{vghost1D})--(\ref{vghost2D}) for the 1D and 2D ``ghost'' RSA.
    (j) Effective potentials (\ref{ahu1D})--(\ref{ahu3D}) for 1D, 2D and 3D antihyperuniform states.
    (k) Effective potentials for a 3D polymer model with the total correlation function (\ref{polymerhr}) at $\rho = 0.5$.
    }
    \label{fig:potentials}
\end{figure*}
    
\begin{itemize}
    \item 1D Gaussian pair statistics:
\begin{equation}
    v(r) = -2\pi r - \varepsilon_1^{(1)}\exp(-(r/\sigma_1)^2)(\ln(r) + \varepsilon_1^{(2)}) + \sum_{j=2}^4 \varepsilon_j\cos(r/\sigma_j^{(1)} + \theta_j)\exp(-(r/\sigma_j^{(2)})^2),
    \label{1Dgaussian}
\end{equation}
where the optimized values of the parameters are given in Table \ref{tab:1Dgaussian}.

\begin{table}[htp]
    \centering
        \caption{Parameters of the effective pair potential (\ref{1Dgaussian}) for 1D Gaussian pair statistics}
    \begin{tabular}{||c|c||c|c||c|c||c|c||}
        $\varepsilon_1^{(1)}$ &  7.368 & $\varepsilon_2$ & 20.15 & $\varepsilon_3$ & -0.2078 & $\varepsilon_4$ & 0.02070\\
        $\sigma_1$ & 5.433 & $\sigma_2^{(1)}$ & 2.311 & $\sigma_3^{(1)}$ & 0.1564 & $\sigma_3^{(1)}$ & $1/(2\pi)$ \\
        $\varepsilon_1^{(2)}$ & 0.3824 & $\theta_2$ & 4.536 & $\theta_3$ & 2.911 & $\theta_4$ & 0 \\
        & & $\sigma_2^{(2)}$ & 0.7983 & $\sigma_3^{(2)}$ & 3.305 & $\sigma_4^{(2)}$ & 86.07 \\
    \end{tabular}
    \label{tab:1Dgaussian}
\end{table}

\item 2D Gaussian pair statistics:
\begin{equation}
    v(r) = -2\ln(r).
\end{equation}

\item 3D Gaussian pair statistics:
\begin{equation}
    v(r) = (1/r)(1 - \exp(-(r/\sigma_1)^2)) - \varepsilon_2\exp(-(r/\sigma_2)^2)\ln(r), 
    \label{3Dgaussian}
\end{equation}
where $\sigma_1 = 0.4672, \varepsilon_2^{(1)} = 1.877, \sigma_2 = 0.6176$.

\item 1D OCP pair statistics:
\begin{equation}
    v(r) = -2r - \varepsilon_1^{(1)}\exp(-(r/\sigma_1)^2)(\ln(r) + \varepsilon_1^{(2)}) + \sum_{j=2}^3 \varepsilon_j\cos(r/\sigma_j^{(1)} + \theta_j)\exp(-(r/\sigma_j^{(2)})^2),
    \label{1Docpv}
\end{equation}
where the optimized values of the parameters are given in Table \ref{tab:1Docpv}.

\begin{table}[htp]
    \centering
        \caption{Parameters of the effective pair potential (\ref{1Docpv}) for the 1D OCP pair statistics.}
    \begin{tabular}{||c|c||c|c||c|c||}
        $\varepsilon_1^{(1)}$ &  1.727 & $\varepsilon_2$ & 35.72 & $\varepsilon_3$ & -8.917 \\
        $\sigma_1$ & 14.22 & $\sigma_2^{(1)}$ & 6.501 & $\sigma_3^{(1)}$ & 1.915 \\
        $\varepsilon_1^{(2)}$ & 1.947 & $\theta_2$ & 4.438 & $\theta_3$ & 3.520 \\
        & & $\sigma_2^{(2)}$ & 1.560 & $\sigma_3^{(2)}$ & 1.393 \\
    \end{tabular}
    \label{tab:1Docpv}
\end{table}

\item 3D OCP pair statistics:
\begin{equation}
    v(r) = \frac{9}{2}\frac{3^{5/6} \Gamma \left(\frac{4}{3}\right)}{2^{2/3} \pi ^{4/3}r}(1 - \exp(-(r/\sigma_2)^2)) - \varepsilon_2\exp(-(r/\sigma_1)^2)\ln(r),
    \label{3docpv}
\end{equation}
where $\sigma_1 =  0.4149$, $\varepsilon_2 = 3.008, \sigma_2 = 0.5624$.

\item Fourier dual of 1D OCP pair statistics:
\begin{equation}
    v(r) = -\ln(r)(1 - \exp(-(r/\sigma_1)^2)) - \varepsilon_2\exp(-(r/\sigma_2)^2)\ln(r),
    \label{dualocp1Dv}
\end{equation}
where $\sigma_1 = 0.2331, \varepsilon_2 = 1.375, \sigma_2 = 0.2331$.

\item Fourier dual of 3D OCP pair statistics:
\begin{equation}
    v(r) = -3\ln(r) + \ln(r/\sigma_1^{(1)})\exp(-(r/\sigma_1^{(2)})^2),
    \label{dualocp3Dv}
\end{equation}
where $\sigma_1^{(1)} = 0.4791$, $\sigma_1^{(2)} = 0.4038$.

\item 1D hyperbolic secant function $g_2(r)$:
\begin{equation}
    v(r) = -4r - \varepsilon_1^{(1)}\exp(-(r/\sigma_1)^2)(\ln(r) + \varepsilon_1^{(2)}) + \sum_{j=2}^3 \varepsilon_j\cos(r/\sigma_j^{(1)} + \theta_j)\exp(-(r/\sigma_j^{(2)})^2),
    \label{1Dsechv}
\end{equation}
where the optimized values of the parameters are given in Table \ref{tab:1Dsechv}.

\begin{table}[htp]
    \centering
        \caption{Parameters of the effective pair potential (\ref{1Dsechv}) for the 1D hyperbolic secant function $g_2(r)$}
    \begin{tabular}{||c|c||c|c||c|c||}
        $\varepsilon_1^{(1)}$ &  5.172 & $\varepsilon_2$ & 22.28 & $\varepsilon_3$ & -0.06283 \\
        $\sigma_1$ & 7.728 & $\sigma_2^{(1)}$ & 8.071 & $\sigma_3^{(1)}$ & 0.1610 \\
        $\varepsilon_1^{(2)}$ & 0.4883 & $\theta_2$ & 4.517 & $\theta_3$ & 3.241 \\
        & & $\sigma_2^{(2)}$ & 0.6085 & $\sigma_3^{(2)}$ & 4.170 \\
    \end{tabular}
    \label{tab:1Dsechv}
\end{table}

\item 2D hyperbolic secant function $g_2(r)$:
\begin{equation}
    v(r) = -\frac{1}{G}\ln(r)(1 - \exp(-(\frac{r}{\sigma_1})^2)) - \varepsilon_2\exp(-(\frac{r}{\sigma_2})^2)\ln(r),
    \label{2dsechv}
\end{equation}
where $\sigma_1 = 2.269, \varepsilon_2 = 1.367, \sigma_2 = 2.269$.

\item 3D hyperbolic secant function $g_2(r)$:
\begin{equation}
    v(r) = \frac{3 \sqrt[3]{\frac{2}{\pi }}}{5r}(1 - \exp(-(r/\sigma_1)^2)) - \varepsilon_2\exp(-(r/\sigma_2)^2)\ln(r),
    \label{3dsechv}
\end{equation}
where $\sigma_1 = 0.1557, \varepsilon_2 = 1.922, \sigma_2 = 0.1174$.

\item 2D Gaussian-damped polynomial pair statistics:
\begin{equation}
    v(r) = -\frac{6}{11} \ln (r) \left(1 - e^{-\left(\frac{r}{\sigma_1}\right)^2}\right) 
    - \varepsilon_2 \ln (r) e^{-\left(\frac{r}{\sigma_2}\right)^2} 
    + \varepsilon_3 \cos \left(\frac{r}{\sigma_3^{(1)}}+\theta\right)e^{-\left(\frac{r}{\sigma_3^{(2)}}\right)^2},
    \label{poly2D}
\end{equation}
where the optimized values of the parameters are given in Table \ref{tab:poly2D}.

\begin{table}[htp]
    \centering
        \caption{Parameters of the effective pair potential (\ref{poly2D}) for the 2D Gaussian-damped polynomial pair statistics.}
    \begin{tabular}{||c|c||c|c||}
        $\sigma_1$ & 1.081 & $\sigma_3^{(1)}$ & 0.1843 \\
        $\varepsilon_2$ & 0.8532 & $\theta$ & 0.3920 \\
        $\sigma_2$ & 0.4884 & $\sigma_3^{(2)}$ & 0.9862\\
        $\varepsilon_3$ & 0.7009 & & \\
    \end{tabular}
    \label{tab:poly2D}
\end{table}

\item 1D Hermite-Gaussian pair statistics with $\lambda = 1/20$:
\begin{equation}
    v(r) = \sum_{j = 1}^2 \varepsilon_j\cos(r/\sigma_j^{(1)} + \theta_j)\exp(-r/\sigma_j^{(2)}) + \varepsilon_3\exp(-r/\sigma_3),
    \label{1Dhermite}
\end{equation}
where the optimized values of the parameters are given in Table \ref{tab:1Dhermite}.
\begin{table}[htp]
    \centering
        \caption{Parameters of the effective pair potential (\ref{1Dhermite}) for the 1D Hermite-Gaussian pair statistics with $\lambda = 1/20$.}
    \begin{tabular}{||c|c||c|c||c|c||}
        $\varepsilon_1$ & 0.4143 & $\varepsilon_2$ & 3.174 & $\varepsilon_3$ & 0.2663 \\
        $\sigma_1^{(1)}$ & 0.4250 & $\sigma_2^{(1)}$ & 14.24 & $\sigma_3$ & 2.392\\
        $\theta_1$ & 0.2047 & $\theta_2$ & 4.609 & & \\
        $\sigma_1^{(2)}$ & 2.829 & $\sigma_2^{(2)}$ & 1.544 & & \\
    \end{tabular}
    \label{tab:1Dhermite}
\end{table}

\item 3D hyposurficial state:
\begin{equation}
    v(r) = \sum_{j = 1}^2 \varepsilon_j\exp(-r/\sigma_j) + \varepsilon_3\exp(-((r - \sigma_3^{(1)})/\sigma_3^{(2)})^2),
    \label{3Dhypo}
\end{equation}
where the optimized values of the parameters are given in Table \ref{tab:3Dhypo}.
\begin{table}[htp]
    \centering
        \caption{Parameters of the effective pair potential (\ref{3Dhypo}) for the 3D hyposurficial state.}
    \begin{tabular}{||c|c||c|c||}
        $\varepsilon_1$ & 1.646 & $\varepsilon_3$ & -0.1241 \\
        $\sigma_1$ & 0.4070 & $\sigma_3^{(1)}$ & 0.2020 \\
        $\varepsilon_2$ & 0.8742 & $\sigma_3^{(2)}$ & 1.312\\
        $\sigma_2$ & 0.8884 & & \\
    \end{tabular}
    \label{tab:3Dhypo}
\end{table}

\item 1D ghost RSA:
\begin{equation}
v(r) = 
    \begin{cases}
        \infty, \quad r < 0.5\\
        \varepsilon_1 \exp(-(r/\sigma_1^{(1)})^2)\cos(r/\sigma_1^{(2)} + \theta), \quad r\geq 0.5
    \end{cases}
    \label{vghost1D}
\end{equation}
where $\varepsilon_1 = -2.341, \sigma_1^{(1)} = 0.9045, \sigma_1^{(2)} = -0.4406, \theta = -1.686$.

\item 2D ghost RSA:
\begin{equation}
v(r) = 
    \begin{cases}
        \infty, \quad r < 1/\sqrt{\pi}\\
        \varepsilon_1 \exp(-(r/\sigma_1^{(1)})^2)\cos(r/\sigma_1^{(2)} + \theta), \quad r\geq 1/\sqrt{\pi}
    \end{cases}
    \label{vghost2D}
\end{equation}
where $\varepsilon_1 = 0.8107, \sigma_1^{(1)} = 0.7860, \sigma_1^{(2)} = 0.2867, \theta = -1.982$.

\item 1D antihyperuniform state:
\begin{equation}
    v(r) = 
    \begin{cases}
        \infty, \quad r < D\\
        -\varepsilon_1/\sqrt{r/D} + \varepsilon_2 \exp(-(r/D - 1)/\sigma_2) +  \varepsilon_3 \exp(-((r/D - 1)/\sigma_3)^2), \quad r> D,
    \end{cases}
    \label{ahu1D}
\end{equation}
where $\varepsilon_1 = 0.3311, \varepsilon_2 = 0.0552, \sigma_2 = 70.21, \varepsilon_3 = 0.4299, \sigma_3 = 0.7560$.

\item 2D antihyperuniform state:
\begin{equation}
    v(r) = 
    \begin{cases}
        \infty, \quad r < D\\
        -\varepsilon_1/(r/D)^{3/2} + \varepsilon_2 \exp(-(r/D - 1)/\sigma_2^{(1)})\cos(r/(D\sigma_2^{(2)}) + \theta), \quad r> D,
    \end{cases}
    \label{ahu2D}
\end{equation}
where $\varepsilon_1 = 0.1483, \varepsilon_2 = 1.647, \sigma_2 = 0.7530, \sigma_2^{(2)} = -4.120, \theta = 4.987$.

\item 3D antihyperuniform state:
\begin{equation}
    v(r) = 
    \begin{cases}
        \infty, \quad r < D\\
        -\varepsilon_1/(r/D)^{5/2} + \varepsilon_2 \exp(-(r/D - 1)/\sigma_2^{(1)})\cos(r/(D\sigma_2^{(2)}) + \theta), \quad r> D,
    \end{cases}
    \label{ahu3D}
\end{equation}
where $\varepsilon_1 = 0.2100, \varepsilon_2 = 0.6116, \sigma_2 = 0.7927, \sigma_2^{(2)} = -0.5982, \theta = 7.203$.

\item 3D hypothetical polymer motivated by the PRISM model:
\begin{equation}
v(r) = \varepsilon_1 \exp(-(r/\sigma_1^{(1)}))\cos(r/\sigma_1^{(2)} + \theta) + \varepsilon_2 \exp(-r/\sigma_2),
\label{vguenza}
\end{equation}
where $\varepsilon_1 = 3.635, \sigma_1^{(1)} = 0.9891, \sigma_1^{(2)} = 1.663, \theta = -0.4998, \varepsilon_2 = -1.133, \sigma_2 = 0.3150$.
\end{itemize}
\end{widetext}

\clearpage

\begin{thebibliography}{115}
\bibitem [{pfq()}]{pfq}%
  \BibitemOpen
  \href@noop {} {}\bibinfo {note} {Here $_pF_q(a_1, \dots, a_p; b_1, \dots,
  b_q; z)$ is the generalized hypergeometric function and
  $\operatorname{ber}_v(z)$ is the Kelvin function.}\BibitemShut {Stop}%
\bibitem [{cat()}]{catalan}%
  \BibitemOpen
  \href@noop {} {}\bibinfo {note} {Here $G = 0.915\dots$ is the Catalan's
  constant.}\BibitemShut {Stop}%
\bibitem [{lam()}]{lambda}%
  \BibitemOpen
  \href@noop {} {}\bibinfo {note} {Here $\lambda$ is a parameter controlling
  the amplitude of oscillations in the pair statistics. The condition that
  $S(k) \geq 0$ for all $k$ implies that $\lambda \leq \exp(\sqrt{11/2}/2 +
  7/4)/(8 \left(\sqrt{22}+4\right) \sqrt{2 \pi }) = 0.1067\dots$.}\BibitemShut
  {Stop}%
\bibitem [{cis()}]{cisi}%
  \BibitemOpen
  \href@noop {} {}\bibinfo {note} {Here Ci is the cosine integral and Si is the
  sine integral.}\BibitemShut {Stop}%
\bibitem [{big()}]{bigd}%
  \BibitemOpen
  \href@noop {} {}\bibinfo {note} {Here $D$ is the hard-core diameter and $A +
  1$ is the peak height of $g_2(r)$ as $r \rightarrow D^+$.}\BibitemShut
  {Stop}%
\end{thebibliography}
%

\end{document}